\documentclass[nohypdvips,onefignum,onetabnum]{siamart171218}

\usepackage{amssymb}
\usepackage{amsmath}
\usepackage{lipsum}
\usepackage{amsfonts}
\usepackage{graphicx}
\usepackage{commath}
\usepackage{caption}
\usepackage{pifont}
\usepackage{epstopdf}
\usepackage{algorithm,algorithmicx,algpseudocode}
\usepackage{stmaryrd}
\usepackage{multirow}
\usepackage{upgreek}
\usepackage{url}
\usepackage{bbold}
\usepackage{afterpage}
\usepackage[figuresright]{rotating}
\usepackage{subfigure}
\usepackage{ifthen}
\usepackage{xspace}
\usepackage{tikz}
\usepackage{cite}  
\usepackage{latexsym}

\usetikzlibrary{matrix,decorations.pathmorphing,calc,decorations.pathreplacing,topaths,fit,positioning,automata,patterns,arrows,shapes,chains}
\urldef\urlnew\url{https://mvono.github.io}
\ifpdf
  \DeclareGraphicsExtensions{.eps,.pdf,.png,.jpg}
\else
  \DeclareGraphicsExtensions{.eps}
\fi

\usetikzlibrary{arrows.meta}
\tikzset{%
  >={Latex[width=2mm,length=2mm]},
            base/.style = {rectangle, rounded corners, draw=black,
                           minimum width=4cm, minimum height=1cm,
                           text centered, font=\sffamily},
  storage/.style = {base, fill=blue!30},
       startstop/.style = {base, fill=red!30},
    activityRuns/.style = {base, fill=green!30},
         process/.style = {base, minimum width=2.5cm, fill=orange!15,
                           font=\ttfamily},
}

\newcommand{\B}[1]{\mathbf{#1}} 
\newcommand{\Bs}[1]{\boldsymbol{#1}} 
\newcommand{\pr}[1]{\left(#1\right)} 
\newcommand{\br}[1]{\left[#1\right]} 
\newcommand{\bbr}[1]{\left\{#1\right\}} 
\newcommand{\nr}[1]{\left\|#1\right\|} 
\newcommand{\xmark}{\ding{55}}%
\def\eqsp{\;}

\makeatletter
\newcommand{\removelatexerror}{\let\@latex@error\@gobble}
\makeatother

\makeatletter
\newcommand{\thickhline}{%
    \noalign {\ifnum 0=`}\fi \hrule height 1pt
    \futurelet \reserved@a \@xhline
}

\definecolor{green}{rgb}{0.0, 0.5, 0.0}
\definecolor{Burgundy}{RGB}{144,0,32}

\usepackage{pgf}
\usepackage{tikz}
\usetikzlibrary{arrows,automata}
\usetikzlibrary{positioning}
\tikzset{
    state/.style={
           rectangle,
           rounded corners,
           draw=black, very thick,
           minimum height=2em,
           inner sep=2pt,
           text centered,
           },
}

\newcommand{\blockmatrix}[9]{
  \draw[draw=#4,fill=#5] (0,0) rectangle( #1,#2);
  \ifthenelse{\equal{#6}{true}}
  {
    \draw[draw=#7,fill=#8] (0,#2) -- (#9,#2) -- ( #1,#9) -- ( #1,0) -- ( #1 - #9,0) -- (0,#2 -#9) -- cycle;
  }
  {}
  \draw ( #1/2, #2/2) node { #3};
}

\newcommand{\mblockmatrix}[4][none]{
  \begin{tikzpicture} 
  \ifthenelse{\equal{#1}{none}}
  {
    \blockmatrix{#2}{#3}{#4}{none}{none}{false}{none}{none}{0.0}
  }
  {
    \definecolor{fillcolor}{rgb}{#1}
    \blockmatrix{#2}{#3}{#4}{none}{fillcolor}{false}{none}{none}{0.0}
  }
  \end{tikzpicture}
}

\newcommand{\fblockmatrix}[4][none]{
  \begin{tikzpicture} 
  \ifthenelse{\equal{#1}{none}}
  {
    \blockmatrix{#2}{#3}{#4}{black}{none}{false}{none}{none}{0.0}
  }
  {
    \definecolor{fillcolor}{rgb}{#1}
    \blockmatrix{#2}{#3}{#4}{black}{fillcolor}{false}{none}{none}{0.0}
  }
  \end{tikzpicture}
}

\newcommand{\dblockmatrix}[4][none]{
  \begin{tikzpicture} 
  \ifthenelse{\equal{#1}{none}}
  {
    \blockmatrix{#2}{#3}{#4}{black}{none}{true}{black}{none}{0.35cm}
  }
  {
    \definecolor{fillcolor}{rgb}{#1}
    \blockmatrix{#2}{#3}{#4}{black}{none}{true}{black}{fillcolor}{0.35cm}
  }
  \end{tikzpicture}
}

\newcommand{\diagonalblockmatrix}[5][none]{
  \begin{tikzpicture} 

  \ifthenelse{\equal{#1}{none}}
  {
    \blockmatrix{#2}{#3}{#4}{black}{none}{true}{black}{none}{#5}
  }
  {
    \definecolor{fillcolor}{rgb}{#1}
    \blockmatrix{#2}{#3}{#4}{black}{none}{true}{black}{fillcolor}{#5}
  }

  \end{tikzpicture}
}

\newenvironment{blockmatrixtabular}
{
  \begin{tabular}{
  @{}l@{}l@{}l@{}l@{}l@{}l@{}l@{}l@{}l@{}l@{}l@{}l@{}l@{}l@{}l@{}l@{}l@{}l@{}l
  @{}l@{}l@{}l@{}l@{}l@{}l@{}l@{}l@{}l@{}l@{}l@{}l@{}l@{}l@{}l@{}l@{}l@{}l@{}l
  @{}l@{}l@{}l@{}l@{}l@{}l@{}l@{}l@{}l@{}l@{}l@{}l@{}l@{}l@{}l@{}l@{}l@{}l@{}l
  @{}
  }
}
{
  \end{tabular}
}

\newsiamremark{remark}{Remark}
\newsiamremark{hypothesis}{Hypothesis}
\crefname{hypothesis}{Hypothesis}{Hypotheses}
\newsiamthm{claim}{Claim}
\newsiamremark{example}{Example}

\headers{High-dimensional Gaussian sampling: a review...}{M. Vono, N. Dobigeon, and P. Chainais}

\title{High-dimensional Gaussian sampling: a review \\and a unifying approach based on a stochastic proximal point algorithm}

\author{Maxime Vono\thanks{Lagrange Mathematics and Computing Research Center, Huawei, 75007 Paris, France 
  (\email{maxime.vono@gmail.com}, \urlnew).}
  \and Nicolas Dobigeon\thanks{Univ. of Toulouse, IRIT/INP-ENSEEIHT, Toulouse, France and Institut Universitaire de France (IUF), France
  (\email{nicolas.dobigeon@enseeiht.fr}).}
\and Pierre Chainais\thanks{Univ. Lille, CNRS, Centrale Lille, UMR 9189 CRIStAL, F-59000 Lille, France 
  (\email{pierre.chainais@centralelille.fr}).}}

\usepackage{amsopn}

\usepackage{color}

\ifpdf
\hypersetup{
  pdftitle={HIGH-DIMENSIONAL GAUSSIAN SAMPLING: A REVIEW AND A UNIFYING APPROACH BASED ON A STOCHASTIC PROXIMAL POINT ALGORITHM},
  pdfauthor={M. Vono, N. Dobigeon, P. Chainais}
}
\fi

\begin{document}

\maketitle

\begin{abstract}
  Efficient sampling from a high-dimensional Gaussian distribution is an old but high-stake issue.
  Vanilla Cholesky samplers imply a computational cost and memory requirements which can rapidly become prohibitive in high dimension. 
  To tackle these issues, multiple methods have been proposed from different communities ranging from iterative numerical linear algebra to Markov chain Monte Carlo (MCMC) approaches.
  Surprisingly, no complete review and comparison of these methods have been conducted.
  This paper aims at reviewing all these approaches by pointing out their differences, close relations, benefits and limitations.
  In addition to this state of the art, this paper proposes a unifying Gaussian simulation framework by deriving a stochastic counterpart of the celebrated proximal point algorithm in optimization.
  This framework offers a novel and unifying revisit of most of the existing MCMC approaches while extending them.
  Guidelines to choose the appropriate Gaussian simulation method for a given sampling problem in high dimension are proposed and illustrated with numerical examples.
\end{abstract}



\section{Introduction}
\label{sec:intro}

If there was only one continuous probability distribution to know, it would certainly be the Gaussian (also known as \textit{normal}) distribution.
Many nice properties of the Gaussian distribution can be listed such as its infinite divisibility, maximum entropy property or its  description thanks to the use of the first two cumulants only (mean and variance).
However, its popularity and ubiquity certainly result from two essential properties, namely the central limit theorem and the statistical interpretation of ordinary least squares, which often motivate its use to describe random noises or residual errors in various applications (e.g., inverse problems in signal and image processing).
The first one originates from the gambling theory.
The binomial distribution, that models the probabilities of successes and failures after a given number of trials, was approximated by a Gaussian distribution in the seminal work by de Moivre \cite{Moivre1718}.
This famous approximation is a specific instance of the central limit theorem which states that the sum of a sufficiently large number of independent and identically distributed (i.i.d.) random variables with finite variance converges in distribution towards a Gaussian random variable.
Capitalizing on this theorem, a lot of complex random events have been approximated by using the Gaussian distribution, sometimes called the \emph{bell curve}.
Another well-known reason for using the Gaussian distribution has been the search for an error distribution in empirical sciences.
For instance, since the end of the 16th century, astronomers have been interested in data summaries to describe their observations.
They found that the estimate defined by the arithmetic mean of the observations was related to the resolution of a least-mean square problem under the assumption of Gaussian measurement errors \cite{Havil2003}. The assumption of Gaussian noise has now become so usual that it is sometimes implicit in many applications.

Motivated by all these features, the Gaussian distribution is omnipresent in problems far beyond the statistics community itself.
In statistical machine learning and signal processing, Gaussian posterior distributions commonly appear when  hierarchical Bayesian models are derived \cite{Park2008,Polson2013,Gilavert2015,Marnissi2018}, in particular when the exponential family is involved.
As archetypal examples, models based on Gaussian Markov random fields or conditional auto-regressions assume that parameters of interest (associated to observations) come from a joint Gaussian distribution with a structured covariance matrix reflecting their interactions \cite{Rue2005}.
Such models have found applications in spatial statistics \cite{Cressie1993,Besag1995}, image analysis \cite{Geman1984,Kittler1984}, graphical structures \cite{Giudici1999} or semi-parametric regression and splines \cite{Fahrmeir2001}.
We can also cite the discretization schemes of stochastic differential equations involving Brownian motions which led to a Gaussian sampling step \cite{Duane1987,Roberts1996,Welling2011}, texture synthesis \cite{Galerne2011} and time series prediction \cite{Brahim2004}. Indeed, the Gaussian distribution is also intimately connected to diffusion processes and statistical physics.

When the dimension of the problem is small or moderate, sampling from this distribution is an old solved problem that raises no particular difficulty.
In high-dimensional settings this multivariate sampling task can become computationally demanding, which may prevent us from using statistically sound methods for real-world applications.
Therefore, even recently, a host of works have focused on the derivation of \textit{efficient} high-dimensional Gaussian sampling methods. Before summarizing our contributions and main results, in what follows, we discuss what we mean by \emph{complexity} and \emph{efficiency}, in light of the most common sampling technique, i.e., exploiting the Cholesky factorization. \\

\noindent\textbf{Computational and storage complexities: notations.}
In the following, we will use the mathematical notations $\Theta(\cdot)$ and $\mathcal{O}(\cdot)$ to refer to the complexities of the computation and the storage required by the sampling algorithms. We recall that $f(d) = \mathcal{O}(g(d))$ if there exists $c > 0$ such that $f(d) \le c g(d)$ when $d \rightarrow \infty$. We use $f(d) = \Theta(g(d))$ if there exist $c_1,c_2 > 0$ such that $c_1 g(d) \le f(d) \le c_2 g(d)$ when $d \rightarrow \infty$. 

\begin{table}[H]
{\footnotesize
  \caption{Vanilla Cholesky sampling. Computational and storage requirements to produce one sample from an arbitrary $d$-dimensional Gaussian distribution.}
  \label{table:complexity}
  \begin{center}
  {\renewcommand{\arraystretch}{1.5}
      \begin{tabular}{|c|c|c|}
    \hline
    \multirow{2}{*}{$d$} &
      \multicolumn{2}{c|}{Vanilla Cholesky sampler} \\
      \cline{2-3}
    & $\Theta(d^3)$ flops & $\Theta(d^2)$ memory requirement \\
    \hline
    $10^3$ & $3.34 \times 10^8$ & $4$ MB   \\
    $10^4$ & $3.33 \times 10^{11}$ & $0.4$ GB \\
    $10^5$ & $3.33 \times 10^{14}$ & $40$ GB  \\
    $10^6$ & $3.33 \times 10^{17}$ & $4$ TB  \\
    \hline
    \end{tabular}}
  \end{center}
}
\end{table}

\noindent\textbf{Algorithmic efficiency: definition.}
To sample from a given $d$-dimensional Gaussian distribution, the most common sampling algorithm is based on the Cholesky factorization \cite{Rue2001}.
Let us recall that the Cholesky factorization of a symmetric positive-definite matrix $\mathbf{Q} \in \mathbb{R}^{d \times d}$ is a decomposition of the form \cite[Section 2.4]{Rue2005},
\begin{equation}
    \label{choleskyfact}
    \mathbf{Q} =\mathbf{CC}^{\top},
\end{equation}
where $\B{C} \in \mathbb{R}^{d \times d}$ is a lower triangular matrix with real and positive diagonal entries.
The computation of the Cholesky factor $\B{C}$ requires $\Theta(d^3)$ floating point operations (flops), that is arithmetic operations such as additions, subtractions, multiplications or divisions \cite{Golub1989}, see also \cref{subsubsec:cholesky} for details.
In addition, the Cholesky factor which involves at most $d(d+1)/2$ non-zero entries must be stored.
In the general case, this implies a global memory requirement of $\Theta(d^2)$.
In high-dimensional settings ($d \gg 1$), both these numerical complexity and storage requirement rapidly become prohibitive for standard computers.
\Cref{table:complexity} illustrates this claim by showing the number of flops (using 64-bit numbers also called \emph{double precision}) and storage space required by the vanilla Cholesky sampler in high dimension.
Note that for $d \ge 10^5$, which is for instance classical in image processing problems, the memory requirement of the Cholesky sampler exceeds the memory capacity of nowadays standard laptops.
To mitigate these computational issues, many works have focused on the derivation of surrogate high-dimensional Gaussian sampling methods.
Compared to Cholesky sampling, these surrogate samplers involve additional sources of approximation in finite-time sampling and as such intend to trade-off \emph{computation} and  \emph{storage} requirements against sampling \emph{accuracy}.
Throughout this review, we will say that a Gaussian sampling procedure is ``efficient'' if, in order to produce a sample with reasonable accuracy, the number of flops and memory required are significantly lower than that of the Cholesky sampler.
For the sake of clarity, at the end of each section presenting an existing Gaussian sampler, we will highlight its theoretical relative efficiency with respect to (w.r.t.) vanilla Cholesky sampling with a dedicated paragraph.
As a typical example, some approaches reviewed in this paper will only require $\mathcal{O}(d^2)$ flops and $\Theta(d)$ storage space, which are lower than Cholesky complexities by an order of magnitude.

\noindent\textbf{Contributions.}
Up to authors' knowledge, no systematic comparison of existing Gaussian sampling approaches is available in the literature. This is probably due to the huge number of contributions from distinct communities related to this task.
Hence, it is not always clear which method is best suited to a given Gaussian sampling task in high dimensions, and what are the main similarities and differences between them.
To this purpose, this paper both reviews the main sampling techniques dedicated to an arbitrary high-dimensional Gaussian distribution (see \Cref{table:overview} p.\pageref{table:overview} for a synthetic overview) and derives general guidelines for practitioners to choose the appropriate sampling approach when Cholesky sampling is not possible (see \Cref{fig:guidelines} p.\pageref{fig:guidelines} for a schematic overview).
On top of that review, we propose to put most of the state-of-the-art Markov chain Monte Carlo (MCMC) methods under a common umbrella by deriving a unifying Gaussian sampling framework.

\noindent\textbf{Main results.}
Our main results are summarized hereafter.
\begin{itemize}

    \item At the expense of some approximation, we show that state-of-the-art Gaussian sampling algorithms are indeed more efficient than Cholesky sampling in high dimension.
    Their computational complexity, memory requirement and accuracy are summarized in \Cref{table:overview}.
    
    \item Among existing Gaussian samplers, some provide i.i.d. samples while others (e.g., MCMC-based ones) produce correlated samples.
    Interestingly, we show in \cref{sec:applications} on several experiments that MCMC approaches might perform better than samplers providing i.i.d. samples.
    This relative efficiency is particularly important in the case where many samples are required.
    
    \item In \cref{sec:PPA}, we show that most of existing MCMC approaches can be seen as special instances of a unifying framework which stands for a stochastic counterpart of the proximal point algorithm (PPA) \cite{Rockafellar1976}.
    The proposed Gaussian sampling framework also allows to both extend existing algorithms by proposing new sampling alternatives, and to draw a one-to-one equivalence between MCMC samplers proposed by distinct communities such as those based on matrix splitting \cite{Johnson2013,Fox2017} and data augmentation \cite{Marnissi2018,Marnissi2019}.
    
    \item Finally, we show that the choice of an appropriate Gaussian sampling approach stands for a subtle compromise between several quantities such as the need to obtain accurate samples or the existence of a natural decomposition of the precision matrix.
    We provide in \Cref{fig:guidelines} simple guidelines for practitioners in the form of a decision tree to choose the appropriate Gaussian sampler based on these parameters.
    
\end{itemize}

\noindent\textbf{Structure of the paper.}
This paper is structured as follows.
In \cref{sec:Gaussian_sampling}, we present the considered multivariate Gaussian sampling problem along with its simple and more complicated instances.
In particular, we will list and illustrate the main difficulties associated to the sampling from a high-dimensional Gaussian distribution with an arbitrary covariance matrix. These difficulties motivated many works to propose surrogate sampling approaches.
The latter are presented in \cref{sec:PIGauss_M_zero} and \cref{sec:PIGauss_M_nonzero}.
More precisely, \cref{sec:PIGauss_M_zero} presents Gaussian sampling schemes which have been derived by adapting ideas from numerical linear algebra. 
In \cref{sec:PIGauss_M_nonzero}, we review another class of sampling techniques, namely MCMC approaches, which build a Markov chain admitting the Gaussian distribution of interest (or a close approximation) as stationary distribution.
In \cref{sec:PPA}, we propose to shed new light on most of these MCMC methods by embedding them into a unifying framework based on a stochastic version of the PPA.
In \cref{sec:applications}, we illustrate and compare the reviewed approaches w.r.t. archetypal experimental scenarios.
Finally, \cref{sec:conclusion} draws concluding remarks.
A guide to the notation used in this paper and technical details associated to each section are given in the appendix p.\pageref{appendix_notations}.

\noindent\textbf{Software.}
All the methods reviewed in this paper have been implemented and made available in a companion package written in \textsc{Python}\xspace called $\texttt{PyGauss}$.
In addition, $\texttt{PyGauss}$ contains a \textsc{Jupyter} notebook which allows to reproduce all the figures and numerical results in this paper.
This package and its associated documentation can be found online\footnote{\url{http://github.com/mvono/PyGauss}}.

\section{Gaussian sampling: problem, instances and issues}
\label{sec:Gaussian_sampling}

This section highlights the considered Gaussian sampling problem, its already-surveyed special instances and its main issues. 
By recalling these specific instances, this section will also define the limits of this paper in terms of reviewed approaches.
Note that for the sake of simplicity, we shall abusively use the same notation for a random variable and its realization.

\subsection{Definitions and notation}
\label{subsec:motivations}

Throughout this review, we will use capital letters (e.g., $\Pi$) to denote a probability distribution and corresponding small letters (e.g., $\pi$) to refer to its associated probability density function (p.d.f.).
We address the problem of sampling from a $d$-dimensional Gaussian distribution $\Pi \triangleq \mathcal{N}\pr{\Bs{\mu},\B{\Sigma}}$ where $d$ may be large.
Its p.d.f. with respect to the $d$-dimensional Lebesgue measure, for all $\Bs{\theta} \in \mathbb{R}^d$, writes

\begin{equation}
  \pi(\Bs{\theta}) = \dfrac{1}{(2\uppi)^{d/2}\mathrm{det}(\B{\Sigma})^{1/2}}\exp\pr{-\dfrac{1}{2}(\Bs{\theta}-\Bs{\mu})^{\top}\B{\Sigma}^{-1}(\Bs{\theta}-\Bs{\mu})}\eqsp, \label{eq:Gaussian_target_Sigma}
\end{equation} 
where $\Bs{\mu} \in \mathbb{R}^d$ and $\B{\Sigma} \in \mathbb{R}^{d \times d}$ respectively stand for the mean vector and the covariance matrix of the considered Gaussian distribution.
Assume throughout that the covariance matrix $\B{\Sigma}$ is positive definite, that is for all $\Bs{\theta} \in \mathbb{R}^d \setminus \{\B{0}_d\}$, $\Bs{\theta}^{\top}\B{\Sigma}\Bs{\theta}>0$. 
Hence, its inverse $\B{Q}=\B{\Sigma}^{-1}$, called the \textit{precision} matrix, exists and is also positive definite.
Unless explicitly specified, we assume in the sequel that the covariance matrix has full rank. When $\B{\Sigma}$ is not of full rank, the distribution $\Pi$ is said to be degenerate and does not admit a density w.r.t. the $d$-dimensional Lebesgue measure.

For some approaches and applications, working with the precision $\B{Q}$ rather than with the covariance $\B{\Sigma}$ will be more convenient (e.g., for conditional auto-regressive models or hierarchical Bayesian  models; see also \cref{sec:PIGauss_M_nonzero}).
In this paper, we choose to present existing approaches by working directly with $\B{Q}$ for the sake of simplicity.
When $\B{Q}$ is unknown but $\B{\Sigma}$ is available instead, simple and straightforward algebraic manipulations can be used to implement the same approaches without increasing their computational complexity.
Sampling from $\mathcal{N}\pr{\Bs{\mu},\B{Q}^{-1}}$ raises several important issues which are mainly related to the structure of $\B{Q}$.
In the following paragraphs, we will detail some special instances of \cref{eq:Gaussian_target_Sigma} and well-known associated sampling strategies before focusing on the general Gaussian sampling problem considered in this paper.

\subsection{Usual special instances}
\label{subsec:special_instances}
For completeness, this subsection recalls special cases of Gaussian sampling tasks that will not be detailed later but are usual common building blocks.
Instead, we point out appropriate references for the interested reader. These special instances include basic univariate sampling and the scenarios where $\B{Q}$ is (i) a diagonal matrix, (ii) a band-matrix or (iii)  a circulant Toeplitz matrix.
Again, with basic algebraic manipulations, the same samplers can be used when $\B{\Sigma}$ has one of these specific structures.

\subsubsection{Univariate Gaussian sampling}
The most simple Gaussian sampling problem boils down to drawing univariate Gaussian random variables with mean $\mu \in \mathbb{R}$ and precision $q > 0$.
Generating the latter quickly and with high accuracy has been the topic of much research works in the last 70 years.
Such methods can be loosely speaking divided into four groups namely (i) cumulative density function (c.d.f.) inversion, (ii) transformation, (iii) rejection and (iv) recursive methods; they are now well-documented.
Interested readers are invited to refer to the comprehensive overview in \cite{Thomas2007} for more details. For instance, 
\cref{algo:Box_Muller} details the well-known Box-Muller method which transforms a pair of independent uniform random variables into a pair of Gaussian random variables by exploiting the radial symmetry of the two-dimensional normal distribution. 
\begin{algorithm}
  \caption{Box-Muller sampler}
  \label{algo:Box_Muller}
  \begin{algorithmic}[1]
    \State Draw $u_1$, $u_2$ $\overset{\mathrm{i.i.d.}}{\sim} \mathcal{U}((0,1])$.
    \State Set $\tilde{u}_1 = \sqrt{-2\log(u_1)}$.
    \State Set $\tilde{u}_2 = 2\uppi u_2$.\\ 
    \Return $(\theta_1,\theta_2) = \pr{\mu + \frac{\tilde{u}_1}{\sqrt{q}} \sin(\tilde{u}_2),\mu + \frac{\tilde{u}_1}{\sqrt{q}} \cos(\tilde{u}_2)}$.
  \end{algorithmic}
\end{algorithm}

\subsubsection{Multivariate Gaussian sampling with diagonal precision matrix}
Let us extend the previous sampling problem and now assume that one wants to generate a $d$-dimensional Gaussian vector $\Bs{\theta}$ with mean $\Bs{\mu}$ and diagonal precision matrix $\B{Q} = \mathrm{diag}(q_1,\cdots,q_d)$.
The $d$ components of $\Bs{\theta}$ being independent, this problem is as simple as the univariate one since we can sample the $d$ components in parallel independently.
A pseudo-code of the corresponding sampling algorithm is given in \cref{algo:multi_diag}.
\begin{algorithm}
  \caption{Sampler when $\B{Q}$ is a diagonal matrix}
  \label{algo:multi_diag}
  \begin{algorithmic}[1]
    \For{$i \in [d]$} \Comment{\textcolor{green}{In some programming languages, this loop can be vectorized.}} 
    \State Draw $\displaystyle{\theta_i \sim \mathcal{N}\pr{\mu_i,1/q_i}}$.
    \EndFor\\
    \Return $\Bs{\theta} = (\theta_1,\cdots,\theta_d)^{\top}$.
  \end{algorithmic}
\end{algorithm}

\paragraph{Algorithmic efficiency}
By using for instance \cref{algo:Box_Muller} for step 2, \cref{algo:multi_diag} admits a computational complexity of $\Theta(d)$ and a storage capacity of $\Theta(d)$.
In this specific scenario, these requirements are significantly lesser than that of vanilla Cholesky sampling whose complexities are recalled in \Cref{table:complexity}.

When $\B{Q}$ is not diagonal, we can no longer sample the $d$ components of $\Bs{\theta}$ independently.
Thus more sophisticated sampling methods must be used. 
For well-structured matrices $\B{Q}$, we show in the following sections that it is possible to draw the random vector of interest more efficiently than vanilla Cholesky sampling.

\subsubsection{Multivariate Gaussian sampling with sparse or band matrix $\B{Q}$}
\begin{figure}
  \centering
  \raisebox{-8ex}{\mbox{{\includegraphics[scale=0.15]{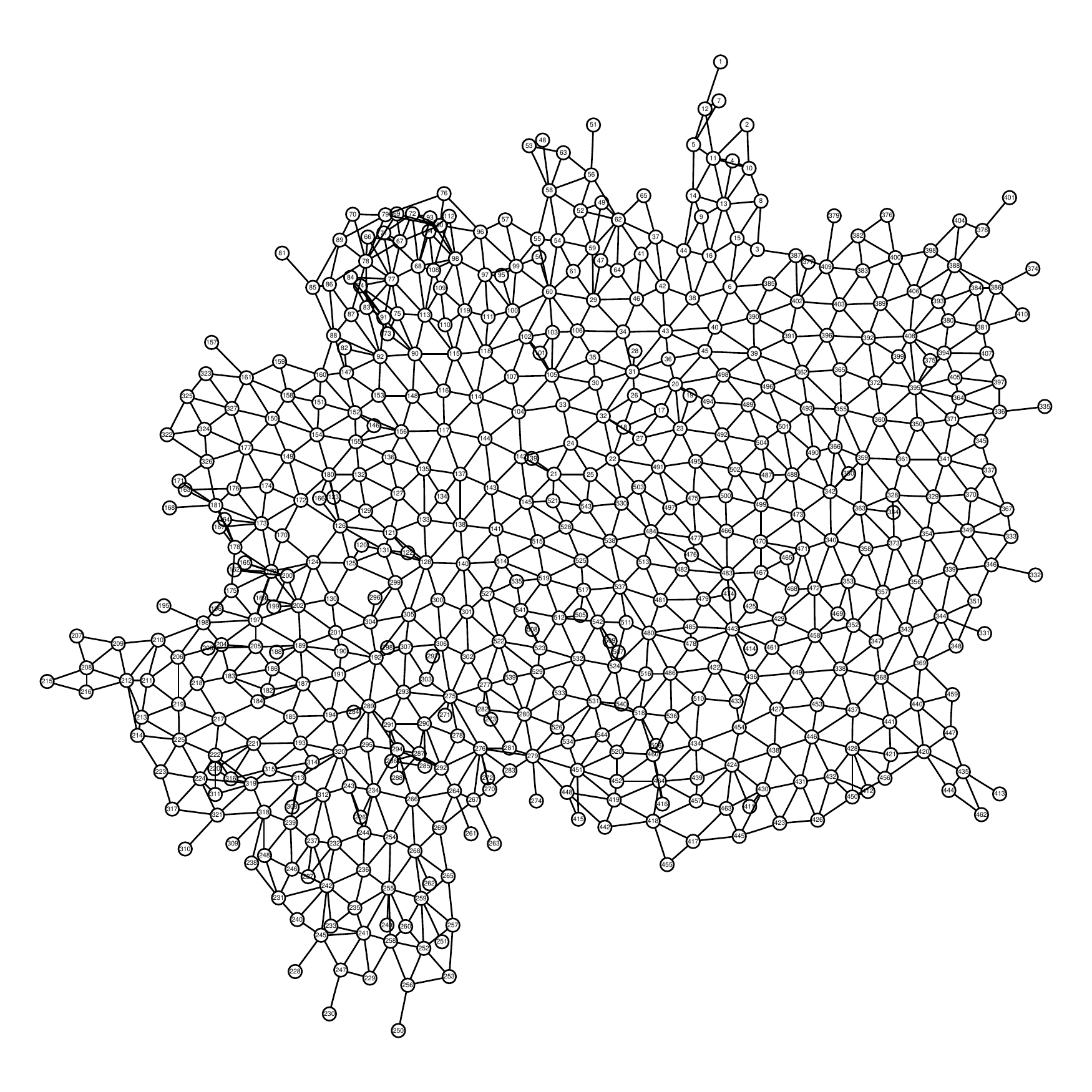}}}}
  \raisebox{-8ex}{\mbox{{\includegraphics[scale=0.15]{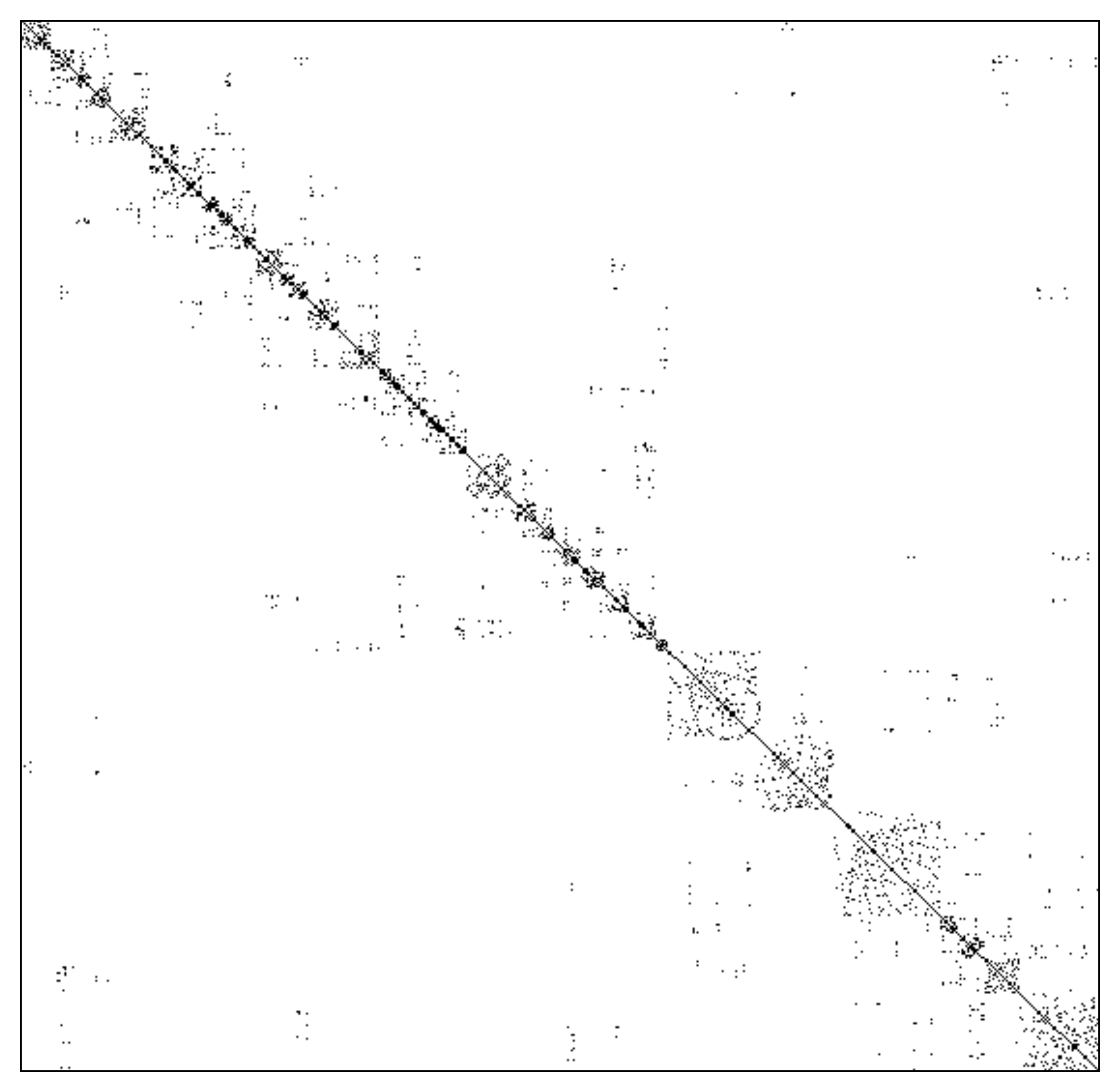}}}}
  \raisebox{-8ex}{\mbox{{\includegraphics[scale=0.15]{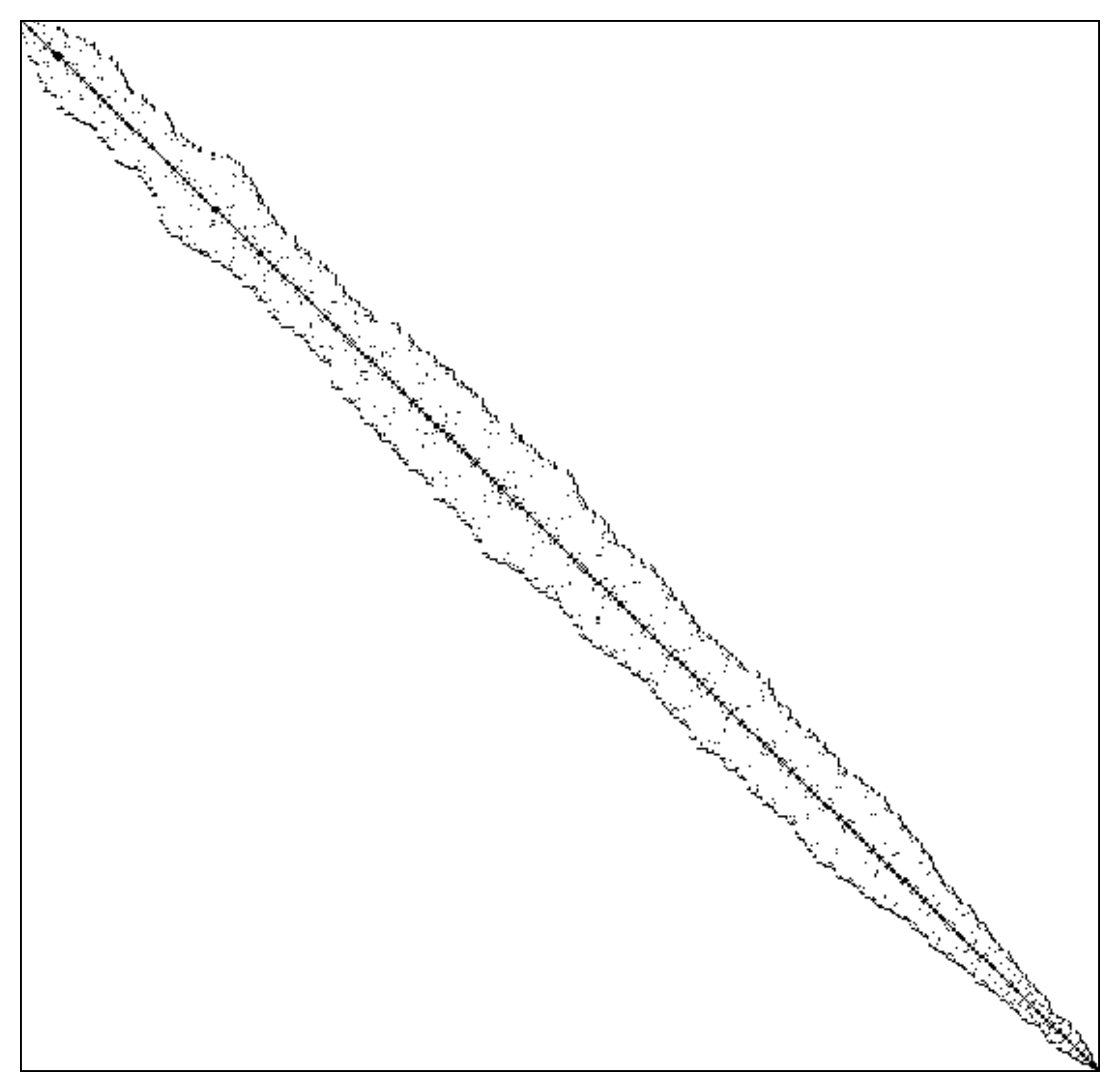}}}}
  \mbox{
  \begin{blockmatrixtabular}
    \raisebox{1.5ex}{\diagonalblockmatrix[0.8,1.0,0.8]{1.13in}{1.13in}{band}{0.25in}}
  \end{blockmatrixtabular}}
  \caption{From left to right: example of an undirected graph defined on the 544 regions of Germany where those sharing a common border are considered as neighbors, its associated precision matrix $\B{Q}$ (bandwidth $b=522$), its re-ordered precision matrix $\B{PQP}^{\top}$ ($b=43$) where $\B{P}$ is a permutation matrix and a drawing of a band matrix.
  For the three matrices, the white entries are equal to zero.}
  \label{fig:germany}
\end{figure}

A lot of standard Gaussian sampling approaches leverage on the sparsity of the matrix $\B{Q}$.
Sparse precision matrices appear for instance when Gaussian Markov random fields (GMRFs) are considered, as illustrated in \cref{fig:germany}.
In this figure, German regions are represented graphically where each edge between two regions stands for a common border.
These edges can then be described by an adjacency matrix which plays the role of the precision matrix $\B{Q}$ of a GMRF. 
Since there are few neighbors for each region, $\B{Q}$ is symmetric and sparse. 
By permuting the rows and columns of $\B{Q}$, one can build a so-called \emph{band matrix} with minimal bandwidth $b$ where $b$ is the smallest integer $b<d$ such that $Q_{ij} = 0, \ \forall i > j + b$ \cite{Rue2001}.
Note that band matrices also naturally appear in specific applications, e.g., when the latter involve finite impulse response linear filters \cite{Idier2008}.
Problems with such structured (sparse or band) matrices have been extensively studied in the literature and as such this paper will not cover them explicitly. 
We provide in \cref{algo:multi_band} the main steps to obtain a Gaussian vector $\Bs{\theta}$ from $\mathcal{N}(\Bs{\mu},\B{Q}^{-1})$ in this scenario and refer the interested reader to \cite{Rue2005} for more details.

\paragraph{Algorithmic efficiency}
\cref{algo:multi_band} is a specific instance of Cholesky sampling for band precision matrices.
In this specific scenario, \cref{algo:multi_band} admits a computational complexity of $\Theta(b^2d)$ flops and a memory space of $\Theta(bd)$ since the band matrix $\B{Q}$ can be stored in a so-called ``$\B{Q}$.array'' \cite[Section 1.2.5]{Golub1989}.
When $b \ll d$, one observes that these computational and storage requirements are smaller than those of vanilla Cholesky sampling by an order of magnitude w.r.t. $d$.
Similar computational savings can be obtained in the sparse case \cite{Rue2005}.

\begin{algorithm}
  \caption{Sampler when $\B{Q}$ is a band matrix}
  \label{algo:multi_band}
  \begin{algorithmic}[1]
    \State Set $\B{C} = \mathrm{chol}(\B{Q})$.
    \Comment{\textcolor{green}{Build the Cholesky factor $\B{C}$ of $\B{Q}$, see \cite[Section 2.4]{Rue2005}.}}
    \State Draw $\B{z} \sim \mathcal{N}(\B{0}_d,\B{I}_d)$.
      \For{$i \in [d]$} \Comment{\textcolor{green}{Solve $\B{C}^{\top}\B{w} = \B{z}$ w.r.t. $\B{w}$ by backward substitution.}}
        \State Set $j = d - i + 1$.
        \State Set $m_1 = \min\{j + b,d\}$.
        \State Set $w_j = \dfrac{1}{C_{jj}}\pr{z_j - \displaystyle\sum_{k=j+1}^{m_1}C_{kj}w_k}$.
      \EndFor\\
    \Return $\Bs{\theta} = \Bs{\mu} + \B{w}$.
  \end{algorithmic}
\end{algorithm}

\subsubsection{Multivariate Gaussian sampling with block circulant (Toeplitz) matrix $\B{Q}$ with circulant (Toeplitz) blocks} 
\begin{figure}
  \centering
  \mbox{{\includegraphics[scale=0.17]{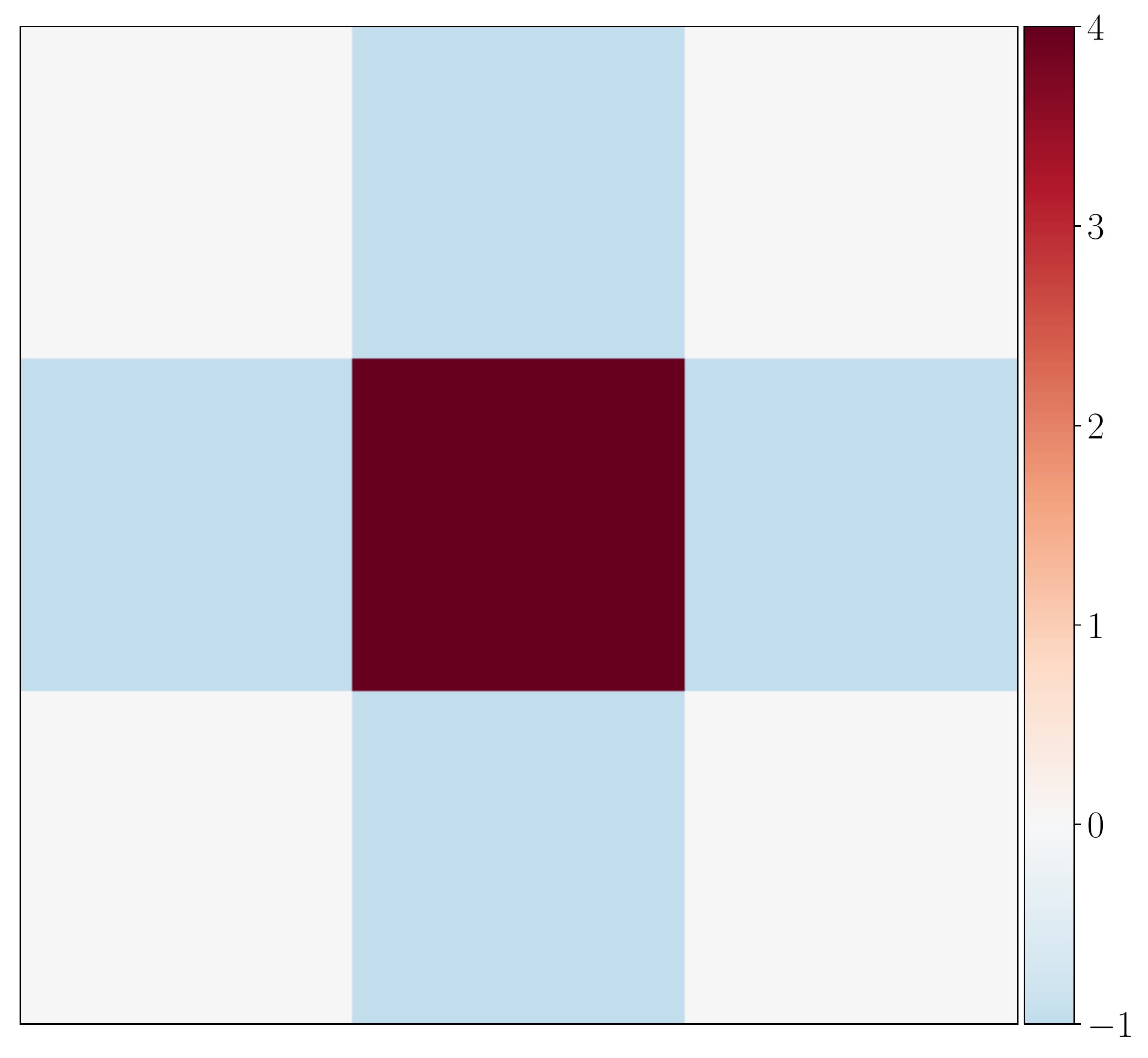}}}
  \mbox{{\includegraphics[scale=0.17]{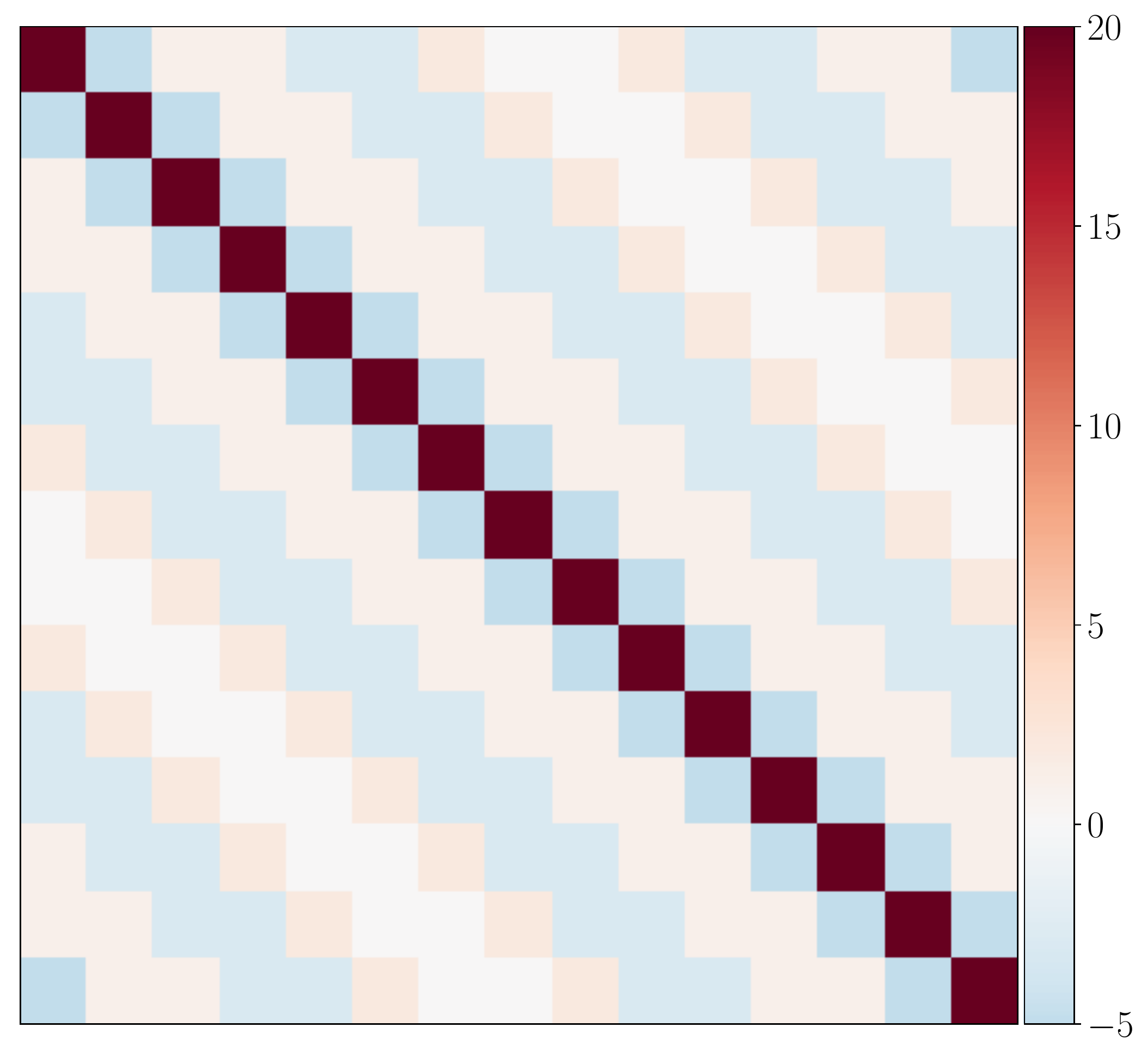}}}
  \mbox{{\includegraphics[scale=0.17]{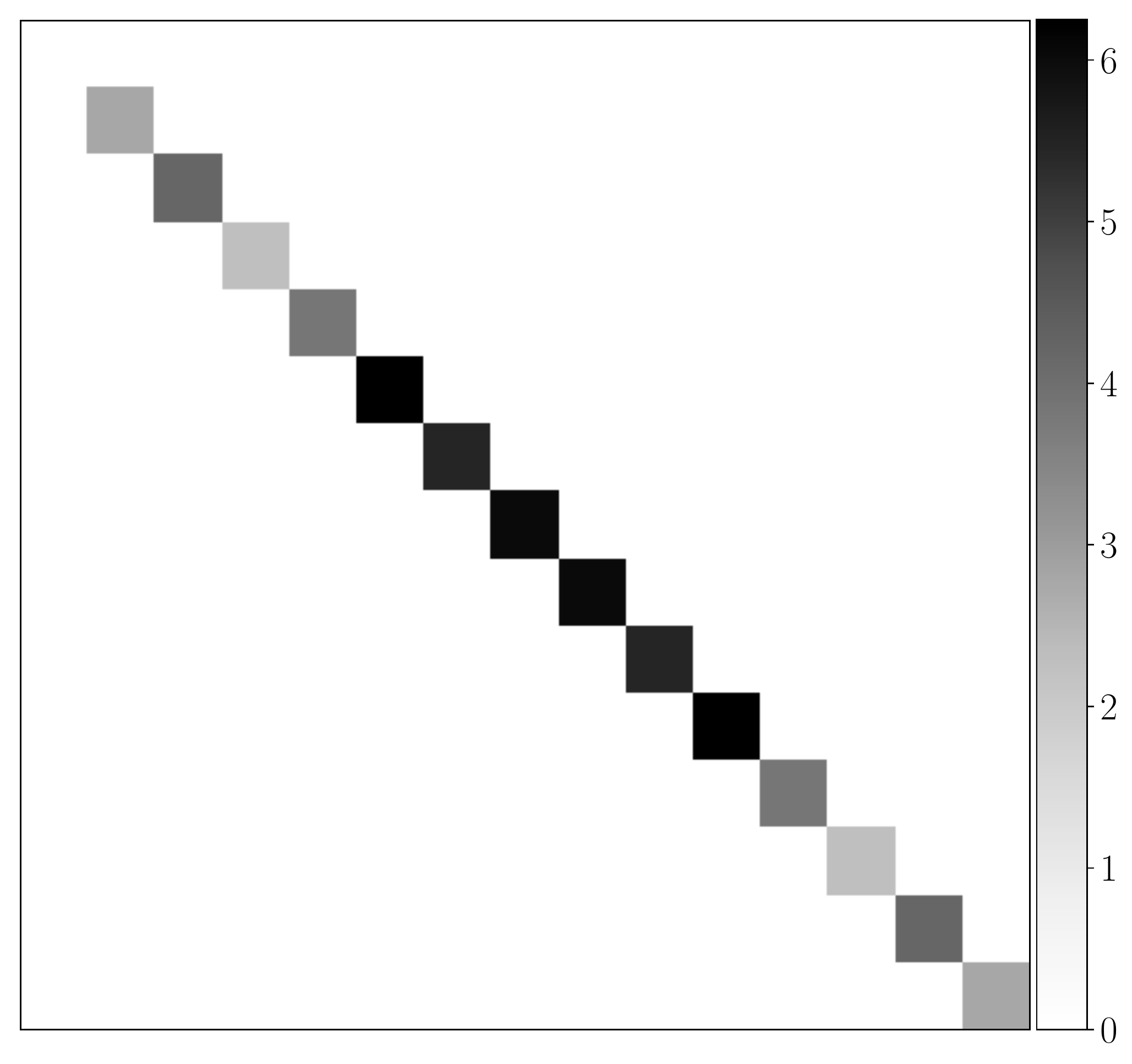}}}
  \caption{From left to right: example of a $3 \times 3$ Laplacian filter, the associated circulant precision matrix $\B{Q} = \B{\Delta}^{\top}\B{\Delta}$ for convolution with periodic boundary conditions and its counterpart diagonal matrix $\B{FQF}^{\mathsf{H}}$ in the Fourier domain, where $\B{F}$ and its Hermitian conjugate $\B{F}^{\mathsf{H}}$ are unitary matrices associated with the Fourier and inverse Fourier transforms.}
  \label{fig:circulant}
\end{figure}

An important special case of \cref{eq:Gaussian_target_Sigma} which has already been surveyed \cite{Rue2005} is when $\B{Q}$ is a block circulant matrix with circulant blocks, that is
\begin{equation}
  \B{Q} = \begin{pmatrix}
          \B{Q}_1 & \B{Q}_2 & \hdots & \B{Q}_M \\
          \B{Q}_M & \B{Q}_1 & \hdots & \B{Q}_{M-1} \\
          \vdots & \vdots & \vdots & \vdots \\
          \B{Q}_2 & \B{Q}_{3} & \hdots & \B{Q}_1
          \end{pmatrix}\eqsp,
\end{equation}
where $\{\B{Q}_{i}\}_{i \in [M]}$ are $M$ circulant matrices. 
Such structured matrices frequently appear in image processing problems since they translate the convolution operator corresponding to linear and shift-invariant filters.
As an illustration, \cref{fig:circulant} shows the circulant structure of the precision matrix associated to the Gaussian distribution with density $\pi(\Bs{\theta}) \propto \exp(-\nr{\B{\Delta}\Bs{\theta}}^2/2)$.
Here, the vector $\Bs{\theta} \in \mathbb{R}^d$ stands for an image reshaped in lexicographic order and $\B{\Delta}$ stands for the Laplacian differential operator with periodic boundaries, also called Laplacian filter. In this case the precision matrix $\B{Q} = \B{\Delta}^{\top}\B{\Delta}$ is a circulant matrix \cite{Orieux2010} so that it is diagonalizable in the Fourier domain. Therefore, sampling from $\mathcal{N}(\Bs{\mu},\B{Q}^{-1})$ can be performed in this domain as shown in \cref{algo:circulant}. 
For Gaussian distributions with more general Toeplitz precision matrices, $\B{Q}$ can be replaced by its circulant approximation and then \cref{algo:circulant} can be used, see \cite{Rue2005} for more details.
Although not considered in this paper, other approaches dedicated to generate stationary Gaussian processes \cite{Li2009} have been considered, such as the spectral \cite{Shinozuka1972,Mejia1974} and turning bands \cite{Mantoglou1982} methods.

\paragraph{Algorithmic efficiency}
Thanks to the use of the fast Fourier transform \cite{Wood1994,Dietrich1997}, \cref{algo:circulant} admits a computational complexity of $\mathcal{O}(d\log(d))$ flops.
In addition, note that only $d$-dimensional vectors have to be stored which implies a memory requirement of $\Theta(d)$.
Overall, these complexities are significantly smaller than those of vanilla Cholesky sampling and as such \cref{algo:circulant} can be considered as ``efficient''.
\begin{algorithm}
  \caption{Sampler when $\B{Q}$ is a block circulant matrix with circulant blocks}
  \label{algo:circulant}
  \hspace*{\algorithmicindent} \textbf{Input:} $M$ and $N$, the number of blocks and the size of each block, respectively.
  \begin{algorithmic}[1]
    \State Compute $\B{F} = \B{F}_M \otimes \B{F}_N$. \Comment{\textcolor{green}{$\B{F}_M$ is the $M \times M$ unitary matrix associated to the Fourier transform and $\otimes$ denotes the tensor product.}}
    \State Draw $\B{z} \sim \mathcal{N}(\B{0}_d,\B{I}_d)$.
    \State Set $\B{\Lambda}_{\B{q}} = \mathrm{diag}(\B{q})$. \Comment{\textcolor{green}{$\B{q}$ is the $d$-dimensional vector built by stacking the first columns of each circulant block of $\B{Q}$.}}
      \State Set $\Bs{\theta} = \Bs{\mu} + \B{F}^{\mathsf{H}}\B{\Lambda}_{\B{q}}^{-1/2}\B{Fz}$.\\
    \Return $\Bs{\theta}$.
  \end{algorithmic}
\end{algorithm}

\subsubsection{Truncated and intrinsic Gaussian distributions}
Eventually, note that several works have focused on sampling from various probability distributions closely related to the Gaussian distribution on $\mathbb{R}^d$.
Two cases are worth being mentioned here, namely the \emph{truncated} and so-called \textit{intrinsic} Gaussian distributions.
Truncated Gaussian distributions on $\mathcal{D} \subset \mathbb{R}^d$ admit, for any $\Bs{\theta} \in \mathbb{R}^d$, probability density functions of the form
\begin{equation}
    \pi_{\mathcal{D}}(\Bs{\theta}) = \mathbf{1}_{\mathcal{D}}(\Bs{\theta}) \cdot 
    Z_{\mathcal{D}}^{-1}
    \exp\pr{-\dfrac{1}{2}(\Bs{\theta}-\Bs{\mu})^{\top}\B{\Sigma}^{-1}(\Bs{\theta}-\Bs{\mu})}\eqsp,
\end{equation}
where $Z_{\mathcal{D}}$ is the appropriate normalizing constant and $\mathbf{1}_{\mathcal{D}}(\Bs{\theta}) = 1$ if $\Bs{\theta} \in \mathcal{D}$ and $0$ otherwise.
The subset $\mathcal{D}$ is usually defined by equalities and/or inequalities.
As archetypal examples, truncations on the hypercube are such that $\mathcal{D} = \prod_{i=1}^d [a_i,b_i]$, $(a_i,b_i)\in \mathbb{R}^2$, $1\leq i\leq d$ or $\mathcal{D} = \{\Bs{\theta} \in \mathbb{R}^d \mid \sum_{i=1}^d \theta_d=1\}$ that limits the domain to the simplex.
Rejection and Gibbs sampling algorithms dedicated to these distributions can be found in \cite{Altmann_IEEE_SSP_2014, Li2015,Wilhelm2010}.
Intrinsic Gaussian distributions are such that $\B{Q}$ is not of full rank, that is $\B{Q}$ may have eigenvalues equal to zero.
This yields an improper Gaussian distribution often used as a prior in GMRFs to remove trend components \cite{Rue2005}.
Sampling from the latter can be done by identifying an appropriate subspace of $\mathbb{R}^d$ where the target distribution is proper and then sampling from the proper Gaussian distribution on this subspace \cite{Besag1995, Parker2012}.

All the usual special sampling problems above will not be considered in the following since they have already been exhaustively reviewed in the literature.

\subsection{Problem statement: sampling from a Gaussian distribution with an arbitrary precision  matrix $\B{Q}$}\label{subsec:problem}

From now on, we will consider and review approaches aiming at sampling from an {\em arbitrary non-intrinsic multivariate} Gaussian distribution $\mathcal{N}(\Bs{\mu},\B{Q}^{-1})$ with density defined in \cref{eq:Gaussian_target_Sigma}, i.e., without assuming any particular structure of the precision or covariance matrix.
If $\B{Q}$ is diagonal or well-structured, we saw in \cref{subsec:special_instances} that sampling can be performed more efficiently than vanilla Cholesky sampling, even in high dimension.
When this matrix is unstructured and possibly dense, these methods with reduced numerical and storage complexities cannot be used anymore. 
In such settings the main challenges for Gaussian sampling are directly related to handling the precision $\B{Q}$ (or covariance $\B{\Sigma}$) matrix in high dimension.
Typical issues include the storage of the $d^2$ entries of the matrix $\B{Q}$ (or $\B{\Sigma}$) and expensive operations of order $\Theta(d^3)$ flops such as inversion or square root which become prohibitive when $d$ is large.
These challenges are illustrated below with an example that typically arises in statistical learning.

\begin{example}[Bayesian ridge regression]\label{ex:bayesian_ridge}
  Let us consider a ridge regression problem from a Bayesian perspective \cite{Bishop2006}.
  For the sake of simplicity and without loss of generality, assume that the observations $\B{y} \in \mathbb{R}^n$ and the known predictor matrix $\B{X} \in \mathbb{R}^{n \times d}$ are such that 
  \begin{align}
  \sum_{i=1}^ny_i = 0\eqsp, \quad \sum_{i=1}^nX_{ij} = 0\eqsp, \quad \text{ and } \quad \sum_{i=1}^nX_{ij}^2 = 1\eqsp, \quad \text{ for $j \in [d]$}\eqsp.
  \end{align}
  Under these assumptions, we consider the following statistical model associated with observations $\B{y}$ which writes
  \begin{equation}
    \B{y} = \B{X}\Bs{\theta} + \boldsymbol{\varepsilon}\eqsp,
  \end{equation} 
  where $\Bs{\theta} \in \mathbb{R}^{d}$ and $\boldsymbol{\varepsilon} \sim \mathcal{N}(\B{0}_n,\sigma^2\B{I}_n)$.
 In this example, the standard deviation $\sigma$ is known and fixed.
  The conditional prior distribution for $\Bs{\theta}$ is chosen as Gaussian i.i.d., that is  
    \begin{align}
    p(\Bs{\theta}\mid\tau) &\propto \exp\pr{-\frac{1}{2\tau}||\Bs{\theta}||^2}\eqsp, \\ 
    p(\tau) &\propto \dfrac{1}{\tau}\Bs{1}_{\mathbb{R}_+\setminus\{0\}}(\tau)\eqsp, 
  \end{align}
  where $\tau > 0$ stands for an unknown variance parameter which is given a diffuse and improper (i.e., non-integrable) Jeffrey's prior \cite{Jeffreys1946,Robert94}. 
 The Bayes' rule then leads to the target joint posterior distribution with density
  \begin{align}
    p(\Bs{\theta},\tau \mid \B{y}) \propto \dfrac{1}{\tau}\Bs{1}_{\mathbb{R}_+\setminus\{0\}}(\tau) \ \exp\Big(-\frac{1}{2\tau}||\Bs{\theta}||^2 - \frac{1}{2\sigma^2}||\B{y}-\B{X}\Bs{\theta}||^2\Big)\eqsp. \label{eq:example_Lasso_posterior}
  \end{align}
Sampling from this joint posterior distribution can be conducted using a Gibbs sampler \cite{Geman1984,Robert2004} which sequentially samples from the conditional posterior distributions.
In particular, the conditional posterior distribution associated to $\Bs{\theta}$ is Gaussian with precision matrix and mean vector
  \begin{align}
    &\B{Q} = \frac{1}{\sigma^{2}}\B{X}^{\top}\B{X} + \tau^{-1}\B{I}_d\eqsp, \\
    &\Bs{\mu} = \frac{1}{\sigma^{2}} \B{Q}^{-1}\B{X}^{\top}\B{y}\eqsp. \label{eq:ex1_mean}
  \end{align} 
  Challenges related to handling the matrix $\B{Q}$ already appear in this classical and simple regression problem.
  Indeed, $\B{Q}$ is possibly high-dimensional and dense which potentially rules out its storage, see \Cref{table:complexity}. 
  The inversion required to compute the mean \cref{eq:ex1_mean} may be very expensive as well. 
  In addition, since $\tau$ is unknown, its value changes at each iteration of the Gibbs sampler used to sample from the joint distribution with density \cref{eq:example_Lasso_posterior}.
  Hence, pre-computing the matrix $\B{Q}^{-1}$ is not possible.
  As an illustration on real data, \cref{fig:LASSO_example} represents three examples of precision matrices\footnote{When considering the dataset itself, $\B{X}^{\top}\B{X}$ is usually interpreted as the empirical covariance of the data $\B{X}$. The reader should not be disturbed by the fact that, turning to the variable $\Bs{\theta}$ to infer, $\B{X}^{\top}\B{X}$ will however play the role of a precision matrix.} $\B{X}^{\top}\B{X}$ for the MNIST \cite{LeCun1998}, leukemia \cite{Armstrong2002} and CoEPrA \cite{CoEPrA} datasets.
  One can note that these precision matrices are potentially both high-dimensional and dense penalizing their numerical inversion at each iteration of the Gibbs sampler.
  \begin{figure}
  \centering
  \mbox{{\includegraphics[scale=0.18]{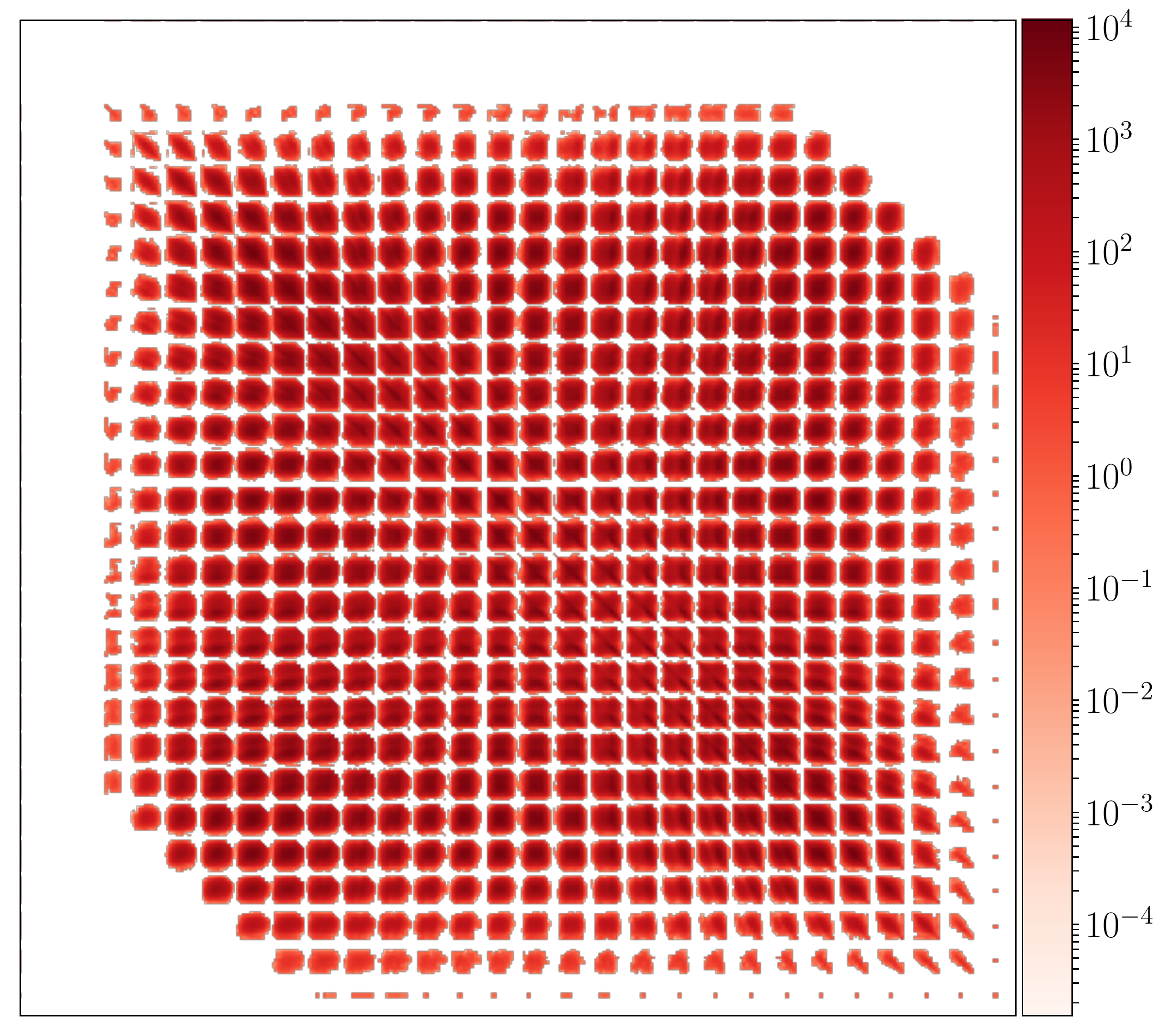}}}
  \mbox{{\includegraphics[scale=0.18]{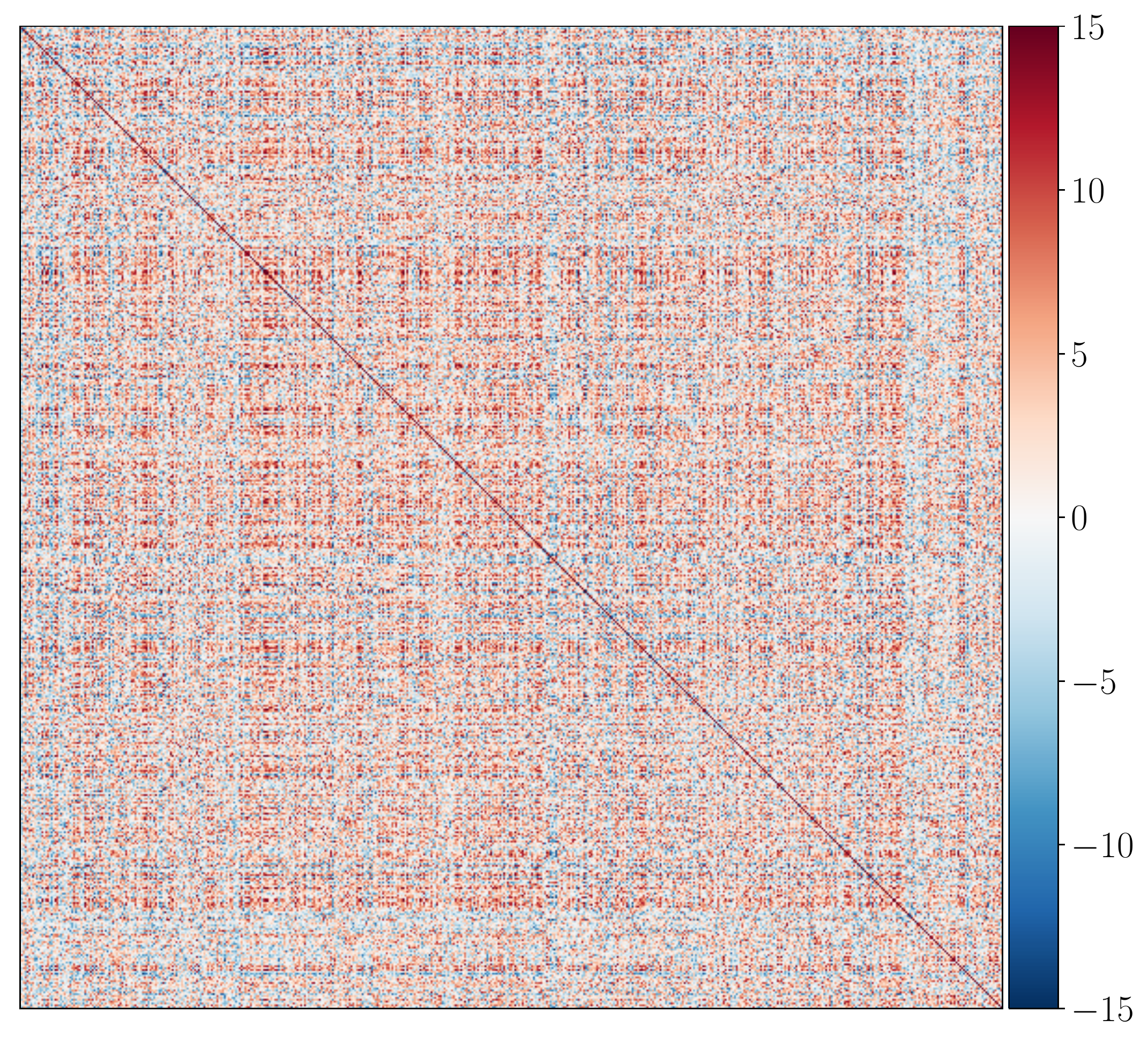}}}
  \mbox{{\includegraphics[scale=0.18]{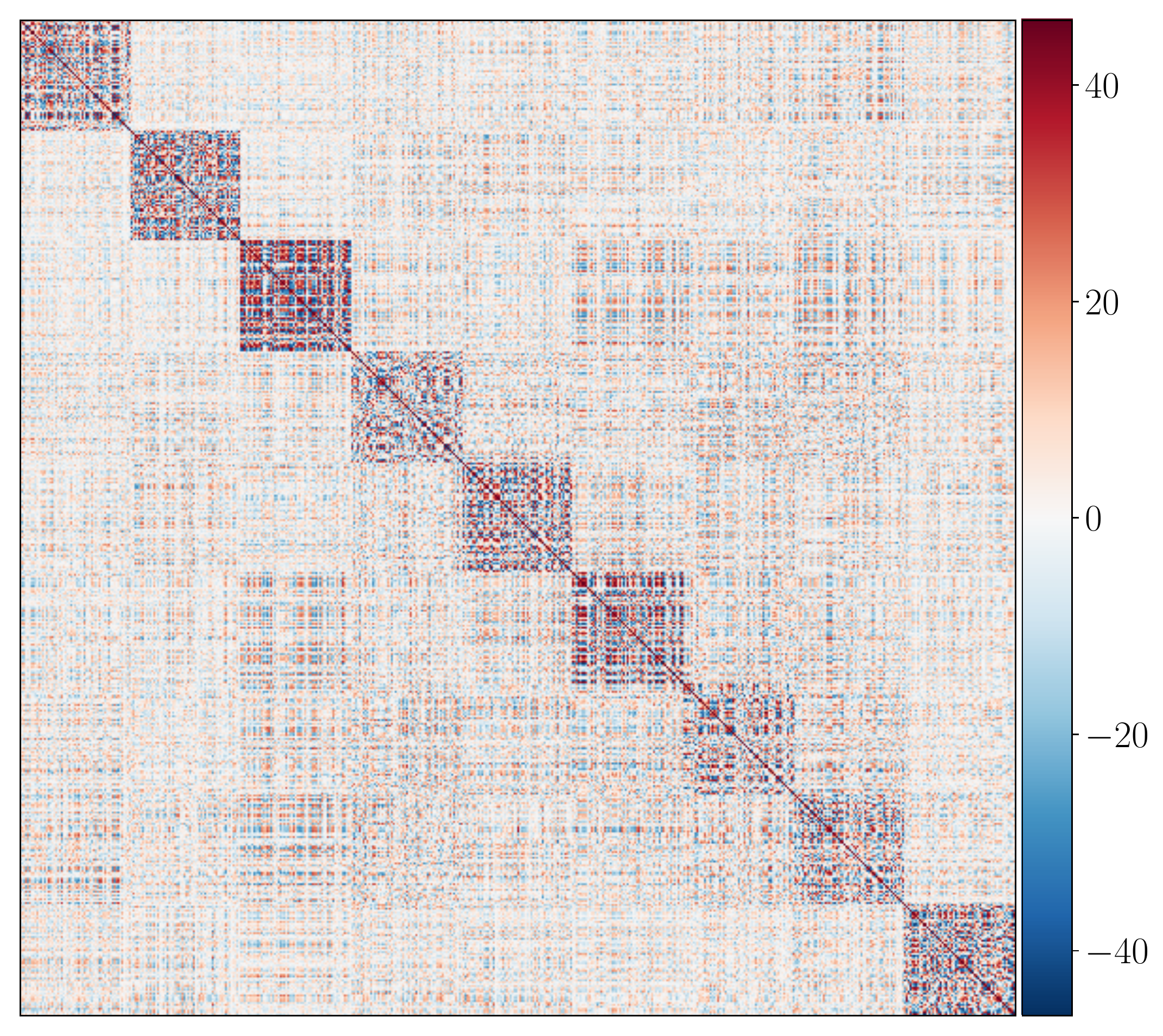}}}
  \caption{Examples of precision matrices $\B{X}^{\top}\B{X}$ for three datasets.
  Left: MNIST dataset \cite{LeCun1998}. Only the predictors associated to the digits 5 and 3 have been taken into account for the MNIST dataset \cite{LeCun1998}.
  Middle: leukemia dataset \cite{Armstrong2002}. For the leukemia dataset \cite{Armstrong2002}, only the first 5,000 predictors (out of 12,600) have been taken into account.
  Right: CoEPrA dataset \cite{CoEPrA}.}
  \label{fig:LASSO_example}
  \end{figure}
\end{example}

Hosts of contributions have been related to high-dimensional Gaussian sampling: it is impossible to  cite them in an exhaustive manner.
As far as possible, the following review aims at gathering and citing the main contributions. 
We refer the reader to references therein for more details.
Next \cref{sec:PIGauss_M_zero} and \cref{sec:PIGauss_M_nonzero} review the two main families of approaches that deal with the sampling issues raised above. 
In \cref{sec:PIGauss_M_zero}, we deal with approaches derived from numerical linear algebra. 
On the other hand, \cref{sec:PIGauss_M_nonzero} deals with Markov chain Monte Carlo (MCMC)  sampling  approaches. 
Later, \cref{sec:PPA} will propose a unifying revisit of Gibbs samplers thanks to a stochastic counterpart of the proximal point algorithm (PPA).
Similarly to \cref{subsec:special_instances}, computational costs, storage requirements and accuracy of the reviewed Gaussian sampling approaches will be detailed for each method in a dedicated paragraph entitled \emph{Algorithmic efficiency}.
For a synthetic summary and comparison of these metrics, we refer the interested reader to \cref{table:overview}.

\section{Sampling algorithms derived from numerical linear algebra}
\label{sec:PIGauss_M_zero}

This section presents sampling approaches which stand for direct adaptations of classical techniques used in numerical linear algebra \cite{Golub1989}.
They can be divided into three main groups: (i) factorization methods which consider appropriate decompositions of $\B{Q}$, (ii) inverse square-root approximation approaches where approximations of $\B{Q}^{-1/2}$ are used to obtain samples from $\mathcal{N}(\Bs{\mu},\B{Q}^{-1})$ at a reduced cost compared to factorization approaches and (iii) conjugate gradient-based methods.

\subsection{Factorization methods}
\label{subsec:factorization_methods}
We begin this review with the most basic but computationally involved sampling techniques namely factorization approaches which have been already introduced in \cref{sec:intro}.
These methods exploit the positive definiteness of $\B{Q}$ to decompose it as a product of simpler matrices and are essentially based on the celebrated Cholesky factorization \cite{Cholesky1910}.
Albeit helpful for problems in small or moderate dimension, these basic sampling approaches fail to address, in high-dimensional scenarios, the computational and memory issues raised in \cref{subsec:problem}.

\subsubsection{Cholesky factorization}
\label{subsubsec:cholesky}

Since $\B{Q}$ is symmetric and positive definite, we noted in \cref{sec:intro}, eq.~\eqref{choleskyfact}, that there exists a unique lower triangular matrix $\B{C} \in \mathbb{R}^{d \times d}$, called Cholesky factor, with positive diagonal entries such that $\B{Q} = \B{C}\B{C}^{\top}$ \cite{Golub1989}.
It is also known as the LU decomposition since $\B{Q}$ is expressed as the product of a lower triangular matrix $\B{C}$ and an upper one $\B{C}^{\top}$. 
\cref{algo:factorization_sampler} details how such a decomposition\footnote{When working with the covariance matrix rather than with the precision matrix, the Cholesky decomposition $\B{\Sigma}=\B{L}\B{L}^{\top}$ leads to the simpler step 3: $\B{w}= \B{Lz}$.} can be used to obtain a sample $\boldsymbol{\theta}$ from $\mathcal{N}(\Bs{\mu},\B{Q}^{-1})$.
\paragraph{Algorithmic efficiency}
In the general case where $\B{Q}$ presents no particular structure, the computational cost is $\Theta(d^3)$ and the storage requirement is $\Theta(d^2)$, see also \cref{sec:intro} and \cref{table:complexity}.
If the dimension $d$ is large but the matrix $\B{Q}$ has a sparse structure, the computational and storage requirements of the Cholesky factorization can be reduced by re-ordering the components of $\B{Q}$ to design an equivalent band matrix \cite{Rue2001}, see \cref{subsec:special_instances} and \cref{algo:multi_band}.

\begin{algorithm}
\caption{Cholesky sampler}
\label{algo:factorization_sampler}
\begin{algorithmic}[1]
\State Set $\B{C} = \mathrm{chol}(\B{Q})$.
\Comment{\textcolor{green}{Build the Cholesky factor $\B{C}$ of $\B{Q}$, see \cite[Section 2.4]{Rue2005}.}}
\State Draw $\B{z} \sim \mathcal{N}(\B{0}_d,\B{I}_d)$.
\State Solve $\B{C}^{\top}\B{w} = \B{z}$ w.r.t. $\B{w}$.\\
\Return $\Bs{\theta} = \Bs{\mu} + \B{w}$.
\end{algorithmic}
\end{algorithm}

\subsubsection{Square root factorization}
The Cholesky factorization in the previous paragraph was used to decompose $\B{Q}$ into a product of a triangular matrix $\B{C}$ and its transpose. 
Then, a Gaussian sample was obtained by solving a triangular linear system.
An extension of this approach has been considered in \cite{Davis1987b} by performing a singular value decomposition (SVD) of the Cholesky factor $\B{C}$ which yields $\B{Q} = \B{B}^2$ with $\B{B} = \B{U}\B{\Lambda}^{1/2}\B{U^{\top}}$ where $\B{\Lambda}$ is diagonal and  $\B{U}$ is orthogonal.
Similar to the Cholesky factor and given $\B{z} \sim \mathcal{N}(\B{0}_d,\B{I}_d)$, this square root can then be used to obtain an arbitrary Gaussian sample by solving $\B{B}\B{w} = \B{z}$ w.r.t. $\B{w}$ and computing $\Bs{\theta} = \Bs{\mu} + \B{w}$.

\paragraph{Algorithmic efficiency}
Although the square root factorization is interesting for establishing the existence of $\B{B}$, the associated sampler is generally as computationally demanding as \Cref{algo:factorization_sampler} since the eigendecomposition of $\B{Q}$ is not cheaper than finding its Cholesky factor.
To avoid these computational problems and since samplers based on $\B{B}$ boil down to compute $\Bs{\theta} = \Bs{\mu} + \B{B}^{-1}\B{z}$, some works focused on approximations of the inverse square root $\B{B}^{-1}$ of $\B{Q}$ which require smaller computational and storage complexities.

\subsection{Inverse square root approximations}
\label{subsec:polynomial_approx}
This idea of finding an efficient way (compared to the costs associated to factorization approaches) to approximate the inverse square root $\B{B}^{-1}$ dates back at least to the work of Davis \cite{Davis1987b}, in the 1980s, who derived a polynomial approximation of the function $x \mapsto x^{1/2}$ to approximate the square root of a given covariance matrix.
Other works used Krylov-based approaches building on the Lanczos decomposition to approximate directly any matrix-vector product $\B{B}^{-1}\B{z}$ involving the square-root $\B{B}$.
The two following paragraphs review these methods.

\subsubsection{Polynomial approximation}
In \cref{subsec:factorization_methods}, we showed that the square root of $\B{Q}$ writes $\B{B} = \B{U}\B{\Lambda}^{1/2}\B{U^{\top}}$, which implies that $\B{Q} = \B{U}\B{\Lambda}\B{U^{\top}}$.
If $f$ stands for a real continuous function, $\B{Q}$ and $f(\B{Q}) = \B{U}f(\B{\Lambda})\B{U}^{\top}$ are diagonalizable with respect to the same eigenbasis $\B{U}$, where $f(\B{\Lambda}) \triangleq \mathrm{diag}(f(\lambda_1),\hdots,f(\lambda_d))$. This is a well-known result coming from the Taylor expansion of a real continuous function $f$.
Hence, a function $f$ such that $f(\B{Q})$ is a good approximation of $\B{B}^{-1} = \B{Q}^{-1/2}$ has to be such that 
\begin{align*}
  f(\lambda_i) \approx 1/\sqrt{\lambda_i}\eqsp, \quad \forall i \in [d]\eqsp.
\end{align*}
Since $f$ only needs to be evaluated at the points corresponding to the eigenvalues $\{\lambda_{i}\}_{i\in[d]}$ of $\B{Q}$, it suffices to find a good approximation of $\B{B}^{-1}$ on the interval $[\lambda_{\min},\lambda_{\max}]$ whose extremal values can be lower and upper-bounded easily using the Gershgorin circle theorem \cite[Theorem 7.2.1]{Golub1989}. 
In the literature \cite{Davis1987b,Ilic2004,Pereira2019}, the function $f$ has been built using Chebyshev polynomials \cite{Mason2002} which are a family $(T_k)_{k \in \mathbb{N}}$ of polynomials defined over [-1,1] by 
\begin{align*}
  &T_k(x) = \cos(k\alpha)\eqsp, \quad \text{ where }\ \forall \alpha \in \mathbb{R},\  x = \cos(\alpha)\eqsp,
\end{align*}
or by the recursion
\begin{equation}
\label{eq:cheby}
\left\{
 \begin{array}{rcl}
     T_0(x) &=& 1  \\
     T_1(x) &=& x \\
     T_{k+1}(x) &=& 2xT_k(x) - T_{k-1}(x) \quad (\forall k \geq 1)\eqsp.
 \end{array}
\right.
\end{equation}
This family of polynomials $(T_k)_{k \in \mathbb{N}}$ exhibits several interesting properties: uniform convergence of the Chebyshev series towards an arbitrary Lipschitz-continuous function over [-1,1] and near minimax property \cite{Mason2002}, along with fast computation of the coefficients of the series via the fast Fourier transform \cite{Press2007}. 
\cref{algo:Chebychev} describes the steps to generate arbitrary Gaussian vectors using this polynomial approximation.

\paragraph{Algorithmic efficiency}
Contrary to factorization methods detailed in \cref{subsec:factorization_methods}, \cref{algo:Chebychev} does not require the storage of $\B{Q}$ since only the computation of matrix-vector products of the form $\B{Qv}$ with $\B{v} \in \mathbb{R}^d$ is necessary.
Assuming that these operations can be computed efficiently in $\mathcal{O}(d^2)$ flops with some black-box routine, e.g., a fast wavelet transform \cite{Mallat_book}, \cref{algo:Chebychev} admits an overall computational cost of
$\mathcal{O}(K_{\mathrm{cheby}}d^2)$ flops and storage capacity of $\Theta(d)$, where $K_{\mathrm{cheby}}$ is a truncation parameter standing for the order of the polynomial approximation. 
When $K_{\mathrm{cheby}} \ll d$, \cref{algo:Chebychev} becomes more computationally efficient than vanilla Cholesky sampling while admitting a reasonable memory overhead.
For a sparse precision matrix $\B{Q}$ composed of $n_{\mathrm{nz}}$ non-zero entries, the computational complexity can be reduced down to $\mathcal{O}(K_{\mathrm{cheby}}n_{\mathrm{nz}})$ flops. 
Note that compared to factorization approaches, \cref{algo:Chebychev} involves an additional source of approximation coming from the order of the Chebyshev series $K_{\mathrm{cheby}}$.
Choosing this parameter in an adequate manner involves some hand-tuning or additional computationally intensive statistical tests \cite{Pereira2019}. 

\begin{algorithm}
\caption{Approx. square root sampler using Chebyshev polynomials}
\label{algo:Chebychev}
\begin{algorithmic}[1] 
\State Set $\lambda_l = 0$ and $\lambda_u = \displaystyle\max_{i \in [d]} \sum_{j \in [d]} |Q_{ij}|$.
\For{$j \in \llbracket0,K_{\mathrm{cheby}}\rrbracket$}\Comment{\textcolor{green}{Do the change of interval.}}
\State Set $g_j = \br{\cos\pr{\uppi\frac{2j+1}{2K_{\mathrm{cheby}}}}\dfrac{(\lambda_u-\lambda_l)}{2} + \dfrac{\lambda_u+\lambda_l}{2}}^{-1/2}$. 
\EndFor
\For{$k \in \llbracket0,K_{\mathrm{cheby}}\rrbracket$}\Comment{\textcolor{green}{Compute coefficients of the $K_{\mathrm{cheby}}$-truncated Chebyshev series.}}
  \State Compute $c_k = \dfrac{2}{K_{\mathrm{cheby}}} \displaystyle \sum_{j=0}^{K_{\mathrm{cheby}}} g_j \cos\pr{\uppi k\frac{2j+1}{2K_{\mathrm{cheby}}}}$. 
\EndFor
  \State Draw $\B{z} \sim \mathcal{N}(\B{0}_d,\B{I}_d)$.
  \State Set $\alpha = \dfrac{2}{\lambda_u-\lambda_l}$ and $\beta = \dfrac{\lambda_u+\lambda_l}{\lambda_u-\lambda_l}$.
  \State Initalize $\B{u}_1 = \alpha\B{Q z} - \beta\B{z}$ and $\B{u}_0 = \B{z}$. 
  \State Set $\B{u} = \dfrac{1}{2}c_0\B{u}_0 + c_1\B{u}_1$ and $k=2$.
  \While{$k \leq K_{\mathrm{cheby}}$} \Comment{\textcolor{green}{Compute the $K_{\mathrm{cheby}}$-truncated Chebyshev series.}}
    \State Compute $\B{u}' = 2(\alpha\B{Q}\B{u}_1 - \beta\B{u}_1) - \B{u}_0$.
    \State Set $\B{u} = \B{u} + c_k\B{u}'$.
    \State Set $\B{u}_0 = \B{u}_1$ and $\B{u}_1 = \B{u}'$.
    \State $k = k + 1$.
  \EndWhile
  \State Set $\Bs{\theta}= \Bs{\mu} + \B{u}$. \Comment{\textcolor{green}{Build the Gaussian vector of interest.}}\\ 
\Return $\Bs{\theta}$.

\end{algorithmic}
\end{algorithm}

\subsubsection{Lanczos approximation}
Instead of approximating the inverse square-root $\B{B}^{-1}$, some approaches approximate directly the matrix-vector product $\B{B}^{-1}\B{z}$ by building on the Lanczos decomposition \cite{Ilic2004,Simpson2008,Ilic2009,Aune2013,Simpson2013,Chow2014}.
The corresponding simulation-based algorithm is described in \cref{algo:Lanczos}.
It iteratively builds an orthonormal basis $\B{H} = \{\B{h}_1,\hdots,\B{h}_{K_{\mathrm{kryl}}}\}$ $\in \mathbb{R}^{d \times K_{\mathrm{kryl}}}$ with $K_{\mathrm{kryl}} \leq d$ for the Krylov subspace $\mathcal{K}_{K_{\mathrm{kryl}}}(\B{Q},\B{z}) \triangleq \mathrm{span}\{\B{z},\B{Qz},\hdots,\B{Q}^{K_{\mathrm{kryl}}-1} \B{z}\}$, and a tridiagonal matrix $\B{T} \approx \B{H}^{\top}\B{Q}\B{H} \in \mathbb{R}^{K_{\mathrm{kryl}} \times K_{\mathrm{kryl}}}$.
Using the orthogonality of $\B{H}$, $\B{z} = \B{H}\B{e}_1$ where $\B{e}_1$ is the first canonical vector of $\mathbb{R}^{K_{\mathrm{kryl}}}$, and the final approximation is 
\begin{equation}
    \label{eq:lanczos}
    \B{B}^{-1}\B{z} = \B{Q}^{-1/2}\B{z} \approx \nr{\B{z}}\B{H}\B{T}^{-1/2}\B{H}^{\top}\B{H}\B{e}_1 = \nr{\B{z}}\B{H}\B{T}^{-1/2}\B{e}_1\eqsp.
\end{equation}
As highlighted in \cref{eq:lanczos}, the main idea of the Lanczos approximation is to transform the computation of $\B{Q}^{-1/2}\B{z}$ into the computation of $\B{T}^{-1/2}\B{e}_1$ which is expected to be simpler since $\B{T}$ is tridiagonal and of size $K_{\mathrm{kryl}} \times K_{\mathrm{kryl}}$.

\begin{algorithm}
\caption{Approx. square root sampler using Lanczos decomposition}
\label{algo:Lanczos}
\begin{algorithmic}[1]
  \State Draw $\B{z} \sim \mathcal{N}(\B{0}_d,\B{I}_d)$.
  \State Set $\B{r}^{(0)} = \B{z}$, $\B{h}^{(0)} = \B{0}_d$, $\beta^{(0)} = \nr{\B{r}^{(0)}}$ and $\B{h}^{(1)} = \B{r}^{(0)}/\beta^{(0)}$.
  \For{$k \in [K_{\mathrm{kryl}}]$} 
  \State Set $\B{w} = \B{Q h}^{(k)} - \beta^{(k-1)}\B{h}^{(k-1)}$.
    \State Set $\alpha^{(k)} = \B{w}^{\top}\B{h}^{(k)}$.
    \State Set $\B{w} = \B{w} - \alpha^{(k)}\B{h}^{(k)}$.\Comment{\textcolor{green}{Gram–Schmidt orthogonalization process.}}
    \State Set $\beta^{(k)} = \nr{\B{w}}$.
    \State Set $\B{h}^{(k+1)} = \B{w}/\beta^{(k)}$.
  \EndFor
  \State Set $\B{T} = \mathrm{tridiag}(\Bs{\beta},\Bs{\alpha},\Bs{\beta})$.
  \State Set $\B{H} = (\B{h}^{(1)},\hdots,\B{h}^{(K_{\mathrm{kryl}})})$.\\
  \State Set $\Bs{\theta} = \Bs{\mu} + \beta^{(0)}\B{HT}^{-1/2}\B{e}_1$, where $\B{e}_1 = (1,0,\hdots,0)^{\top} \in \mathbb{R}^{K_{\mathrm{kryl}}}$.\\
  \Return $\Bs{\theta}$.

\end{algorithmic}
\end{algorithm}

\paragraph{Algorithmic efficiency}
Numerous approaches have been proposed to compute $\B{T}^{-1/2}\B{e}_1$ exactly or approximately and they generally require $\mathcal{O}(K_{\mathrm{kryl}}^2)$ flops \cite{tridiag1,tridiag2}, see also \Cref{algo:Chebychev}.
By using such approaches, \cref{algo:Lanczos} admits a computational complexity of $\mathcal{O}(K_{\mathrm{kryl}}d^2)$ and a memory requirement of $\mathcal{O}(K_{\mathrm{kryl}}d)$.
Similarly to $K_{\mathrm{cheby}}$ in \cref{algo:Chebychev}, one can note that $K_{\mathrm{kryl}}$ stands for a trade-off between computation, storage and accuracy.
As emphasized in \cite{Aune2013, Simpson2008}, adjusting this truncation parameter can be achieved by using the conjugate gradient (CG) algorithm.
In addition to stand for an approximate sampling technique when $K_{\mathrm{kryl}}<d$, the main and well-known drawback of \cref{algo:Lanczos} is that the basis $\B{H}$ loses orthogonality in floating point arithmetic due to round-off errors.
Some possibly complicated procedures to cope with this problem are surveyed in \cite{Stewart2001}.
Finally, one major problem of the Lanczos decomposition is the construction and storage of the basis $\B{H} \in \mathbb{R}^{d \times K_{\mathrm{kryl}}}$ which becomes as large as $\B{Q}$ when $K_{\mathrm{kryl}}$ tends to $d$.
Two main approaches have been proposed to deal with this problem, namely a so-called 2-pass strategy and a restarting strategy both reviewed in \cite{Aune2013,Ilic2009, Simpson2008}.
We eventually point out that preconditioning methods have been proposed to reduce the computational burden of Lanczos samplers \cite{Chow2014}.

\subsubsection{Other square-root approximations}
At least two other methods have been proposed to approximate the inverse square-root $\B{B}^{-1}$.
Since these approaches have been less used than others, only their main principle is given and we refer the interested reader to the corresponding references.
The first one is the rational approximation of $\B{B}^{-1}$ based on numerical quadrature of a contour integral \cite{Hale2008b} while the other one stands for a continuous deformation method based on a system of ordinary differential equations \cite{Allen2000}.
These two approaches are reviewed and illustrated on numerical examples in \cite{Aune2013}.

\subsection{Conjugate gradient-based samplers}
\label{subsec:CG}

Instead of building on factorization results, some approaches start from the finding that Gaussian densities with invertible precision matrix $\B{Q}$ can be re-written in a so-called \textit{information form}, that is
\begin{equation}
  \pi(\Bs{\theta}) \propto \exp\pr{-\dfrac{1}{2}\Bs{\theta}^{\top}\B{Q}\Bs{\theta} + \B{b}^{\top}\Bs{\theta}}\eqsp, \label{eq:implicit_gaussian}
\end{equation}
where $\B{b} = \B{Q}\Bs{\mu}$ is called the potential vector. 
If one is able to draw a Gaussian vector $\B{z}' \sim \mathcal{N}(\B{0}_d,\B{Q})$, then a sample $\Bs{\theta}$ from $\mathcal{N}(\Bs{\mu},\B{Q}^{-1})$ is obtained by solving the linear system
\begin{equation}
  \B{Q}\Bs{\theta} = \B{b} + \B{z}'\eqsp,
  \label{eq:perturbed_linear_pb}
\end{equation}
where $\B{Q}$ is positive definite so that conjugate gradient methods are relevant.
This approach uses the affine transformation of a Gaussian vector $\B{u} = \B{b} + \B{z}'$: if $\B{u} \sim \mathcal{N}(\B{Q}\Bs{\mu},\B{Q})$, then $\B{Q}^{-1}\B{u} \sim \mathcal{N}(\Bs{\mu},\B{Q}^{-1})$.

\subsubsection{Perturbation before optimization}
A first possibility to handle the perturbed linear problem \cref{eq:perturbed_linear_pb} consists in first computing the potential vector $\B{b}$, then perturbing this vector with the Gaussian vector $\B{z}'$, and finally solving the linear system with numerical algebra techniques. 
This approach is detailed in \cref{algo:PO}.
While the computation of $\B{b}$ is not difficult in general, drawing $\B{z}'$ might be computationally involved.
Hence, this sampling approach is of interest only if we are able to draw efficiently (i.e., in $\mathcal{O}(d^2)$ flops) the Gaussian vector $\B{z}'$.
This is for instance the case when $\B{Q} = \B{Q}_1 + \B{Q}_2$ with $\B{Q}_i = \B{G}_i^{\top}\B{\Lambda}_i^{-1}\B{G}_i$ ($i \in [2]$), provided that the symmetric and positive definite matrices $\{\B{\Lambda}_i\}_{i \in [2]}$ have simple structures, see \cref{subsec:special_instances}. 
Such situations often arise when Bayesian hierarchical models are considered \cite[Chap. 10]{Robert94}.
In these scenarios, an efficient way to compute $\B{b} + \B{z}'$ has been proposed in \cite{Papandreou2011} and is based on a local perturbation of the mean vectors $\{\Bs{\mu}_i\}_{i \in [2]}$.
\begin{algorithm}
\caption{Perturbation-optimization sampler}
\label{algo:PO}
\begin{algorithmic}[1]
\State Draw $\B{z}' \sim \mathcal{N}(\B{0}_d,\B{Q})$.\Comment{\textcolor{green}{with local perturbation as in \cite{Papandreou2011}.}}
\State Set $\Bs{\eta} = \B{b} + \B{z}'$. 
\State Solve $\B{Q}\Bs{\theta} = \Bs{\eta}$ w.r.t. $\Bs{\theta}$. \Comment{\textcolor{green}{with the CG solver for instance \cite{HestenesStiefel1952}.}}\\
\Return $\Bs{\theta}$.
\end{algorithmic}
\end{algorithm}
Such an approach has been coined \textit{perturbation-optimization} (PO) since it draws perturbed versions of the mean vectors involved in the hierarchical model before using them to define the linear system to solve \cite{Papandreou2011}.

\paragraph{Algorithmic efficiency}
If $K \in \mathbb{N}^*$ iterations of an appropriate linear solver (e.g., the conjugate gradient (CG) method) are used for step 3 in \cref{algo:PO}, the global computational and storage complexities of this algorithm are of order $\mathcal{O}(Kd^2)$ and $\Theta(d)$.
Regarding sampling accuracy, while \cref{algo:PO} in theory stands for an exact approach, the $K$-truncation procedure implies an approximate sampling scheme \cite{Orieux2012}.
A solution to correct this bias has been proposed in \cite{Gilavert2015} by building upon a reversible-jump approach \cite{Green1995}.

\subsubsection{Optimization with perturbation}
Alternatively, \cref{eq:perturbed_linear_pb} can be seen as well as a perturbed version of the linear system $\B{Q}\Bs{\theta} = \B{b}$.
Thus some works have focused on modified versions of well-known linear solvers such as the conjugate gradient (CG) \cite{Rue2001,Schneider2003,Parker2012}.
Actually, only one additional line of code standing for an univariate Gaussian sampling step (perturbation) is required to turn out the classical CG solver into a CG sampler \cite{Schneider2003,Parker2012}, see step 8 in  \cref{algo:CG}. 
This perturbation step sequentially builds a Gaussian vector with a covariance matrix being the best $k$-rank approximation of $\B{Q}^{-1}$ in the Krylov subspace $\mathcal{K}_k(\B{Q},\B{r}^{(0)})$ \cite{Parker2012}. Then a perturbation vector $\B{y}^{(K_{\mathrm{CG}})}$ is simulated before addition to $\Bs{\mu}$ so that finally $\Bs{\theta} = \Bs{\mu} + \B{y}^{(K_{\mathrm{CG}})}$. 
\begin{algorithm}
\caption{Conjugate gradient sampler}
\label{algo:CG}
\hspace*{\algorithmicindent} \textbf{Input:} Threshold $\epsilon > 0$, fixed initialization $\Bs{\omega}^{(0)}$ and random vector $\B{c} \in \mathbb{R}^d$.
\begin{algorithmic}[1]
  \State Set $k=1$, $\B{r}^{(0)} = \B{c} - \B{Q}\Bs{\omega}^{(0)}$, $\B{h}^{(0)} = \B{r}^{(0)}$, $d^{(0)} = \B{h}^{(0)\top}\B{Qh}^{(0)}$ and $\B{y}^{(0)} = \Bs{\omega}^{(0)}$. 
  \While{$\nr{\B{r}^{(k)}} \geq \epsilon$} 
    \State Set $\gamma^{(k-1)} = \dfrac{\B{r}^{(k-1)\top}\B{r}^{(k-1)}}{d^{(k-1)}}$.
    \State Set $\B{r}^{(k)} = \B{r}^{(k-1)} - \gamma^{(k-1)}\B{Qh}^{(k-1)}$.
    \State Set $\eta^{(k)} = -\dfrac{\B{r}^{(k)\top}\B{r}^{(k)}}{\B{r}^{(k-1)T}\B{r}^{(k-1)}}$.
    \State Set $\B{h}^{(k)} = \B{r}^{(k)} - \eta^{(k)}\B{h}^{(k-1)}$.
    \State Set $d^{(k)} = \B{h}^{(k)\top}\B{Qh}^{(k)}$.
      \State Set $\B{y}^{(k)} = \B{y}^{(k-1)} + \dfrac{z}{\sqrt{d^{(k-1)}}}\B{h}^{(k-1)}$ where $z \sim \mathcal{N}(0,1)$.
    \State $k = k + 1$.
  \EndWhile
  \State Set $\Bs{\theta} = \Bs{\mu} + \B{y}^{(K_{\mathrm{CG}})}$ where $K_{\mathrm{CG}}$ is the number of CG iterations.\\
\Return $\Bs{\theta}$.

\end{algorithmic}
\end{algorithm}

\paragraph{Algorithmic efficiency}
From a computational point of view, the CG sampler inherits the benefits of the CG solver: only matrix-vector products involving $\B{Q}$ and the storage of two $d$-dimensional vectors are needed, and one exact sample from $\mathcal{N}(\Bs{\mu},\B{Q}^{-1})$ is obtained after at most $K_{\mathrm{CG}} = d$ iterations.
This yields an approximate computational cost of $\mathcal{O}(K_{\mathrm{CG}}d^2)$ flops and a storage requirement of $\Theta(d)$ where $K_{\mathrm{CG}}$ is the number of CG iterations \cite{HestenesStiefel1952}.
The CG sampler belongs to the family of Krylov-based samplers (e.g., Lanczos). As such, it suffers from the same numerical problem due to finite machine precision and the $K_{\mathrm{CG}}$-truncation procedure.
In addition, the covariance of the generated samples depends on the distribution of the eigenvalues of the matrix $\B{Q}$.
Actually, if these eigenvalues are not well spread out, \cref{algo:CG} stops after $K_{\mathrm{CG}} < d$ iterations which yields an approximate sample with the best $K_{\mathrm{CG}}$-rank approximation of $\B{Q}^{-1}$ as the actual covariance matrix.
In order to correct this approximation, re-orthogonalization schemes can be employed but could become as computationally prohibitive as Cholesky sampling when $d$ is large \cite{Schneider2003}.
These sources of approximation are detailed in \cite{Parker2012}.
A generalization of \cref{algo:CG} has been considered in \cite{Feron2016} where a random set of $K'$ mutually conjugate directions \{$\B{h}^{(k)}\}_{k \in [K']}$ is considered at each iteration of a Gibbs sampler.

\section{Sampling algorithms based on MCMC}
\label{sec:PIGauss_M_nonzero}

The previous section presented existing Gaussian sampling approaches by directly adapting ideas and techniques from numerical linear algebra such as matrix decompositions and matrix approximations. 
In this section, we will present another family of sampling approaches, namely MCMC approaches, which build a discrete-time Markov chain $(\boldsymbol{\theta}^{(t)})_{t \in \mathbb{N}}$ having $\mathcal{N}(\Bs{\mu},\B{Q}^{-1})$ (or a close approximation of $\mathcal{N}(\Bs{\mu},\B{Q}^{-1})$) as its invariant distribution \cite{Robert2004}.
In the sequel, we state that a MCMC approach is exact if the associated sampler admits an invariant distribution which  coincides with $\mathcal{N}(\Bs{\mu},\B{Q}^{-1})$.
Contrary to the approaches reviewed in \cref{sec:PIGauss_M_zero} which produce i.i.d. samples from $\mathcal{N}(\Bs{\mu},\B{Q}^{-1})$ or a close approximation to it, MCMC approaches produce correlated samples that are asymptotically distributed according to their invariant distribution.
Hence, at a first glance, it seems natural to think that MCMC samplers are less trustworthy since the number of iterations required until convergence is very difficult to assess in practice \cite{HobertJones2001}.
Interestingly, we will show numerically in \cref{sec:applications} that MCMC methods might perform better than i.i.d. samplers in some cases and as such might stand for serious contenders for the most efficient Gaussian sampling algorithms, see also \cite{Fox2017}.
On top of that review, we will also show that most of these MCMC approaches can be unified via a stochastic version of the proximal point algorithm \cite{Rockafellar1976}. 
This framework will be presented and detailed in \cref{sec:PPA}.

\subsection{Matrix splitting}
\label{subsec:matrix_split}

We begin the review of MCMC samplers by detailing so-called \textit{matrix splitting} (MS) approaches that build on the decomposition $\B{Q} = \B{M} - \B{N}$ of the precision matrix.
As we shall see, both exact and approximate MS samplers have been proposed in the existing literature.
These methods embed one of the simplest MCMC method, namely the component-wise Gibbs sampler \cite{Geman1984}.
Similarly to \cref{algo:factorization_sampler} for samplers in \cref{sec:PIGauss_M_zero}, it can be viewed as one of the simplest and straightforward approaches to sample from a target Gaussian distribution.

\subsubsection{Exact matrix splitting} \label{subsec:matrix_split_exact}
Given the multivariate Gaussian distribution $\mathcal{N}(\Bs{\mu},\B{Q}^{-1})$ with density $\pi$ in \cref{eq:Gaussian_target_Sigma}, an attractive and simple option is to sequentially draw one component of $\Bs{\theta}$ given the others.
This is the well-known component-wise Gibbs sampler, see \cref{algo:component_wise_Gibbs} \cite{Geman1984,Gelman95,Rue2005}.
The main advantage of \cref{algo:component_wise_Gibbs} is its simplicity and the low cost per \textit{sweep} (i.e., internal iteration) of $\mathcal{O}(d^2)$ flops which is comparable with Cholesky applied to Toeplitz covariance matrices \cite{Trench1964}.
More generally, one can also consider random sweeps over the $d$ components of $\Bs{\theta}$ or block-wise strategies which update simulteanously several components of $\Bs{\theta}$.
The analysis of these strategies and their respective convergence rates are detailed in \cite{RobertsSahu1997}.
\begin{algorithm}
\caption{Component-wise Gibbs sampler}
\label{algo:component_wise_Gibbs}
\hspace*{\algorithmicindent} \textbf{Input:} Number $T$ of iterations and initialization $\Bs{\theta}^{(0)}$.
\begin{algorithmic}[1]
\State Set $t = 1$.
\While{$t\leq T$}
  \For{$i \in [d]$}
    \State Draw $z \sim \mathcal{N}(0,1)$.
    \State Set $\theta^{(t)}_i = \dfrac{[\B{Q}\Bs{\mu}]_i}{Q_{ii}} + \dfrac{z}{\sqrt{Q_{ii}}} - \dfrac{1}{Q_{ii}}\pr{\displaystyle\sum_{j > i}Q_{ij}\theta_j^{(t-1)} + \displaystyle\sum_{j < i}Q_{ij}\theta_j^{(t)}}$.
  \EndFor
  \State Set $t = t + 1$.
\EndWhile\\
\Return $\Bs{\theta}^{(T)}$.
\end{algorithmic}
\end{algorithm}

In \cite{Adler1981,Barone1990,Goodman1989}, the authors showed by rewriting \cref{algo:component_wise_Gibbs} using a matrix formulation that it actually stands for a stochastic version of the Gauss-Seidel linear solver that relies on the decomposition $\B{Q}=\B{L}+\B{D}+\B{L}^{\top}$ where $\B{L}$ and $\B{D}$ are the strictly lower triangular and diagonal parts of $\B{Q}$, respectively. Indeed, each iteration solves the linear system
\begin{equation}
 (\B{L} + \B{D})\Bs{\theta}^{(t)} = \B{Q}\Bs{\mu} + \B{D}^{1/2}\B{z} - \B{L}^{\top}\Bs{\theta}^{(t-1)}\eqsp,
  \label{eq:Gauss_Seidel}
\end{equation}
where $\B{z} \sim \mathcal{N}(\B{0}_d,\B{I}_d)$.
By setting $\B{M} = \B{L}+\B{D}$ and $\B{N} = -\B{L}^{\top}$ so that $\B{Q} = \B{M} - \B{N}$, the updating rule \cref{eq:Gauss_Seidel} can be written as solving the usual Gauss-Seidel linear system
\begin{equation}
  \B{M}\Bs{\theta}^{(t)} = \B{Q}\Bs{\mu} +  \tilde{\B{z}}+\B{N}\Bs{\theta}^{(t-1)}\eqsp,
  \label{eq:matrix_splitting}
\end{equation}
where $\B{N} = -\B{L}^{\top}$ is strictly upper triangular and $\tilde{\B{z}} \sim \mathcal{N}(\B{0}_d,\B{D})$ is easy to sample.

Interestingly, \cref{eq:matrix_splitting} stands for a perturbed instance of MS schemes which are a class of linear iterative solvers  based on the splitting of $\B{Q}$ into $\B{Q} = \B{M} - \B{N}$ \cite{Golub1989,Saad2003}.
Capitalizing on this one-to-one equivalence between samplers and linear solvers, the authors in \cite{Fox2017} extended \cref{algo:component_wise_Gibbs} to other MCMC samplers based on different matrix splittings $\B{Q} = \B{M} - \B{N}$. They are reported in \cref{table:matrix_splitting} and yield \cref{algo:matrix_splitting}.
The acronym SOR stands for \textit{successive over-relaxation}.
\begin{algorithm}
\caption{MCMC sampler based on exact matrix splitting}
\label{algo:matrix_splitting}
\hspace*{\algorithmicindent} \textbf{Input:} Number $T$ of iterations, initialization $\Bs{\theta}^{(0)}$ and splitting $\B{Q} = \B{M} - \B{N}$.
\begin{algorithmic}[1]
\State Set $t=1$.
\While{$t\leq T$}
  \State Draw $\tilde{\B{z}} \sim \mathcal{N}(\B{0}_d,\B{M}^{\top} + \B{N})$.
  \State Solve $\B{M}\Bs{\theta}^{(t)} = \B{Q}\Bs{\mu} + \tilde{\B{z}} + \B{N}\Bs{\theta}^{(t-1)} $ w.r.t. $\Bs{\theta}^{(t)}$.
  \State Set $t = t + 1$.
\EndWhile\\
\Return $\Bs{\theta}^{(T)}$.
\end{algorithmic}
\end{algorithm}
\paragraph{Algorithmic efficiency}
Similarly to linear solvers, \cref{algo:matrix_splitting} is guaranteed to converge towards the correct distribution $\mathcal{N}(\Bs{\mu},\B{Q}^{-1})$ if $\rho\pr{\B{M}^{-1}\B{N}} < 1$ where $\rho(\cdot)$ stands for the spectral radius of a matrix.
In practice, \cref{algo:matrix_splitting} is stopped after $T$ iterations and the error between the distribution of $\Bs{\theta}^{(T)}$ and $\mathcal{N}(\Bs{\mu},\B{Q}^{-1})$ can be assessed quantitatively, see \cite{Fox2017}.
The computational efficiency of \cref{algo:matrix_splitting} is directly related to the complexity of solving the linear systems $\B{M}\Bs{\theta}^{(t)} = \B{Q}\Bs{\mu} +  \tilde{\B{z}}+\B{N}\Bs{\theta}^{(t-1)}$, similar to \eqref{eq:matrix_splitting}, and the difficulty of sampling $\tilde{\B{z}}$ with covariance $\B{M}^{\top}+\B{N}$.   
As pointed out in \cite{Fox2017}, the simpler $\B{M}$, the denser $\B{M}^{\top} + \B{N}$ and the more difficult the sampling of $\tilde{\B{z}}$.
For instance, Jacobi and Richardson schemes yield a simple diagonal linear system requiring $\mathcal{O}(d)$ flops but one has to sample from a Gaussian distribution with an arbitrary covariance matrix, see step 3 of \cref{algo:matrix_splitting}.
Iterative samplers requiring at least $K$ steps such as those reviewed in \cref{sec:PIGauss_M_zero} can be used.
This yields a computational burden of $\mathcal{O}(KTd^2)$ flops for step 3 and as such Jacobi and Richardson samplers admit a computational cost of $\mathcal{O}(KTd^2)$ and a storage requirement of $\Theta(d)$.
On the other hand, both Gauss-Seidel and SOR schemes are associated to a simple sampling step which can be performed in $\mathcal{O}(d)$ flops with \cref{algo:multi_diag} but one has to solve a lower triangular system which can be done in $\mathcal{O}(d^2)$ flops via forward substitution.
In order to mitigate the trade-off between steps 3 and 4, approximate MS approaches have been proposed recently \cite{Barbos2017,Johnson2013}, see \cref{subsec:matrix_split_approx}.

\begin{table}
{\footnotesize
  \caption{Examples of MS schemes for $\B{Q}$ which can be used in \cref{algo:matrix_splitting}. The matrices $\B{D}$ and $\B{L}$ denote the diagonal and strictly lower triangular parts of $\B{Q}$, respectively. The vector $\tilde{\B{z}}$ is the one appearing in step 3 of \cref{algo:matrix_splitting} and $\omega$ is a positive scalar.}
  \label{table:matrix_splitting}
  \begin{center}
  {\renewcommand{\arraystretch}{1.5}
    \begin{tabular}{|l|c|c|c|c|} 
      \hline
      \textbf{Sampler} & $\B{M}$ & $\B{N}$ & $\mathrm{cov}(\tilde{\B{z}}) = \B{M}^{\top} + \B{N}$ & convergence\\
      \hline 
      Richardson & $\B{I}_d/\omega$ & $\B{I}_d/\omega - \B{Q}$ & $2\B{I}_d/\omega - \B{Q}$ & $0 < \omega < 2/\nr{\B{Q}}$\\ 
      Jacobi & $\B{D}$ & $\B{D} - \B{Q}$ & $2\B{D} - \B{Q}$ & $|Q_{ii}| > \sum_{j\neq i}|Q_{ij}|$ $\forall i \in [d]$\\
      Gauss-Seidel & $\B{D} + \B{L}$ & $-\B{L}^{\top}$ & $\B{D}$ & always\\
      SOR & $\B{D}/\omega + \B{L}$ & $\frac{1-\omega}{\omega}\B{D} - \B{L}^{\top}$ & $\frac{2-\omega}{\omega}\B{D}$ & $0 < \omega < 2$\\[1em]
      \hline
    \end{tabular}}
  \end{center}
}
\end{table}
\noindent{\textbf{Polynomial accelerated Gibbs samplers.}}
When the splitting $\B{Q}=\B{M}-\B{N}$ is symmetric, that is both $\B{M}$ and $\B{N}$ are symmetric matrices, the rate of convergence of \cref{algo:matrix_splitting} can be improved by using polynomial preconditioners \cite{Fox2017}.
For the ease of presentation, we will first explain how such a preconditioning accelerates linear solvers based on matrix splitting, before building upon the one-to-one equivalence between linear solvers and Gibbs samplers to show how \cref{algo:matrix_splitting} can be accelerated.
Given a linear system $\B{Q}\Bs{\theta} = \B{v}$ for $\B{v} \in \mathbb{R}^d$, linear solvers based on the matrix splitting $\B{Q} = \B{M}-\B{N}$ consider the recursion, for any $t \in \mathbb{N}$ and $\Bs{\theta}^{(0)} \in \mathbb{R}^d$, $\Bs{\theta}^{(t+1)} = \Bs{\theta}^{(t)} +  \B{M}^{-1}(\B{v}-\B{Q}\Bs{\theta}^{(t)})$.
The error at iteration $t$ defined by $\B{e}^{(t+1)} = \Bs{\theta}^{(t+1)} - \B{Q}^{-1}\B{v}$ can be shown to be equal to $(\B{I}_d-\B{M}^{-1}\B{Q})^t\B{e}^{(0)}$ \cite{Golub1989}.
Since this error is a $t$-th order polynomial of $\B{M}^{-1}\B{Q}$, it is then natural to wonder whether one can find another $t$-th order polynomial $\mathsf{P}_t$ that achieves a lower error that is $\rho(\mathsf{P}_t(\B{M}^{-1}\B{Q})) < \rho((\B{I}_d-\B{M}^{-1}\B{Q})^t)$.
This can be accomplished by considering the second-order iterative scheme defined, for any $t \in \mathbb{N}$, by \cite{axelsson_1994}
$$
\Bs{\theta}^{(t+1)} = \alpha_t\Bs{\theta}^{(t)} + (1-\alpha_t)\Bs{\theta}^{(t-1)} + \beta_t \B{M}^{-1}(\B{v}-\B{Q}\Bs{\theta}^{(t)})\eqsp,
$$
where $(\alpha_t,\beta_t)_{t \in \mathbb{N}}$ are a set of acceleration parameters.
This iterative method yields an error at step $t$ given by  $\B{e}^{(t+1)}=\mathsf{P}_t(\B{M}^{-1}\B{Q})\B{e}^{(0)}$ where $\mathsf{P}_t$ stands for a scaled Chebyshev polynomial, see \Cref{eq:cheby}.
Optimal values for $(\alpha_t,\beta_t)_{t \in \mathbb{N}}$ are given by \cite{axelsson_1994}
$$
\alpha_t = \tau_1\beta_t  \quad \text{and} \quad \beta_t = \pr{\tau_1 - \tau_2^2\beta_{t-1}}^{-1} \ , 
$$
$\tau_1 = [\lambda_{\mathrm{min}}(\B{M}^{-1}\B{Q}) + \lambda_{\mathrm{max}}(\B{M}^{-1}\B{Q})]/2$ and $\tau_2 = [\lambda_{\mathrm{max}}(\B{M}^{-1}\B{Q}) - \lambda_{\mathrm{min}}(\B{M}^{-1}\B{Q})]/4$.
Note that these optimal choices suppose that the minimal and maximal eigenvalues of $\B{M}^{-1}\B{Q}$ are real-valued and available.
The first requirement is for instance satisfied if the splitting $\B{Q}=\B{M}-\B{N}$ is symmetric while the second one is met by using the CG algorithm as explained in \cite{Fox2017}.
In the literature \cite{RobertsSahu1997,Fox2017}, a classical symmetric splitting scheme that has been considered is derived from the SOR splitting and as such called \textit{symmetric SOR} (SSOR).
Denote by $\B{M}_{\mathrm{SOR}}$ and $\B{N}_{\mathrm{SOR}}$ the matrices involved in the SOR splitting such that $\B{Q} = \B{M}_{\mathrm{SOR}} - \B{N}_{\mathrm{SOR}}$, see row 4 of \Cref{table:matrix_splitting}.
Then for any $0<\omega<2$, the SSOR splitting is defined by $\B{Q} = \B{M}_{\mathrm{SSOR}} - \B{N}_{\mathrm{SSOR}}$ with 
$$
\B{M}_{\mathrm{SSOR}} = \frac{\omega}{2-\omega}\B{M}_{\mathrm{SOR}}\B{D}^{-1}\B{M}_{\mathrm{SOR}}^{\top} \quad \text{and} \quad \B{N}_{\mathrm{SSOR}} = \frac{\omega}{2-\omega}\B{N}_{\mathrm{SOR}}\B{D}^{-1}\B{N}_{\mathrm{SOR}}^{\top} \ .
$$
By resorting to the one-to-one equivalence between linear solvers and Gibbs samplers, \cite{FoxParker14,Fox2017} showed that the above acceleration via Chebyshev polynomials can be applied to Gibbs samplers based on a symmetric splitting.
In this context, the main challenge when dealing with accelerated Gibbs samplers compared to accelerated linear solvers is the calibration of the noise covariance to ensure that the invariant distribution coincides with $\mathcal{N}(\Bs{\mu},\B{Q}^{-1})$.
For the sake of completeness, the pseudo-code associated to an accelerated version of \cref{algo:matrix_splitting} based on the SSOR splitting is detailed in \cref{algo:acceleratedSSOR}.
Associated convergence results and numerical studies associated to this algorithm can be found in \cite{FoxParker14,Fox2017}.
\begin{algorithm}
\caption{Chebyshev accelerated SSOR sampler}
\label{algo:acceleratedSSOR}
\hspace*{\algorithmicindent} \textbf{Input:} SSOR tuning parameter $0<\omega<2$, extreme eigenvalues $\lambda_{\mathrm{min}}(\B{M}^{-1}_{\mathrm{SSOR}}\B{Q})$ and $\lambda_{\mathrm{max}}(\B{M}^{-1}_{\mathrm{SSOR}}\B{Q})$ of $\B{M}^{-1}_{\mathrm{SSOR}}\B{Q}$, number $T$ of iterations, initialization $\B{w}^{(0)}$, diagonal $\B{D}$ of $\B{Q}$ and SOR splitting $\B{Q} = \B{M}_{\mathrm{SOR}} - \B{N}_{\mathrm{SOR}}$.
\begin{algorithmic}[1]
\State Set $\B{D}_{\omega} = (2/\omega-1)\B{D}$.
\State Set $\sqrt{\delta}=(\lambda_{\mathrm{max}}(\B{M}^{-1}_{\mathrm{SSOR}}\B{Q})-\lambda_{\mathrm{min}}(\B{M}^{-1}_{\mathrm{SSOR}}\B{Q}))/4$.
\State Set $\tau = 2/(\lambda_{\mathrm{max}}(\B{M}^{-1}_{\mathrm{SSOR}}\B{Q})+\lambda_{\mathrm{min}}(\B{M}^{-1}_{\mathrm{SSOR}}\B{Q}))$.
\State Set $\beta=2\tau$, $\alpha=1$, $e = 2/\alpha - 1$, $c=(2/\tau-1)e$ and $\kappa=\tau$. 
\State Set $t=1$.
\While{$t\leq T$}
  \State Draw $\B{z}_1 \sim \mathcal{N}(\B{0}_d,\B{I}_d)$.
  \State Solve $\B{M}_{\mathrm{SOR}}\B{x}_1 = \B{M}_{\mathrm{SOR}}\B{w}^{(t-1)} + \sqrt{e}\B{D}_{\omega}^{1/2}\B{z}_1 - \B{Q}\B{w}^{(t-1)}$ w.r.t. $\B{x}_1$.
  \State Draw $\B{z}_2 \sim \mathcal{N}(\B{0}_d,\B{I}_d)$.
  \State Solve $\B{M}_{\mathrm{SOR}}^{\top}\B{x}_2 = \B{M}_{\mathrm{SOR}}^{\top}(\B{x}_1-\B{w}^{(t-1)}) + \sqrt{c}\B{D}_{\omega}^{1/2}\B{z}_2 - \B{Q}\B{x}_1$ w.r.t. $\B{x}_2$.
  \If{$t=1$}
    \State Set $\B{w}^{(t)} = \alpha(\B{w}^{(t-1)} + \tau \B{x}_2)$.
    \State Set $\Bs{\theta}^{(t)} = \Bs{\mu} + \B{w}^{(t)}$.
  \Else
    \State Set $\B{w}^{(t)} = \alpha(\B{w}^{(t-1)} - \B{w}^{(t-2)} + \tau \B{x}_2) + \B{w}^{(t-2)}$.
    \State Set $\Bs{\theta}^{(t)} = \Bs{\mu} + \B{w}^{(t)}$.
 \EndIf
  \State Set $\beta=\frac{1}{1/\tau-\beta\delta}$, $\alpha=\frac{\beta}{\tau}$, $e = \frac{2\kappa(1-\alpha)}{\beta} + 1$, $c=\frac{2}{\tau}-1 + (e-1)(\frac{1}{\tau}+\frac{1}{\kappa}-1)$ and $\kappa = \beta + (1-\alpha)\kappa$. 
  \State Set $t = t + 1$.
\EndWhile\\
\Return $\Bs{\theta}^{(T)}$.
\end{algorithmic}
\end{algorithm}

\paragraph{Algorithmic efficiency}
Similarly to \cref{algo:matrix_splitting}, \cref{algo:acceleratedSSOR} is exact in the sense that it admits $\mathcal{N}(\Bs{\mu},\B{Q}^{-1})$ as invariant distribution.
\cite{Fox2017} gave guidelines to choose the truncation parameter $T$ such that the error between the distribution of $\Bs{\theta}^{(T)}$ and $\mathcal{N}(\Bs{\mu},\B{Q}^{-1})$ is sufficiently small.
Regarding computation and storage, since triangular linear systems can be solved in $\mathcal{O}(d^2)$ flops by either back or forward substitution, \cref{algo:acceleratedSSOR} admits a computational cost of $\mathcal{O}(Td^2)$ and a storage requirement of $\Theta(d)$.

\subsubsection{Approximate matrix splitting}  \label{subsec:matrix_split_approx}
Motivated by efficiency and parallel computations, the authors in \cite{Barbos2017} and \cite{Johnson2013} proposed to relax exact MS and introduced two MCMC samplers whose invariant distributions are  approximations of $\mathcal{N}(\Bs{\mu},\B{Q}^{-1})$. 
First, in order to solve efficiently the linear system $\B{M}\Bs{\theta}^{(t)} = \B{Q}\Bs{\mu} + \tilde{\B{z}}+\B{N}\Bs{\theta}^{(t-1)}$ involved in step 4 of \cref{algo:matrix_splitting}, these approximate approaches consider MS schemes with diagonal matrices $\B{M}$.
For exact samplers, e.g., Richardson and Jacobi, we saw in the previous paragraph that such a convenient structure for $\B{M}$ implies that the drawing of the Gaussian vector $\tilde{\B{z}}$ becomes more demanding. 
To bypass this issue, approximate samplers draw Gaussian vectors $\tilde{\B{z}}'$ with simpler covariance matrices $\tilde{\B{M}}$ instead of $\B{M}^{\top} + \B{N}$.
Again, attractive choices for $\tilde{\B{M}}$ are diagonal matrices since the associated sampling task then boils down to \cref{algo:multi_diag}.
This yields \cref{algo:approx_matrix_splitting} which is highly amenable to parallelization since both the covariance matrix $\tilde{\B{M}}$ of $\tilde{\B{z}}'$ and the matrix $\B{M}$ involved in the linear system to solve are diagonal.
\cref{table:approx_matrix_splitting} gathers the respective expressions of $\B{M}$, $\B{N}$ and $\tilde{\B{M}}$ for the two approaches introduced in \cite{Johnson2013} and \cite{Barbos2017} and coined ``Hogwild sampler'' and ``clone MCMC'', respectively.
\begin{algorithm}
\caption{MCMC sampler based on approximate matrix splitting}
\label{algo:approx_matrix_splitting}
\hspace*{\algorithmicindent} \textbf{Input:} Number $T$ of iterations, initialization $\Bs{\theta}^{(0)}$ and splitting $\B{Q} = \B{M} - \B{N}$.
\begin{algorithmic}[1]
\State Set $t = 1$.
\While{$t\leq T$}
  \State Draw $\tilde{\B{z}}' \sim \mathcal{N}(\B{0}_d,\tilde{\B{M}})$.   \Comment{ {\textcolor{green}{$\tilde{\B{M}}=\B{D}$ or $2\pr{\B{D}+ 2\omega\B{I}_d}$, see \cref{table:approx_matrix_splitting}.}}}
  \State Solve $\B{M}\Bs{\theta}^{(t)} = \B{Q}\Bs{\mu} + \tilde{\B{z}}' + \B{N}\Bs{\theta}^{(t-1)}$.
  \State Set $t = t + 1$.
\EndWhile\\
\Return $\Bs{\theta}^{(T)}$.
\end{algorithmic}
\end{algorithm}

\paragraph{Algorithmic efficiency}
Regarding sampling accuracy, the Hogwild sampler and clone MCMC define a Markov chain whose invariant distribution is Gaussian with the correct mean $\Bs{\mu}$ but with precision matrix $\widetilde{\B{Q}}_{\mathrm{MS}}$, where 
\begin{equation*}
  \widetilde{\B{Q}}_{\mathrm{MS}} =
    \begin{cases}
      \B{Q}\pr{\B{I}_d - \B{D}^{-1}(\B{L}+\B{L}^{\top})} & \text{for the Hogwild sampler}\\
      \B{Q}\pr{\B{I}_d - \frac{1}{2}(\B{D} + 2\omega^{-1}\B{I}_d)^{-1}\B{Q}} & \text{for clone MCMC.}
    \end{cases}       
\end{equation*}
Contrary to the Hogwild sampler, clone MCMC is able to sample exactly from $\mathcal{N}(\Bs{\mu},\B{Q}^{-1})$ in the asymptotic scenario $\omega \rightarrow 0$ since in this case $\widetilde{\B{Q}}_{\mathrm{MS}} \rightarrow \B{Q}$.
While keeping a memory requirement of $\Theta(d)$, the induced approximation yields a highly parallelizable sampler.
Indeed, compared to \cref{algo:matrix_splitting}, the computational complexities associated to step 3 and the solving of the triangular system in step 4 are decreased by an order of magnitude to $\mathcal{O}(d)$.
Note that the overall computational complexity of step 4 is still $\mathcal{O}(d^2)$ because of the matrix-vector product $\B{N}\Bs{\theta}^{(t-1)}$.

\begin{table}
{\footnotesize
  \caption{MS schemes for $\B{Q}$ which can be used in \cref{algo:approx_matrix_splitting}. The matrices $\B{D}$ and $\B{L}$ denote the diagonal and strictly lower triangular parts of $\B{Q}$, respectively. The vector $\tilde{\B{z}}'$ is the one appearing in step 3 of \cref{algo:approx_matrix_splitting} and $\omega > 0$ is a tuning parameter controlling the bias of those methods. Sufficient conditions to guarantee $\rho(\B{M}^{-1}\B{N}) < 1$ are given in \cite{Johnson2013,Barbos2017}.}
  \label{table:approx_matrix_splitting}
  \begin{center}
  {\renewcommand{\arraystretch}{1.5}
    \begin{tabular}{|l|c|c|c|} 
      \hline
      \textbf{Sampler} & $\B{M}$ & $\B{N}$ & $\mathrm{cov}(\tilde{\B{z}}') = \tilde{\B{M}}$ \\
      \hline 
      Hogwild with blocks of size 1 \cite{Johnson2013} & $\B{D}$ & $-\B{L} - \B{L}^{\top}$ & $\B{D}$ \\ 
      Clone MCMC \cite{Barbos2017} & $\B{D} + 2\omega\B{I}_d$ & $2\omega\B{I}_d -\B{L} - \B{L}^{\top}$ & $2\pr{\B{D}+ 2\omega\B{I}_d}$\\[0.3em]
      \hline
    \end{tabular}}
  \end{center}
}
\end{table}

\subsection{Data augmentation}
\label{subsec:data_aug}

Since the precision matrix $\B{Q}$ has been assumed to be arbitrary, the MS schemes $\B{Q} = \B{M} - \B{N}$ in \cref{table:matrix_splitting} were not motivated by its structure but rather by the computational efficiency of the associated samplers.
Hence, inspired by efficient linear solvers, relevant choices for $\B{M}$ and $\B{N}$ given in \cref{table:matrix_splitting} and \cref{table:approx_matrix_splitting} have been considered.
Another line of search explores schemes specifically dedicated to precision matrices $\B{Q}$ of the form
\begin{equation}
  \B{Q} = \B{Q}_1 + \B{Q}_2\eqsp,
  \label{eq:precision_sum}
\end{equation}
where, contrary to the MS schemes discussed in the previous section, the two matrices $\B{Q}_1$ and $\B{Q}_2$ are not chosen by the user but directly result from the statistical model under consideration. In particular, such situations arise when deriving hierarchical Bayesian models (see, e.g., \cite{Rue2005,Idier2008,Orieux2010}). 
By capitalizing on possible specific structures of $\{\B{Q}_i\}_{i \in [2]}$, it may be desirable to separate $\B{Q}_1$ and $\B{Q}_2$ in two different hopefully simpler steps of a Gibbs sampler. To this purpose, this section discusses \textit{data augmentation} (DA) approaches which introduce one (or several) auxiliary variable $\B{u} \in \mathbb{R}^k$ such that the joint distribution of the couple $(\Bs{\theta},\B{u})$ yields simple conditional distributions thus sampling steps within a Gibbs sampler \cite{Barbos2017,Marnissi2018,Vono2019,Marnissi2019}.
Then a straightforward marginalization of the auxiliary variable $\B{u}$ permits to retrieve the target distribution $\mathcal{N}(\Bs{\mu},\B{Q}^{-1})$, either exactly or in an asymptotic regime depending on the nature of the DA scheme.
Both exact and approximate DA methods have been proposed.

\subsubsection{Exact data augmentation} \label{subsec:data_aug_exact}
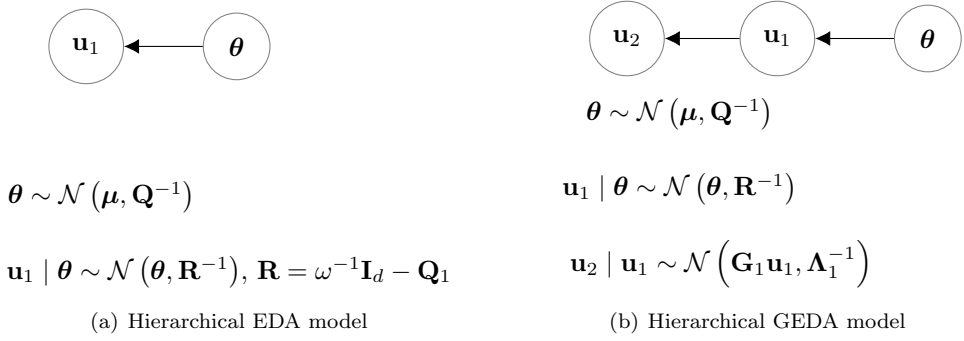
\begin{figure}
\centering
\subfigure[Hierarchical EDA model]
{\begin{tikzpicture}
  \node[circle,draw=gray,inner sep=2mm] (u1) at (1,1) {$\B{u}_1$};
  \node[circle,draw=gray,inner sep=2mm] (theta) at (3,1) {$\boldsymbol{\theta}$};
  \node (pi1) at (1.2,-1) {$ \Bs{\theta} \sim \mathcal{N}\pr{\Bs{\mu},\B{Q}^{-1}}$};
  \node (pi2) at (2.9,-2) {$ \B{u}_1 \mid \Bs{\theta} \sim \mathcal{N}\pr{\Bs{\theta},\B{R}^{-1}}$, $\B{R} = \omega^{-1}\B{I}_d - \B{Q}_1$};
  \draw[->] (theta) -- (u1);
\end{tikzpicture}}
\hspace{1cm}
\subfigure[Hierarchical GEDA model]
{\begin{tikzpicture}
  \node[circle,draw=gray,inner sep=2mm] (u2) at (1,0) {$\B{u}_2$};
  \node[circle,draw=gray,inner sep=2mm] (u1) at (3,0) {$\B{u}_1$};
  \node[circle,draw=gray,inner sep=2mm] (theta) at (5,0) {$\boldsymbol{\theta}$};
  \node (pi1) at (1.7,-1) {$ \Bs{\theta} \sim \mathcal{N}\pr{\Bs{\mu},\B{Q}^{-1}}$};
  \node (pi2) at (1.7,-2) {$ \B{u}_1 \mid \Bs{\theta} \sim \mathcal{N}\pr{\Bs{\theta},\B{R}^{-1}}$};
  \node (pi3) at (2.25,-3) {$ \B{u}_2 \mid \B{u}_1 \sim \mathcal{N}\pr{\B{G}_1\B{u}_1,\B{\Lambda}_1^{-1}}$};
  \draw[->] (theta) -- (u1);
  \draw[->] (u1) -- (u2);
\end{tikzpicture}}
\caption{Hierarchical models proposed in \cite{Marnissi2018,Marnissi2019} where $\omega$ is such that $0 < \omega < \nr{\B{Q}_1}^{-1}$.}
\label{fig:hierarchical_models_Marnissi}
\end{figure}

This paragraph reviews some exact DA approaches to obtain samples from $\mathcal{N}(\Bs{\mu},\B{Q}^{-1})$.
The term \textit{exact} means here that the joint distribution of $(\Bs{\theta},\B{u})$ admits a density $\pi(\Bs{\theta},\B{u})$ which satisfies almost surely
\begin{equation}
  \label{eq:exact_DA}
  \int_{\mathbb{R}^k} \pi(\Bs{\theta},\B{u}) = \pi(\Bs{\theta})\eqsp,
\end{equation}
and yields proper marginal distributions. 
\Cref{fig:hierarchical_models_Marnissi} describes the directed acyclic graphs (DAG) associated with two hierarchical models proposed in \cite{Marnissi2018,Marnissi2019} to decouple $\B{Q}_1$ from $\B{Q}_2$ by involving auxiliary variables.  
In the following, we detail the motivations behind these two data augmentation schemes.
Among the two matrices $\B{Q}_1$ and $\B{Q}_2$ involved in the composite precision matrix $\B{Q}$, without loss of generality, we assume that $\B{Q}_2$ presents a particular and simpler structure (e.g., diagonal or circulant) than $\B{Q}_1$. We want now to benefit from this structure by leveraging the efficient sampling schemes previously discussed in \cref{subsec:special_instances} and well suited to handle a Gaussian distribution with a precision matrix only involving $\B{Q}_2$. This is the aim of the first data augmentation model called EDA which introduces the joint distribution with p.d.f.
\begin{align}
  \pi(\Bs{\theta},\B{u}_1) &\propto \exp\pr{-\dfrac{1}{2}\br{(\Bs{\theta}-\Bs{\mu})^{\top}\B{Q}(\Bs{\theta}-\Bs{\mu}) + (\B{u}_1-\Bs{\theta})^{\top}\B{R}(\B{u}_1-\Bs{\theta})}}\eqsp, \label{eq:EDA}
\end{align}
with $\B{R} = \omega^{-1}\B{I}_d - \B{Q}_1$ and $0< \omega < \nr{\B{Q}_1}^{-1}$, where $\nr{\cdot}$ is the spectral norm. The resulting Gibbs sampler (see \cref{algo:exact_DA}) relies on two conditional Gaussian sampling steps whose associated conditional distributions are detailed in  \Cref{table:exact_DA}. This scheme has the great advantage of decoupling the two precision matrices $\B{Q}_1$ and $\B{Q}_2$ since they are not simultaneously involved in any of the two steps. In particular, introducing the auxiliary variable $\B{u}_1$ permits to remove the dependence in $\B{Q}_1$ when defining the precision matrix of the conditional distribution of $\Bs{\theta}$.
While efficient sampling from this conditional is possible, we have to ensure that sampling the auxiliary variable $\B{u}_1$ can be achieved with a reasonable computational cost.
Again, if $\B{Q}_1$ presents a nice structure, the specific approaches reviewed in \cref{subsec:special_instances} can be employed.
If this is not the case, the authors in \cite{Marnissi2018,Marnissi2019} proposed a generalization of EDA, called GEDA, to simplify the whole Gibbs sampling procedure when $\B{Q}$ arises from a hierarchical Bayesian  model.
In such models, $\B{Q}_1$, as a fortiori $\B{Q}_2$, naturally admits an explicit decomposition which writes $\B{Q}_1 = \B{G}_1^{\top}\Bs{\Lambda}_1\B{G}_1$, where $\Bs{\Lambda}_1$ is a positive definite (and very often diagonal) matrix.
By building on this explicit decomposition, GEDA introduces an additional auxiliary variable $\B{u}_2$ such that the augmented p.d.f. writes
\begin{align}
  \pi(\Bs{\theta},\B{u}_1,\B{u}_2) &\propto \exp\pr{-\dfrac{1}{2}\br{(\Bs{\theta}-\Bs{\mu})^{\top}\B{Q}(\Bs{\theta}-\Bs{\mu}) + (\B{u}_1-\Bs{\theta})^{\top}\B{R}(\B{u}_1-\Bs{\theta})}} \nonumber \\
  &\times \exp\pr{-\dfrac{1}{2}(\B{u}_2-\B{G}_1\B{u}_1)^{\top}\B{\Lambda}_1(\B{u}_2-\B{G}_1\B{u}_1)}\eqsp. \label{eq:GEDA}
\end{align}
The associated joint distribution yields conditional Gaussian distributions with diagonal covariance matrices for both $\B{u}_1$ and $\B{u}_2$ that can be sampled efficiently with \cref{algo:multi_diag}, see \Cref{table:exact_DA}.

\paragraph{Algorithmic efficiency}
First, both EDA and GEDA admit $\mathcal{N}(\Bs{\mu},\B{Q}^{-1})$ as invariant distribution and hence are exact.
Regarding EDA, since the conditional distribution of $\B{u}_1\mid\Bs{\theta}$ might admit an arbitrary precision matrix in the worst-case scenario, its computational and storage complexities are $\mathcal{O}(KTd^2)$ and $\Theta(d)$ where $K$ stands for a truncation parameter associated to one of the algorithms reviewed in \cref{sec:PIGauss_M_zero}.
On the other hand, GEDA benefits from an additional data augmentation which yields reduced computational and storage requirements of $\mathcal{O}(Td^2)$ and $\Theta(d)$.

\begin{table}
{\scriptsize
  \caption{Conditional probability distributions of $\Bs{\theta} \mid \B{u}_1$, $\B{u}_1\mid\Bs{\theta},\B{u}_2$ and $\B{u}_2\mid\B{u}_1$ for the exact data augmentation schemes detailed in \cref{subsec:data_aug_exact}. The parameter $\omega$ is such that $0 < \omega < \nr{\B{Q}_1}^{-1}$.
  For simplicity, the conditioning is notationally omitted.}
  \label{table:exact_DA}
  \begin{center}
   {\renewcommand{\arraystretch}{2}
    \begin{tabular}{|l|c|c|c|} 
      \hline
      \textbf{Sampler} & $\Bs{\theta} \sim \mathcal{N}(\Bs{\mu}_{\Bs{\theta}},\B{Q}_{\Bs{\theta}}^{-1})$ & $\B{u}_1 \sim \mathcal{N}(\Bs{\mu}_{\B{u}_1},\B{Q}_{\B{u}_1}^{-1})$ & $\B{u}_2  \sim \mathcal{N}(\Bs{\mu}_{\B{u}_2},\B{Q}_{\B{u}_2}^{-1})$ \\
      \hline 
      \multirow{3}{*}{EDA} & $\B{Q}_{\Bs{\theta}} = \omega^{-1}\B{I}_d + \B{Q}_2$ & $\B{Q}_{\B{u}_1} = \B{R}$ & - \\
      & $\Bs{\mu}_{\Bs{\theta}} = \B{Q}_{\Bs{\theta}}^{-1}\pr{\B{R}\B{u}_1 + \B{Q}\Bs{\mu}}$ & $\Bs{\mu}_{\B{u}_1} = \Bs{\theta}$ & - \\[0.3em]
      \hline
      \multirow{3}{*}{GEDA} 
      & $\B{Q}_{\Bs{\theta}} = \omega^{-1}\B{I}_d + \B{Q}_2$ 
      & $\B{Q}_{\B{u}_1} = \omega^{-1}\B{I}_d$ & $\B{Q}_{\B{u}_2} = \B{\Lambda}_1$ \\
      & $\Bs{\mu}_{\Bs{\theta}} = \B{Q}_{\Bs{\theta}}^{-1}(\B{R}\B{u}_1 + \B{Q}\Bs{\mu})$ 
      & $\Bs{\mu}_{\B{u}_1} = \Bs{\theta} - \omega(\B{Q}_1\Bs{\theta} -\B{G}_1^{\top}\B{\Lambda}_1^{-1}\B{u}_2)$ & $\Bs{\mu}_{\B{u}_2} = \B{G}_1\B{u}_1$ \\[0.3em]
      \hline
    \end{tabular}}
  \end{center}
}
\end{table}
\begin{algorithm}
\caption{Gibbs sampler based on exact data augmentation (G)EDA}
\label{algo:exact_DA}
\hspace*{\algorithmicindent} \textbf{Input:} Number $T$ of iterations and initialization $\Bs{\theta}^{(0)}$, $\B{u}_1^{(0)}$.
\begin{algorithmic}[1]
\State Set $t = 1$.
\While{$t\leq T$}
  \State Draw $\B{u}_2^{(t)} \sim \mathcal{N}(\Bs{\mu}_{\B{u}_2},\B{Q}_{\B{u}_2}^{-1})$. \Comment{\textcolor{green}{Only if GEDA is considered.}}
  \State Draw $\B{u}_1^{(t)} \sim \mathcal{N}(\Bs{\mu}_{\B{u}_1},\B{Q}_{\B{u}_1}^{-1})$.
  \State Draw $\Bs{\theta}^{(t)} \sim \mathcal{N}(\Bs{\mu}_{\Bs{\theta}},\B{Q}_{\Bs{\theta}}^{-1})$.
  \State Set $t = t + 1$.
\EndWhile\\
\Return $\Bs{\theta}^{(T)}$.
\end{algorithmic}
\end{algorithm}

\subsubsection{Approximate data augmentation} \label{subsec:data_aug_approx}
An approximate data augmentation scheme inspired by variable-splitting approaches in optimization \cite{Boyd2011,Afonso2010,Afonso2011} was proposed in \cite{Vono2019}. This framework, also called asymptotically exact data augmentation (AXDA) \cite{Vono_2019sub}, was initially introduced to deal with any target distributions, not limited to Gaussian ones; it therefore {\em a fortiori} applies to them as well. 
An auxiliary variable $\B{u} \in \mathbb{R}^d$ is introduced such that the joint p.d.f. of ($\Bs{\theta},\B{u}$) writes
{\small
\begin{align}
  \pi(\Bs{\theta},\B{u}) \propto \exp\pr{- \dfrac{1}{2}\br{(\Bs{\theta}-\Bs{\mu})^{\top}\B{Q}_2(\Bs{\theta}-\Bs{\mu}) +  (\B{u}-\Bs{\mu})^{\top}\B{Q}_1(\B{u}-\Bs{\mu}) + \dfrac{\nr{\B{u}-\Bs{\theta}}^2}{\omega}}}\eqsp,
  \label{eq:joint_SGS}
\end{align}
}
where $\omega > 0$.
The main idea behind \cref{eq:joint_SGS} is to replicate the variable of interest $\Bs{\theta}$ in order to sample via a Gibbs sampling strategy two different random variables $\B{u}$ and $\Bs{\theta}$ with covariance matrices involving separately $\B{Q}_1$ and $\B{Q}_2$.
This algorithm, coined split Gibbs sampler (SGS), is detailed in \cref{algo:approx_DA} and sequentially draws from the conditional distributions
\begin{align}
  \B{u} \mid \Bs{\theta} &\sim \mathcal{N}\pr{(\omega^{-1}\B{I}_d + \B{Q}_1)^{-1}(\omega^{-1}\Bs{\theta} + \B{Q}_1\Bs{\mu}),(\omega^{-1}\B{I}_d + \B{Q}_1)^{-1}}\eqsp, \label{eq:SGS_u}\\
  \Bs{\theta} \mid \B{u} &\sim \mathcal{N}\pr{(\omega^{-1}\B{I}_d + \B{Q}_2)^{-1}(\omega^{-1}\B{u} + \B{Q}_2\Bs{\mu}),(\omega^{-1}\B{I}_d + \B{Q}_2)^{-1}}\eqsp. \label{eq:SGS_theta}
\end{align}
Again, this approach has the great advantage of decoupling the two precision matrices $\B{Q}_1$ and $\B{Q}_2$ defining $\B{Q}$ since they are not simultaneously involved in any of the two steps of the Gibbs sampler. In \cite{Marnissi2019}, the authors showed that exact DA schemes (i.e., EDA and GEDA) generally outperform AXDA as far as Gaussian sampling is concerned. This was expected since the AXDA framework proposed is not specifically designed for Gaussian targets but for a wide family of distributions.

\paragraph{Algorithmic efficiency}
The sampling efficiency of \cref{algo:approx_DA} depends upon the parameter $\omega$ which controls the strength of the coupling between $\B{u}$ and $\Bs{\theta}$ as well as the bias-variance trade-off of this method; it yields exact sampling when $\omega \rightarrow 0$. 
Indeed, the marginal distribution of $\boldsymbol{\theta}$ under the joint distribution with density defined in \cref{eq:joint_SGS} is a Gaussian with the correct mean $\Bs{\mu}$ but with an approximate precision matrix $\widetilde{\B{Q}}_{\mathrm{DA}}$ which admits the closed-form expression
\begin{equation}
\widetilde{\B{Q}}_{\mathrm{DA}} = \B{Q}_2 + \pr{\B{Q}_1^{-1} + \omega\B{I}_d}^{-1} \eqsp.
\end{equation}
In the worst-case scenario where $\B{Q}_1$ is arbitrary, sampling from the conditional distribution \cref{eq:SGS_u} can be performed with an iterative algorithm running $K$ iterations as those reviewed in \cref{sec:PIGauss_M_zero}. 
Hence \cref{algo:approx_DA} admits the same computational and storage complexities as EDA (see \cref{algo:exact_DA}), that is $\mathcal{O}(KTd^2)$ and $\Theta(d)$.

\begin{algorithm}
\caption{Gibbs sampler based on approximate data augmentation}
\label{algo:approx_DA}
\hspace*{\algorithmicindent} \textbf{Input:} Number $T$ of iterations and initialization $\Bs{\theta}^{(0)}$.
\begin{algorithmic}[1]
\State Set $t = 1$.
\While{$t\leq T$}
  \State Draw $\B{u}^{(t)} \sim \mathcal{N}(\Bs{\mu}_{\B{u}},\B{Q}_{\B{u}}^{-1})$ {\color{green} in \cref{eq:SGS_u}.}
  \State Draw $\Bs{\theta}^{(t)} \sim \mathcal{N}(\Bs{\mu}_{\Bs{\theta}},\B{Q}_{\Bs{\theta}}^{-1})$ {\color{green} in \cref{eq:SGS_theta}.}
  \State Set $t = t + 1$.
\EndWhile\\
\Return $\Bs{\theta}^{(T)}$.
\end{algorithmic}
\end{algorithm}

\section{A unifying revisit of Gibbs samplers via a stochastic version of the PPA}
\label{sec:PPA}

\Cref{sec:PIGauss_M_zero} and \cref{sec:PIGauss_M_nonzero} showed that numerous  approaches have been proposed to sample from a possibly high-dimensional Gaussian distribution with density  \cref{eq:Gaussian_target_Sigma}.
This section proposes to unify these approaches within a general Gaussian simulation framework which actually stands for a stochastic counterpart of the celebrated proximal point algorithm (PPA) in optimization \cite{Rockafellar1976}.
This viewpoint will shed new light on the connections between the reviewed simulation-based algorithms, and particularly between Gibbs samplers.

\subsection{A unifying proposal distribution}
\label{subsec:PIGauss}

The approaches described in \cref{sec:PIGauss_M_nonzero} use surrogate probability distributions (e.g., conditional or approximate distributions) to make Gaussian sampling easier.
In the following, we show that most of these surrogate distributions can be put under a common umbrella by considering the density

\begin{equation}
  \kappa(\Bs{\theta},\B{u}) \propto \pi(\Bs{\theta})\exp\pr{-\frac{1}{2}(\Bs{\theta}-\B{u})^{\top}\B{R}(\Bs{\theta}-\B{u})}\eqsp, \label{eq:cond_PPA}
\end{equation}
where $\B{u} \in \mathbb{R}^d$ stands for an additional (auxiliary) variable and $\B{R} \in \mathbb{R}^{d \times d}$ is a symmetric matrix acting as a preconditioner such that $\kappa$ defines a proper density on an appropriate state space.
More precisely, in the following, depending on the definition of the variable $\B{u}$, the probability density $\kappa$ in \cref{eq:cond_PPA} shall refer to either a joint p.d.f. $\pi(\Bs{\theta},\B{u})$ or a conditional probability density $\pi(\Bs{\theta}\mid\B{u})$.
Contrary to MCMC samplers detailed in \cref{sec:PIGauss_M_nonzero}, the methods described in \cref{sec:PIGauss_M_zero} do not use explicit surrogate distributions to simplify the sampling procedure.
Instead, they directly perturb deterministic approaches from numerical linear algebra without explicitly defining a simpler surrogate distribution at each iteration.
This feature can be encoded with the choice $\B{R} \rightarrow \B{0}_{d \times d}$ so that these methods can be described by this  unifying model as well.
Then, the main motivation for using the surrogate density $\kappa$ is to \textit{precondition} the initial p.d.f. $\pi$ to end up with simpler sampling steps as in \cref{sec:PIGauss_M_nonzero}.

\subsection{Revisiting MCMC sampling approaches}
\label{subsec:PPA_to_Gibbs}

This section builds on the probability kernel density \cref{eq:cond_PPA} to revisit, unify and extend the exact and approximate approaches reviewed in \cref{sec:PIGauss_M_nonzero}. 
We emphasize that exact approaches indeed target the distribution of interest $\mathcal{N}(\Bs{\mu},\B{Q}^{-1})$ while approximate ones only target an approximation of $\mathcal{N}(\Bs{\mu},\B{Q}^{-1})$.

\subsubsection{From exact data augmentation to exact matrix splitting}
\label{subsubsec:DA_MS}

We assume here that the variable $\B{u}$ refers to an auxiliary variable such that the joint distribution of the couple $(\Bs{\theta},\B{u})$ has a density given by $\pi(\Bs{\theta},\B{u}) \triangleq \kappa(\Bs{\theta},\B{u})$.
In addition, we restrict here $\B{R}$ to be positive definite.
It follows that
\begin{equation}
  \label{eq:exact_DA2}
  \int_{\mathbb{R}^d} \pi(\Bs{\theta},\B{u}) \mathrm{d}\B{u} = Z^{-1}\pi(\Bs{\theta})\int_{\mathbb{R}^d} \exp\pr{-\frac{1}{2}(\Bs{\theta}-\B{u})^{\top}\B{R}(\Bs{\theta}-\B{u})}\mathrm{d}\B{u} = \pi(\Bs{\theta})
\end{equation}
holds almost surely with $Z = \mathrm{det}(\B{R})^{-1/2}(2\uppi)^{d/2}$.
Hence, the joint density \cref{eq:cond_PPA} yields an exact DA scheme whatever the choice of the positive definite matrix $\B{R}$.
We will show that the exact DA approaches schemed by \cref{algo:exact_DA} precisely fit the proposed generic framework with a specific choice for the preconditioning matrix $\B{R}$.
We will then extend this class of exact DA approaches and show a one-to-one equivalence between Gibbs samplers based on exact MS (see \cref{subsec:matrix_split_exact}) and those based on exact DA (see \cref{subsec:data_aug_exact}).

To this purpose, we start by making the change of variable $\B{v} = \B{Ru}$. Combined with the joint probability density \cref{eq:cond_PPA}, it yields the two following conditional probability distributions:
\begin{align}
  \B{v}\mid\Bs{\theta} &\sim \mathcal{N}\pr{\B{R}\Bs{\theta},\B{R}}\eqsp, \label{eq:DA1}\\
  \Bs{\theta}\mid\B{v} &\sim \mathcal{N}\pr{(\B{Q} + \B{R})^{-1}(\B{v}+\B{Q}\Bs{\mu}),(\B{Q} + \B{R})^{-1}}\eqsp.\label{eq:DA2}
\end{align}
As emphasized in \cref{subsec:PIGauss}, the aim of introducing the preconditioning matrix $\B{R}$ is to yield simpler sampling steps.
In the general case where $\B{Q} = \B{Q}_1 + \B{Q}_2$ with $\B{Q}_1$ and $\B{Q}_2$ two matrices that cannot be easily handled jointly (e.g, because not diagonalized in the same basis),
an attractive option is $\B{R} = \omega^{-1}\B{I}_d - \B{Q}_1$.
Indeed, this choice ensures that $\B{Q}_1$ and $\B{Q}_2$ are separated and are not simultaneously involved in any of the two conditional sampling steps.
Note that this choice yields the EDA scheme already discussed in \cref{subsec:data_aug_exact}, see \cref{table:exact_DA}.
Now we relate this exact DA scheme to an exact MS one. 
By re-writing the Gibbs sampling steps associated with the conditional distributions \cref{eq:DA1} and \cref{eq:DA2} as an auto-regressive process of order 1 w.r.t. $\Bs{\theta}$ \cite{Box1994}, it follows that an equivalent sampling strategy writes 
\begin{eqnarray}
  \tilde{\B{z}} &\sim & \mathcal{N}\pr{\B{Q}\Bs{\mu}, 2\B{R} + \B{Q}}\eqsp,\\
  \Bs{\theta}^{(t)} & = & \pr{\B{Q}+\B{R}}^{-1}\pr{\tilde{\B{z}}+\B{R}\Bs{\theta}^{(t-1)}}\eqsp.
\end{eqnarray}
Defining $\B{M} = \B{Q}+\B{R}$ and $\B{N} = \B{R}$, or equivalently $\B{Q} = \B{M} - \B{N}$, it yields  
\begin{eqnarray}
    \tilde{\B{z}} & \sim & \mathcal{N}\pr{\B{Q}\Bs{\mu}, \B{M}^{\top} + \B{N}}\eqsp,\label{eq:z}\\
    \Bs{\theta}^{(t)} & = & \B{M}^{-1}\pr{\tilde{\B{z}} + \B{N}\Bs{\theta}^{(t-1)}}\eqsp,
\end{eqnarray}
which boils down to the Gibbs sampler based on exact MS  discussed in \cref{subsec:matrix_split_exact} (see \cref{algo:matrix_splitting}). 

To illustrate the interest of this rewriting when considering the case of two matrices $\B{Q}_1$ and $\B{Q}_2$ that cannot be efficiently handled in the same basis, \Cref{table:DA_matrix_splitting} presents two possible choices of $\B{R}$ which relate two MS strategies with their DA counterparts. 
First, one particular choice of $\B{R}$ (row 1 of \cref{table:DA_matrix_splitting}) directly shows that the Richardson MS sampler can be rewritten as the EDA sampler. More precisely, the auto-regressive process of order 1 w.r.t. $\Bs{\theta}$ defined by EDA yields a variant of the Richardson sampler.
This finding relates two different approaches proposed by authors from distinct communities (numerical algebra and signal processing).
Secondly, the proposed unifying framework also permits to go beyond existing approaches by proposing a novel exact DA approach via a specific choice for the precision matrix $\B{R}$ driven by an existing MS method. Indeed, following the same rewriting trick with another particular choice of $\B{R}$ (row 2 of \cref{table:DA_matrix_splitting}), an exact DA scheme can be easily derived from the Jacobi MS approach. Up to our knowledge, this novel DA method, referred to as EDAJ in the table, has not been documented in the existing literature. 

Finally, this table reports two particular choices of $\B{R}$ which lead to revisit existing MS and/or DA methods. It is worth noting that other relevant choices may be possible, which would allow to derive new exact DA and MS methods or to draw further analogies between existing approaches. Note also that \cref{table:DA_matrix_splitting} shows the main benefit of an exact DA scheme over its MS counterpart thanks to the decoupling between $\B{Q}_1$ and $\B{Q}_2$ in two separate simulation steps. 
This feature can be directly observed by comparing the two first columns of \cref{table:DA_matrix_splitting} with the third one.
\begin{table}
{\footnotesize
  \caption{Equivalence relations between exact DA and exact MS approaches. The matrices $\B{Q}_1$ and $\B{Q}_2$ are such that $\B{Q} = \B{Q}_1 + \B{Q}_2$. The matrix $\B{D}_1$ denotes the diagonal part of $\B{Q}_1$, and $\omega>0$ is a positive scalar ensuring the positive definiteness of $\B{R}$.
  Bold acronyms refer to novel samplers which derive from the proposed unifying framework.} 
  \label{table:DA_matrix_splitting}
  \begin{center}
   {\renewcommand{\arraystretch}{1.5}
    \begin{tabular}{|c|c|c|c|c|} 
      \hline
      $\B{R}=\mathrm{cov}(\B{v}|\Bs{\theta})$ 
      & $(\B{Q} + \B{R})^{-1}=\mathrm{cov}(\Bs{\theta}|\B{v})$ 
      & $\B{M}^{\top} + \B{N}=\mathrm{cov}(\tilde{\B{z}})$ 
      & DA sampler & MS sampler \\
      \hline 
      $\dfrac{\B{I}_d}{\omega} - \B{Q}_1$ 
      & $\pr{\dfrac{\B{I}_d}{\omega} + \B{Q}_2}^{-1}$ 
      & $\dfrac{2\B{I}_d}{\omega} + \B{Q}_2 - \B{Q}_1$ 
      & EDA \cite{Marnissi2019} & Richardson \cite{Fox2017} \\
      $\dfrac{\B{D}_1}{\omega} - \B{Q}_1$ 
      & $\pr{\dfrac{\B{D}_1}{\omega} + \B{Q}_2}^{-1}$ 
      & $\dfrac{2\B{D}_1}{\omega} + \B{Q}_2 - \B{Q}_1$ 
      & {\bf EDAJ} & Jacobi \cite{Fox2017} \\[1em]
      \hline
    \end{tabular}}
  \end{center}
}
\end{table}

\subsubsection{From approximate matrix splitting to approximate data augmentation}
We now build on the proposed unifying proposal \cref{eq:cond_PPA} to extend the class of samplers based on approximate matrix splitting and reviewed in \cref{subsec:matrix_split_approx}. 
With some abuse of notation, the variable $\B{u}$ in \cref{eq:cond_PPA} now refers to an iterate associated to $\Bs{\theta}$.
More precisely, let define $\B{u} = \Bs{\theta}^{(t-1)}$ to be the current iterate within an MCMC algorithm and $\kappa$ to be 
\begin{equation}
  \kappa(\Bs{\theta},\B{u}) \triangleq p\pr{\Bs{\theta}|\B{u}=\Bs{\theta}^{(t-1)}} \propto \pi(\Bs{\theta})\exp\pr{-\frac{1}{2}\pr{\Bs{\theta}-\Bs{\theta}^{(t-1)}}^{\top}\B{R}\pr{\Bs{\theta}-\Bs{\theta}^{(t-1)}}}. 
  \label{eq:cond_PPA_MH}
\end{equation}
Readers familiar with MCMC algorithms will recognize in \cref{eq:cond_PPA_MH} a proposal density that can be used within Metropolis-Hastings schemes \cite{Robert2004}.
However, unlike the usual random-walk algorithm which considers the Gaussian proposal distribution $\mathcal{N}(\Bs{\theta}^{(t-1)},\lambda\B{I}_d)$ with $\lambda >0$, the originality of \cref{eq:cond_PPA_MH} is to define the proposal by combining the Gaussian target density $\pi$ with a term that is equal to a Gaussian kernel density when $\B{R}$ is positive definite. If we always accept the proposed sample obtained from \cref{eq:cond_PPA_MH} without any correction, that is $\Bs{\theta}^{(t)} = \widetilde{\Bs{\theta}} \sim P(\cdot\mid\B{u}=\Bs{\theta}^{(t-1)})$ with density \cref{eq:cond_PPA_MH}, this directly implies that the associated Markov chain converges in distribution towards a Gaussian random variable with distribution $\mathcal{N}(\Bs{\mu},\widetilde{\B{Q}}^{-1})$ with the correct mean $\Bs{\mu}$ but with precision matrix 
\begin{align}
  \widetilde{\B{Q}} = \B{Q}\pr{\B{I}_d + (\B{R}+\B{Q})^{-1}\B{R}}\eqsp.
\end{align} 
This algorithm is detailed in \cref{algo:MS_PPA}. Note again that one can obtain samples from the initial target distribution $\mathcal{N}(\Bs{\mu},\B{Q}^{-1})$ by replacing step 4 with an acceptance/rejection step, see \cite{Robert2004} for details.
\begin{algorithm}
\caption{MCMC sampler based on \cref{eq:cond_PPA_MH}.}
\label{algo:MS_PPA}
\hspace*{\algorithmicindent} \textbf{Input:} Number $T$ of iterations and initialization $\Bs{\theta}^{(0)}$.
\begin{algorithmic}[1]
\State Set $t = 1$.
\While{$t\leq T$}
  \State Draw $\widetilde{\Bs{\theta}} \sim P\pr{\cdot\mid\B{u}=\Bs{\theta}^{(t-1)}}$ {\color{green} in \cref{eq:cond_PPA_MH}.}
  \State Set $\Bs{\theta}^{(t)} = \widetilde{\Bs{\theta}}$. 
  \State Set $t = t + 1$.
\EndWhile\\
\Return $\Bs{\theta}^{(T)}$.
\end{algorithmic}
\end{algorithm}

Moreover, the instance \cref{eq:cond_PPA_MH} of \cref{eq:cond_PPA} paves the way to an extended class of samplers based on approximate matrix splitting.
More precisely, the draw of a proposed sample $\widetilde{\Bs{\theta}}$ from \cref{eq:cond_PPA_MH} can be replaced by the following two-step sampling procedure:
\begin{eqnarray}
\label{eq:recursion522}
  \tilde{\B{z}}' &\sim & \mathcal{N}\pr{\B{Q}\Bs{\mu}, \B{R} + \B{Q}}\eqsp,\\
  \Bs{\theta}^{(t)} & = & \pr{\B{Q}+\B{R}}^{-1}\pr{\tilde{\B{z}}'+\B{R}\Bs{\theta}^{(t-1)}}\eqsp.
\end{eqnarray}
The matrix splitting form with $\B{M} = \B{Q}+\B{R}$, $\B{N}=\B{R}$ writes
\begin{eqnarray}
  \tilde{\B{z}}' &\sim & \mathcal{N}\pr{\B{Q}\Bs{\mu}, \B{M}}\eqsp,\\
  \Bs{\theta}^{(t)} & = & \B{M}^{-1}\pr{\tilde{\B{z}}' + \B{N}\Bs{\theta}^{(t-1)}}\eqsp. \label{eq:MS_proposal}
\end{eqnarray}
This recursion defines an extended class of approximate MS-based samplers and encompasses the Hogwild sampler reviewed in \cref{subsec:matrix_split_approx} by taking $\B{R} = -\B{L} - \B{L}^{\top}$.
In addition to the existing Hogwild approach, \cref{table:extended_approx_matrix_splitting} lists two other and new approximate MS approaches resulting from specific choices of the preconditioning matrix $\B{R}$.
They are coined \textit{approximate} Richardson and Jacobi samplers since the expressions for $\B{M}$ and $\B{N}$ are very similar to the ones associated to their exact counterparts, see \cref{table:matrix_splitting}.
For those two samplers, note that the approximate precision matrix $\widetilde{\B{Q}}$ tends towards $2\B{Q}$ in the asymptotic regime $\omega \rightarrow 0$.
Indeed, for the approximate Jacobi sampler, we have
\begin{align*}
 \widetilde{\B{Q}} &= \B{Q}\pr{\B{I}_d + \omega\pr{\frac{\B{I}_d}{\omega} - \B{Q}}} \\
 &= \B{Q}\pr{2\B{I}_d - \omega\B{Q}} \\
 &\underset{\omega\rightarrow 0}{\rightarrow} 2\B{Q}\eqsp.
\end{align*}
In order to retrieve the original precision matrix $\B{Q}$ when $\omega \rightarrow 0$, \cite{Barbos2017} proposed an approximate data augmentation strategy which can be related to the work of \cite{Vono_2019sub}. 
\begin{table}
{\footnotesize
  \caption{Extended class of Gibbs samplers based on approximate MS with $\B{Q} = \B{M} - \B{N}$ with $\B{N}=\B{R}$ and approximate DA. The matrices $\B{D}$ and $\B{L}$ denote the diagonal and strictly lower triangular parts of $\B{Q}$, respectively. $\omega$ is a positive scalar.
  Bold names and acronyms refer to novel samplers which derive from the proposed unifying framework.}
  \label{table:extended_approx_matrix_splitting}
  \begin{center}
    \begin{tabular}{|c|c|c|c|c|} 
      \hline
      $\frac{1}{2}\B{M} = \mathrm{cov}(\B{v}'|\Bs{\theta})$ & $\frac{1}{2}\B{M}^{-1} = \mathrm{cov}(\Bs{\theta}|\B{v}')$ & $\B{M} = \mathrm{cov}(\tilde{\B{z}}')$ & MS sampler & DA sampler \\
      \hline 
      $\frac{1}{2}\B{D}$ & $\frac{1}{2}\B{D}^{-1}$ & $\B{D}$ & Hogwild \cite{Johnson2013} & {\bf ADAH}\\[1em]
      $\dfrac{\B{I}_d}{2\omega}$ & $\dfrac{\omega\B{I}_d}{2}$ & $\dfrac{\B{I}_d}{\omega}$ & {\bf approx. Richardson} & {\bf ADAR}\\[1em]
     $\dfrac{\B{D}}{2\omega}$ & $\dfrac{\omega\B{D}^{-1}}{2}$ & $\dfrac{\B{D}}{\omega}$ & {\bf approx. Jacobi} & {\bf ADAJ}\\[1em]
      \hline
    \end{tabular}
  \end{center}
}
\end{table}

In \cref{subsubsec:DA_MS}, we showed that exact DA approaches can be rewritten to recover exact MS approaches. In the following, we will take the opposite path to show that approximate MS approaches admit approximate DA counterparts, which are highly amenable to distributed and parallel computations.
Using the fact that $\B{z}' = \B{Q}\Bs{\mu} + \B{z}_1 + (\B{Q} + \B{R})\B{z}_2$ where $\B{z}_1 \sim \mathcal{N}(\B{0}_d, \frac{1}{2}(\B{R} + \B{Q}))$ and $\B{z}_2 \sim \mathcal{N}(\B{0}_d, \frac{1}{2}(\B{R} + \B{Q})^{-1})$, the recursion \eqref{eq:MS_proposal} can be equivalently written as
$$
\widetilde{\Bs{\theta}} = \pr{\B{Q}+\B{R}}^{-1}\pr{\B{Q}\Bs{\mu} + \B{R}\Bs{\theta}^{(t-1)} + \B{z}_1} + \B{z}_2\eqsp. 
$$
By introducing an auxiliary variable $\B{v}'$ defined by $\B{v}' = \B{R}\Bs{\theta}^{(t-1)} + \B{z}_1$, the resulting two-step Gibbs sampling relies on the conditional sampling steps
\begin{align*}
  \B{v}'\mid\Bs{\theta} &\sim \mathcal{N}\pr{\B{R}\Bs{\theta},\frac{1}{2}(\B{R} + \B{Q})}\eqsp,\\
  \Bs{\theta}\mid\B{v}' &\sim \mathcal{N}\pr{(\B{Q} + \B{R})^{-1}(\B{v}'+\B{Q}\Bs{\mu}),\frac{1}{2}(\B{R} + \B{Q})^{-1}}\eqsp,
\end{align*}
and targets the joint distribution with density $\pi(\Bs{\theta},\B{v}')$. 
Compared to exact DA approaches reviewed in \cref{subsec:data_aug_exact}, the sampling difficulty associated to each conditional sampling step is the same and only driven by the structure of the matrix $\B{M} = \B{R} + \B{Q}$. In particular, this matrix  becomes diagonal for three specific choices listed in \cref{table:extended_approx_matrix_splitting}. 
These choices lead to three new sampling schemes that we name ADAH, ADAR and ADAJ since they stand for the DA counterparts of the approximate MS samplers discussed above. Interestingly, these DA schemes naturally emerge here without assuming any explicit decomposition $\B{Q} = \B{Q}_1 + \B{Q}_2$ or including an additional auxiliary variable (as in GEDA). Finally, as previously highlighted, when compared to their exact counterpart, these DA schemes have the great advantage of leading to Gibbs samplers suited for  parallel computations, hence simplifying the sampling procedure.

\subsection{Gibbs samplers as stochastic versions of the PPA}
\label{subsec:PPA}
This section aims at drawing new connections between optimization and the sampling approaches discussed in this paper. In particular, we will focus on the proximal point algorithm (PPA)\cite{Rockafellar1976}. After briefly presenting this optimization method, we will show that the Gibbs samplers based on the proposal \cref{eq:cond_PPA_MH} can be interestingly interpreted as stochastic counterparts of the PPA. 
Let assume here that $\B{R}$ is positive semi-definite and define the \textit{weighted} norm w.r.t. $\B{R}$ for all $\Bs{\theta} \in \mathbb{R}^d$ by
\begin{equation}
  \nr{\Bs{\theta}}_{\B{R}}^2 \triangleq {\Bs{\theta}^{\top}\B{R}\Bs{\theta}}\eqsp.
\end{equation} 
\noindent\textbf{The proximal point algorithm (PPA).} 
The PPA \cite{Rockafellar1976} is an important and widely used method to find zeros of a maximal monotone operator $\mathsf{K}$, that is to solve problems of the form 
\begin{equation}
  \text{Find $\Bs{\theta}^{\star} \in \mathcal{H}$ such that } \B{0}_d \in \mathsf{K}(\Bs{\theta}^{\star})\eqsp, \label{eq:zeros_operators}
\end{equation}
where $\mathcal{H}$ is a real Hilbert space. For simplicity, we will take here $\mathcal{H} = \mathbb{R}^d$ equipped with the usual Euclidean norm and focus on the particular case $\mathsf{K} = \partial f$ where $f$ is a lower semicontinuous (l.s.c.), proper, coercive and convex function and $\partial$ denotes the subdifferential operator, see \cref{appendix_sec_2_subsec_PPA}.
In this case, the PPA is equivalent to the proximal minimization algorithm  \cite{Martinet1970,Martinet1972} which aims at solving the minimization problem 
\begin{equation}
  \text{Find $\Bs{\theta}^{\star} \in \mathbb{R}^d$ such that } \Bs{\theta}^{\star} = \underset{\Bs{\theta}\in\mathbb{R}^d}{\arg \min} \ f(\Bs{\theta})\eqsp. \label{eq:optimization_problem}
\end{equation}
This algorithm is called the proximal point algorithm in reference to the work by Moreau \cite{Moreau1965}.
For readability reasons, we refer to \cref{appendix_sec_2_subsec_PPA} for details about this algorithm for a general operator $\mathsf{K}$ and refer the interested reader to the comprehensive overview in \cite{Rockafellar1976} for more information.
\begin{algorithm}
\caption{Proximal point algorithm (PPA)}
\label{algo:PPA}
\begin{algorithmic}[1]
\State Choose an initial value $\Bs{\theta}^{(0)}$, a positive semi-definite matrix $\B{R}$ and a maximal number of iterations $T$.
\State Set $t=1$.
\While{$t\leq T$}
\State $\Bs{\theta}^{(t)} = \underset{\Bs{\theta}\in\mathbb{R}^d}{\arg \min}\ f(\Bs{\theta}) + \dfrac{1}{2}\nr{\Bs{\theta}-\Bs{\theta}^{(t-1)}}_{\B{R}}^2$.
\EndWhile\\
\Return $\Bs{\theta}^{(T)}$.
\end{algorithmic}
\end{algorithm}

The PPA is detailed in \cref{algo:PPA}.
Note that instead of directly minimizing the objective function $f$, the PPA adds a quadratic penalty term depending on the previous iterate $\Bs{\theta}^{(t-1)}$ and then solves an approximation of the initial optimization problem at each iteration.
This idea of successive approximations is exactly the deterministic counterpart of \cref{eq:cond_PPA_MH} which proposes a new sample based on successive approximations of the target density $\pi$ via a Gaussian kernel with precision matrix $\B{R}$.
Actually, searching for the maximum a posteriori estimator under the proposal distribution $P(\cdot \mid \Bs{\theta}^{(t-1)})$ with density $p(\cdot \mid \Bs{\theta}^{(t-1)})$ in \cref{eq:cond_PPA_MH} boils down to solving 
\begin{equation}
  \underset{\Bs{\theta}\in\mathbb{R}^d}{\arg \min}\ \underbrace{-\log\pi(\Bs{\theta})}_{f(\Bs{\theta})} + \dfrac{1}{2}\nr{\Bs{\theta}-\Bs{\theta}^{(t-1)}}_{\B{R}}^2\eqsp,
\end{equation} 
which coincides with step 4 in \cref{algo:PPA} by taking $f = -\log \pi$.
This puts a first emphasis on the tight connection between optimization and simulation that we already highlighted in previous sections.\\

\noindent\textbf{The PPA, ADMM and the approximate Richardson Gibbs sampler.} An important motivation of the PPA is also related to the {\em preconditioning} idea used in the unifying model proposed in \cref{eq:cond_PPA}.
Indeed, the PPA has been extensively used within the alternating direction method of multipliers (ADMM) \cite{Glowinski1975,Gabay1976,Boyd2011} as a preconditioner in order to avoid high-dimensional inversions \cite{Esser2010,Zhang2011,Chambolle2011,Li2016,Bredies2017}.
The ADMM \cite{Boyd2011} stands for an optimization approach that solves the minimization problem in \cref{eq:optimization_problem} when $g(\Bs{\theta}) = g_1(\B{A}\Bs{\theta}) + g_2(\Bs{\theta})$, $\B{A} \in \mathbb{R}^{k \times d}$, via the following iterative scheme
\begin{align}
  &\B{z}^{(t)} = \underset{\B{z}\in\mathbb{R}^k}{\arg \min}\ g_1(\B{z}) + \dfrac{1}{2\rho}\nr{\B{z} - \B{A}\Bs{\theta}^{(t-1)} - \B{u}^{(t-1)}}^2 \\
  &\Bs{\theta}^{(t)} = \underset{\Bs{\theta}\in\mathbb{R}^d}{\arg \min}\ g_2(\Bs{\theta}) + \dfrac{1}{2\rho}\nr{\B{A}\Bs{\theta} - \B{z}^{(t)} + \B{u}^{(t-1)}}^2 \label{eq:ADMM_x} \\
  &\B{u}^{(t)} = \B{u}^{(t-1)} + \B{A}\Bs{\theta}^{(t)} - \B{z}^{(t)},
\end{align}
where $\B{z} \in \mathbb{R}^k$ is a splitting variable, $\B{u} \in \mathbb{R}^k$ is a scaled dual variable and $\rho$ is a positive penalty parameter.
Without loss of generality\footnote{If $g_2$ admits a non-quadratic form, an additional splitting variable can be introduced and the following comments would still hold.}, let assume that $g_2$ is a quadratic function, that is for any $\Bs{\theta} \in \mathbb{R}^d$, $g_2(\Bs{\theta}) = (\Bs{\theta}-\bar{\Bs{\theta}})^{\top}(\Bs{\theta}-\bar{\Bs{\theta}})/2$.
Even in this simple case, one can notice that the $\Bs{\theta}$-update \cref{eq:ADMM_x} involves a matrix $\B{A}$ operating directly on $\Bs{\theta}$ preluding an expensive inversion of a high-dimensional matrix associated to $\B{A}$.
To deal with such an issue, \cref{algo:PPA} is considered to solve approximately the minimization problem in \cref{eq:ADMM_x}.
The PPA applied to the minimization problem \cref{eq:ADMM_x} reads
\begin{equation}
  \Bs{\theta}^{(t)} = \underset{\Bs{\theta}\in\mathbb{R}^d}{\arg \min}\ \dfrac{1}{2}(\Bs{\theta}-\bar{\Bs{\theta}})^{\top}(\Bs{\theta}-\bar{\Bs{\theta}}) + \dfrac{1}{2\rho}\nr{\B{A}\Bs{\theta} - \B{z}^{(t)} + \B{u}^{(t-1)}}^2 + \dfrac{1}{2}\nr{\Bs{\theta}-\Bs{\theta}^{(t-1)}}^2_{\B{R}} \label{eq:ADMM_x_PPA}.
\end{equation}
In order to draw some connections with \eqref{eq:cond_PPA_MH}, we set $\B{Q} = \rho^{-1}\B{A}^{\top}\B{A} + \B{I}_d$ and $\Bs{\mu} = \B{Q}^{-1}[\B{A}^{\top}(\B{u}^{(t-1)} - \B{z}^{(t)})/\rho + \bar{\Bs{\theta}}]$ and rewrite \eqref{eq:ADMM_x_PPA} as 
\begin{equation}
  \Bs{\theta}^{(t)} = \underset{\Bs{\theta}\in\mathbb{R}^d}{\arg \min}\ \dfrac{1}{2}(\Bs{\theta}-\Bs{\mu})^{\top}\B{Q}(\Bs{\theta}-\Bs{\mu}) + \dfrac{1}{2}\nr{\Bs{\theta}-\Bs{\theta}^{(t-1)}}^2_{\B{R}} \label{eq:ADMM_x_PPA_2}.
\end{equation}
Note that $(1/2)(\Bs{\theta}-\Bs{\mu})^{\top}\B{Q}(\Bs{\theta}-\Bs{\mu})$ stands for the potential function associated to $\pi$ in \eqref{eq:Gaussian_target_Sigma} and as such \eqref{eq:ADMM_x_PPA_2} can be seen as the deterministic counterpart of \eqref{eq:cond_PPA_MH}.
By defining $\B{R} = \omega^{-1}\B{I}_d - \B{Q}$, where $0<\omega \leq \rho\nr{\B{A}}^{-2}$ ensures that $\B{R}$ is positive semi-definite, the $\Bs{\theta}$-update in \cref{eq:ADMM_x_PPA} becomes (see \cref{appendix_sec_2_subsec_PPA})
\begin{equation}
  \Bs{\theta}^{(t)} = \underset{\Bs{\theta}\in\mathbb{R}^d}{\arg \min} \dfrac{1}{2\omega}\nr{\Bs{\theta} -  \omega \pr{\B{R}\Bs{\theta}^{(t-1)} + \B{Q}\Bs{\mu}}}^2
  \label{eq:ADMM_x_PPA_3}
\end{equation}
Note that \eqref{eq:ADMM_x_PPA_3} boils down to solving $\omega^{-1}\Bs{\theta} = \B{R}\Bs{\theta}^{(t-1)} + \B{Q}\Bs{\mu}$,
which is exactly the deterministic counterpart of the approximate Richardson Gibbs sampler in \cref{table:extended_approx_matrix_splitting}.
This highlights even more the tight links between the proposed unifying framework and the use of the PPA in the optimization literature.
It also paves the way to novel sampling methods inspired by optimization approaches which are not necessarily dedicated to Gaussian sampling; this goes beyond the scope of the present article.

\section{A comparison of Gaussian sampling methods with numerical simulations}
\label{sec:applications}

\begin{sidewaystable}
\renewcommand{\arraystretch}{1.3}
\caption{Taxonomy of existing methods to sample from an arbitrary $d$-dimensional Gaussian distribution $\Pi \triangleq \mathcal{N}(\Bs{\mu},\B{Q}^{-1})$, see \cref{eq:Gaussian_target_Sigma}. 
Notation $K \in \mathbb{N}_*$ and $T_{\mathrm{bi}} \in \mathbb{N}_*$ stand for the number of iterations of a given iterative sampler (e.g., the CG one) and for the number of burn-in iterations for MCMC algorithms, respectively.
The notation ``target $\Pi$'' refers to approaches that target the right distribution $\Pi$ under infinite precision arithmetic.
Note that some i.i.d. samplers using a truncation parameter $K$ might target $\Pi$ for specific choices of $K$ (e.g., the classical CG sampler is exact for $K=d$).
The notation ``finite time sampling'' refers here to approaches which need a truncation procedure to deliver a sample in finite time. 
The matrix $\B{A}$ is a symmetric and positive definite matrix associated to $\B{Q}$, see \cref{sec:PIGauss_M_nonzero}.
The sampling methods highlighted in bold stand for novel approaches which derive from the proposed unifyign framework.
AMS stands for approximate matrix splitting (see \cref{table:extended_approx_matrix_splitting}), EDA for exact data augmentation (see \cref{table:exact_DA}) and ADA for approximate data augmentation (see \cref{table:extended_approx_matrix_splitting}).}
\label{table:overview}
\centering
{\footnotesize
\begin{tabular}{l l c c c c c c c c}
\thickhline
\multirow{2}{*}{\bfseries method} & \multirow{2}{*}{\bfseries instance} & \bfseries target & \bfseries finite time & \bfseries \multirow{2}{*}{comp. cost} & \bfseries \multirow{2}{*}{storage cost} & \bfseries \multirow{2}{*}{sampling} & \multirow{2}{*}{\bfseries linear system} & \multirow{2}{*}{\bfseries no tuning} \\
&&$\Pi$&\bfseries sampling&&&&&& \\
\hline
\multirow{2}{*}{Factorization} & Cholesky \cite{Scheur1962,Rue2001} & \checkmark & \checkmark & $\mathcal{O}(d^2(d+T)$ & $\Theta(d^2)$ & $\mathcal{N}(0,1)$ & triangular & \checkmark\\
& Square root \cite{Davis1987b} & \checkmark & \checkmark & $\mathcal{O}(d^2(d+T))$ & $\Theta(d^2)$ & $\mathcal{N}(0,1)$ & full & \checkmark\\
\hline
\multirow{2}{*}{$\B{Q}^{1/2}$ approx.} & Chebyshev \cite{Davis1987b,Pereira2019} & \xmark & \checkmark & $\mathcal{O}(d^2KT)$ & $\Theta(d)$ & $\mathcal{N}(0,1)$ & \xmark & \checkmark\\
& Lanczos \cite{Simpson2008,Aune2013} & \xmark & \checkmark & $\mathcal{O}(K(K + d^2)T)$ & $\Theta(K(K+d))$ & $\mathcal{N}(0,1)$ & \xmark & \checkmark\\
\hline
\multirow{2}{*}{CG} & Classical \cite{Parker2012} & \xmark & \checkmark & $\mathcal{O}(d^2KT)$ & $\Theta(d)$ & $\mathcal{N}(0,1)$ & \xmark & \checkmark \\
& Gradient scan \cite{Feron2016} & \checkmark & \xmark & $\mathcal{O}(d^2K(T+T_{\mathrm{bi}}))$ & $\Theta(Kd)$ & $\mathcal{N}(0,1)$ & \xmark & \checkmark\\
\hline
\multirow{2}{*}{PO} & Truncated \cite{Papandreou2011,Orieux2012} & \xmark & \checkmark & $\mathcal{O}(d^2KT)$ & $\Theta(d)$ & $\mathcal{N}(0,1)$ & full & \checkmark \\
& Reversible jump \cite{Gilavert2015} & \checkmark & \xmark & $\mathcal{O}(d^2K(T+T_{\mathrm{bi}}))$ & $\Theta(d)$ & $\mathcal{N}(0,1)$ & full & \checkmark\\
\hline
\multirow{7}{*}{MS} & Richardson \cite{Fox2017} & \checkmark & \xmark & $\mathcal{O}(d^2(T+T_{\mathrm{bi}}))$ & $\Theta(d)$ & $\mathcal{N}(\B{0}_d,\B{A})$ & diagonal & \xmark \\
& Jacobi \cite{Fox2017} & \checkmark & \xmark & $\mathcal{O}(d^2(T+T_{\mathrm{bi}}))$ & $\Theta(d)$ & $\mathcal{N}(\B{0}_d,\B{A})$ & diagonal & \checkmark \\
& Gauss-Seidel \cite{Fox2017} & \checkmark & \xmark & $\mathcal{O}(d^2(T+T_{\mathrm{bi}}))$ & $\Theta(d)$ & $\mathcal{N}(0,1)$ & triangular & \checkmark \\
& SOR \cite{Fox2017} & \checkmark & \xmark & $\mathcal{O}(d^2(T+T_{\mathrm{bi}}))$ & $\Theta(d)$ & $\mathcal{N}(0,1)$ & triangular & \xmark \\
& SSOR \cite{Fox2017} & \checkmark & \xmark & $\mathcal{O}(d^2(T+T_{\mathrm{bi}}))$ & $\Theta(d)$ & $\mathcal{N}(0,1)$ & triangular & \xmark \\
& Cheby-SSOR \cite{FoxParker14,Fox2017} & \checkmark & \xmark & $\mathcal{O}(d^2(T+T_{\mathrm{bi}}))$ & $\Theta(d)$ & $\mathcal{N}(0,1)$ & triangular & \xmark \\
& Hogwild \cite{Johnson2013} & \xmark & \xmark & $\mathcal{O}(d^2(T+T_{\mathrm{bi}}))$ & $\Theta(d)$ & $\mathcal{N}(0,1)$ & diagonal & \checkmark\\
& Clone MCMC \cite{Barbos2017} & \xmark & \xmark & $\mathcal{O}(d^2(T+T_{\mathrm{bi}}))$ & $\Theta(d)$ & $\mathcal{N}(0,1)$ & diagonal & \xmark\\
& \textbf{Unifying AMS} & \xmark & \xmark & $\mathcal{O}(d^2(T+T_{\mathrm{bi}}))$ & $\Theta(d)$ & $\mathcal{N}(0,1)$ & diagonal & \checkmark\\
\hline
\multirow{5}{*}{DA} & EDA \cite{Marnissi2019} & \checkmark & \xmark & $\mathcal{O}(d^2(T+T_{\mathrm{bi}}))$ & $\Theta(d)$ & $\mathcal{N}(\B{0}_d,\B{A})$ & \xmark & \checkmark\\
& GEDA \cite{Marnissi2019} & \checkmark & \xmark & $\mathcal{O}(d^2(T+T_{\mathrm{bi}}))$ & $\Theta(d)$ & $\mathcal{N}(0,1)$ & \xmark & \checkmark\\
& SGS \cite{Vono2019} & \xmark & \xmark & $\mathcal{O}(d^2(T+T_{\mathrm{bi}}))$ & $\Theta(d)$ & $\mathcal{N}(\B{0}_d,\B{A})$ & \xmark & \xmark\\
& \textbf{Unifying EDA} & \checkmark & \xmark & $\mathcal{O}(d^2(T+T_{\mathrm{bi}}))$ & $\Theta(d)$ & $\mathcal{N}(\B{0}_d,\B{A})$ & \xmark & \checkmark\\
& \textbf{Unifying ADA} & \xmark & \xmark & $\mathcal{O}(d^2(T+T_{\mathrm{bi}}))$ & $\Theta(d)$ & $\mathcal{N}(0,1)$ & \xmark & \xmark\\
\thickhline
\end{tabular}
}
\end{sidewaystable}

This section aims at providing readers with a detailed comparison of the reviewed Gaussian sampling techniques discussed in \cref{sec:PIGauss_M_zero} and \cref{sec:PIGauss_M_nonzero}. 
In particular, we summarize the benefits and bottlenecks associated to each method, illustrate some of them on numerical applications and propose an up-to-date selection of the most efficient algorithms for archetypal Gaussian sampling tasks. General guidelines resulting from this review and numerical experiments are gathered in \cref{fig:guidelines}.

\subsection{Summary, comparison and discussion of existing approaches}
\label{subsec:summary}

\cref{table:overview} lists and summarizes the main features of the sampling techniques reviewed above.
In particular, for each approach, this table recalls its exactness (or not), its computational and storage costs, the most expensive sampling step to compute, the possible linear system to solve and the presence of tuning parameters. This table aims at making a synthesis of the main pros and cons of each class of samplers.
Regarding MCMC approaches, the computational cost associated to the required sampling step is not taken into account in the column ``compt. cost'' since it depends upon the structure of $\B{Q}$. 
Instead, the column ``sampling'' indicates the type of sampling step required by the sampling approach.

Rather than conducting a one-to-one comparison between samplers, we make the choice of focusing on selected important questions raised by the taxonomy reported in this table. 
Concerning more technical or in-depth comparisons between specific approaches, we refer the interested reader to the appropriate references, see for instance those highlighted in \cref{sec:PIGauss_M_zero} and \cref{sec:PIGauss_M_nonzero}.
These questions of interest will lead to scenarios that will motivate dedicated numerical experiments conducted in \cref{subsec:results}. Then \cref{subsec:discussion} will gather guidelines to choose an appropriate sampler for a given sampling task. 
Here are the typical questions.\vspace{0.25cm}
\noindent\textbf{Question 1:} In which scenarios iterative approaches become interesting compared to factorization approaches?
\vspace{0.25cm} 

In \cref{sec:PIGauss_M_zero}, one can notice that square root, CG and PO approaches bypass the computational and storage costs of factorization thanks to an iterative process of $K$ cheaper steps, with $K \in \mathbb{N}_*$.
A natural question is: in which scenarios does the total cost of $K$ iterations remain efficient when compared to factorization methods?
\Cref{table:overview} tells us that high-dimensional scenarios ($d \gg 1$) are in favour of iterative methods as soon as memory needs of order $\Theta(d^2)$ become prohibitive. 
If this storage is not an issue, iterative samplers become interesting only if the number of iterations $K$ is such that $K \ll (d+T-1)/T$.
This inequality is verified only in cases where a small number of samples $T$ is required ($T \ll d$), which in turn imposes $K \ll d$. 
Note that this condition remains similar when a Gaussian sampling step is embedded within a Gibbs sampler with a varying covariance or precision matrix (see \cref{ex:bayesian_ridge}): the condition on $K$ is $K \ll d$, whatever the number $T$ of samples, since it is no longer possible to pre-compute the factorization of $\B{Q}$.
\vspace{0.25cm}

\noindent\textbf{Question 2:} When should we prefer an iterative sampler from \cref{sec:PIGauss_M_zero} or an MCMC sampler from \cref{sec:PIGauss_M_nonzero}? 
\vspace{0.25cm}

\Cref{table:overview} shows that iterative samplers reviewed in \cref{sec:PIGauss_M_zero} have to perform $K$ iterations to generate one sample. 
In contrast, most of MCMC samplers generate one sample per iteration. However, these samples are distributed according to the target distribution (or an approximation of it) only in an asymptotic regime, i.e., when $T \rightarrow \infty$ and in practice after a burn-in period. 
If one considers a burn-in period of length $T_{\mathrm{bi}}$ whose samples are discarded, MCMC samplers are interesting w.r.t. iterative ones only if $T + T_{\mathrm{bi}} \ll KT$.
Since most often $K\ll T_{\mathrm{bi}}$,
this condition goes in favour of MCMC methods when a large number $T\gtrsim T_{\mathrm{bi}}$ of samples is desired. When a small number $T\lesssim T_{\mathrm{bi}}/K$  of samples is desired, one shall prefer iterative methods. 
In intermediate situations, the choice depends on the precise number of required samples $T$, mixing properties of the MCMC sampler and the number of iterations $K$ of the alternative iterative algorithm. \vspace{0.25cm}
\noindent\textbf{Question 3:} When is it efficient to use a decomposition $\B{Q} = \B{Q}_1 + \B{Q}_2$ of the precision matrix in comparison with other approaches?\vspace{0.25cm}

\Cref{sec:PIGauss_M_zero} and \cref{sec:PIGauss_M_nonzero} have shown that some sampling methods, such as \cref{algo:PO} and \cref{algo:exact_DA}, exploit a decomposition of the form $\B{Q} = \B{Q}_1 + \B{Q}_2$ to simplify the sampling task.
Regarding the pertubation-optimization approaches, the main benefit lies in the cheap computation of the vector $\B{z}' \sim \mathcal{N}(\B{0}_d,\B{Q})$ before solving the linear system $\B{Q}\Bs{\theta} = \B{z}'$, see \cite{Papandreou2011} for more details. On the other hand, MCMC samplers based on exact data augmentation yield simpler sampling steps a priori and do not require to solve any high-dimensional linear system. 
However, the Achille's heel of MCMC methods is that they only produce samples of interest in the asymptotic regime where the number of iterations tends towards infinity.
For a fixed number of MCMC iterations, dependent samples are obtained and their quality highly depends upon the mixing properties of the MCMC sampler. 
Numerical experiments in \cref{subsec:results} allow discussion on this point.

\subsection{Numerical illustrations}
\label{subsec:results}

This section aims at illustrating the main differences between the methods reviewed in \cref{sec:PIGauss_M_zero} and \cref{sec:PIGauss_M_nonzero}.
The main purpose is not to give an exhaustive one-to-one comparison between all approaches listed in \cref{table:overview}.
Instead, these methods are compared in light of three experimental scenarios that address the questions raised in \cref{subsec:summary}. 
More specific numerical simulations can be found in cited works and references therein.
Since the main challenges of Gaussian sampling are related to the properties of the precision matrix $\B{Q}$, or the covariance $\Bs{\Sigma}$ (see \cref{subsec:special_instances}), the mean vector $\Bs{\mu}$ is set to $\B{0}_d$ and only centered distributions are considered.
For the first two scenarios associated with Questions 1 and 2, the unbiased estimate of the empirical covariance matrix will be used to assess the convergence in distribution of the samples generated by each algorithm:
\begin{equation}
  \B{\widehat{\Sigma}}_T = \dfrac{1}{T-1}\sum_{t=1}^T(\Bs{\theta}^{(t)} - \Bs{\bar{\theta}})(\Bs{\theta}^{(t)} - \Bs{\bar{\theta}})^{\top}\eqsp, \label{eq:cov_estimate}
\end{equation}
where $\Bs{\bar{\theta}} = T^{-1}\sum_{t=1}^T\Bs{\theta}^{(t)}$ stands for the empirical mean. 
Note that other metrics (such as the empirical precision matrix) could have been used to assess that these samples are distributed according to a sufficiently close approximation of the target distribution $\mathcal{N}(\Bs{\mu},\B{Q}^{-1})$.
Among available metrics, we chose the one that has been the most used in the literature, that is \cref{eq:cov_estimate} \cite{Gilavert2015,Fox2017,Barbos2017}.

For the scenario 3 associated with the corresponding last question, the considered high-dimensional setting will preclude the computation of exact and empirical covariance matrices. Instead, MCMC samplers will be rather compared in terms of computational efficiency and quality of the generated chains (see \cref{subsubsec:scenario3} for details).

The experimental setting is the following. To ensure fair comparisons, all algorithms have been implemented on equal grounds, with the same quality of optimization.
The programming language is \textsc{Python} and all loops have been carefully replaced by  matrix-vector products as far as possible.
Simulations have run on a computer equipped with an Intel Xeon 3.70 GHz processor with 16.0 GB of RAM, under Windows 7.
Among the infinite set of possible examples, we chose examples of Gaussian sampling problems that often appear in applications and that have been previously considered in the literature so that they stand for good tutorials to answer the question raised by each scenario.
The code to reproduce all the figures of this section is available in a \textsc{Jupyter} notebook format available online at \url{https://github.com/mvono/PyGauss/tree/master/notebooks}.

\subsubsection{Scenario 1}
\label{subsubsec:scenario1}
 This first set of experiments addresses Question 1 about iterative versus factorization approaches.
 We consider a sampling problem also tackled in \cite{Parker2012} to demonstrate the performances of \cref{algo:CG} based on the conjugate gradient.
For the sake of clarity, we divide this scenario into two sub-experiments.\\

\begin{figure}
  \centering
  \mbox{{\includegraphics[scale=0.45]{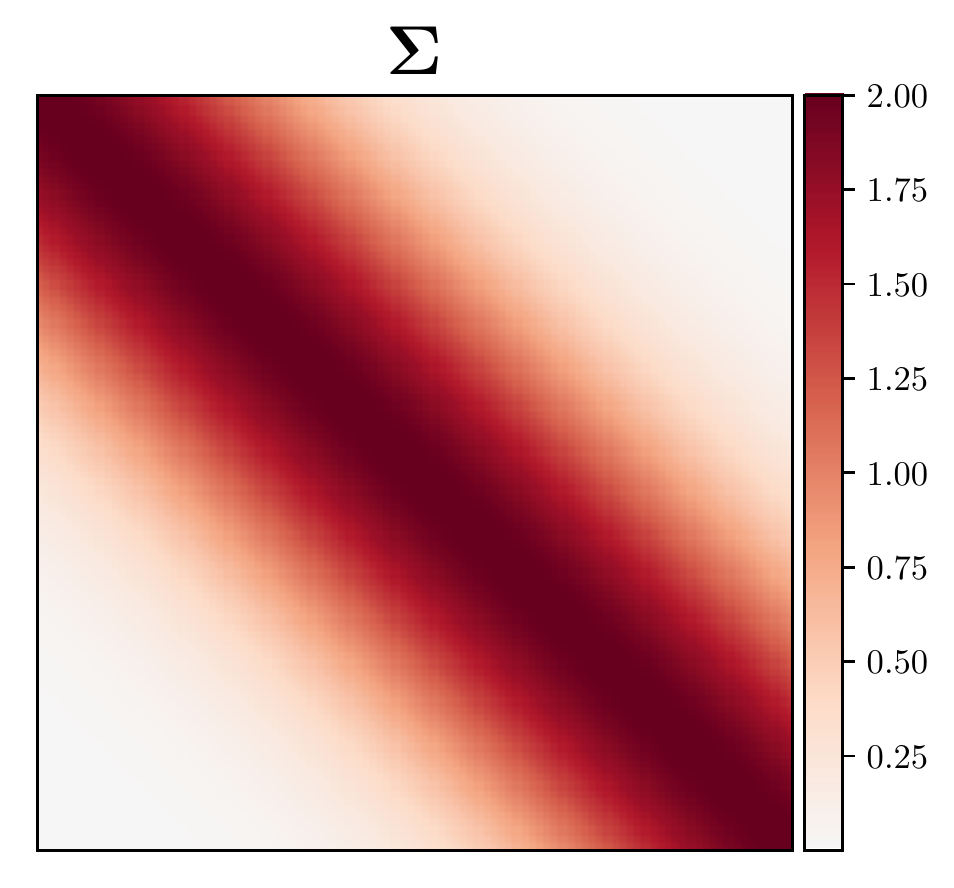}}}
  \mbox{{\includegraphics[scale=0.4]{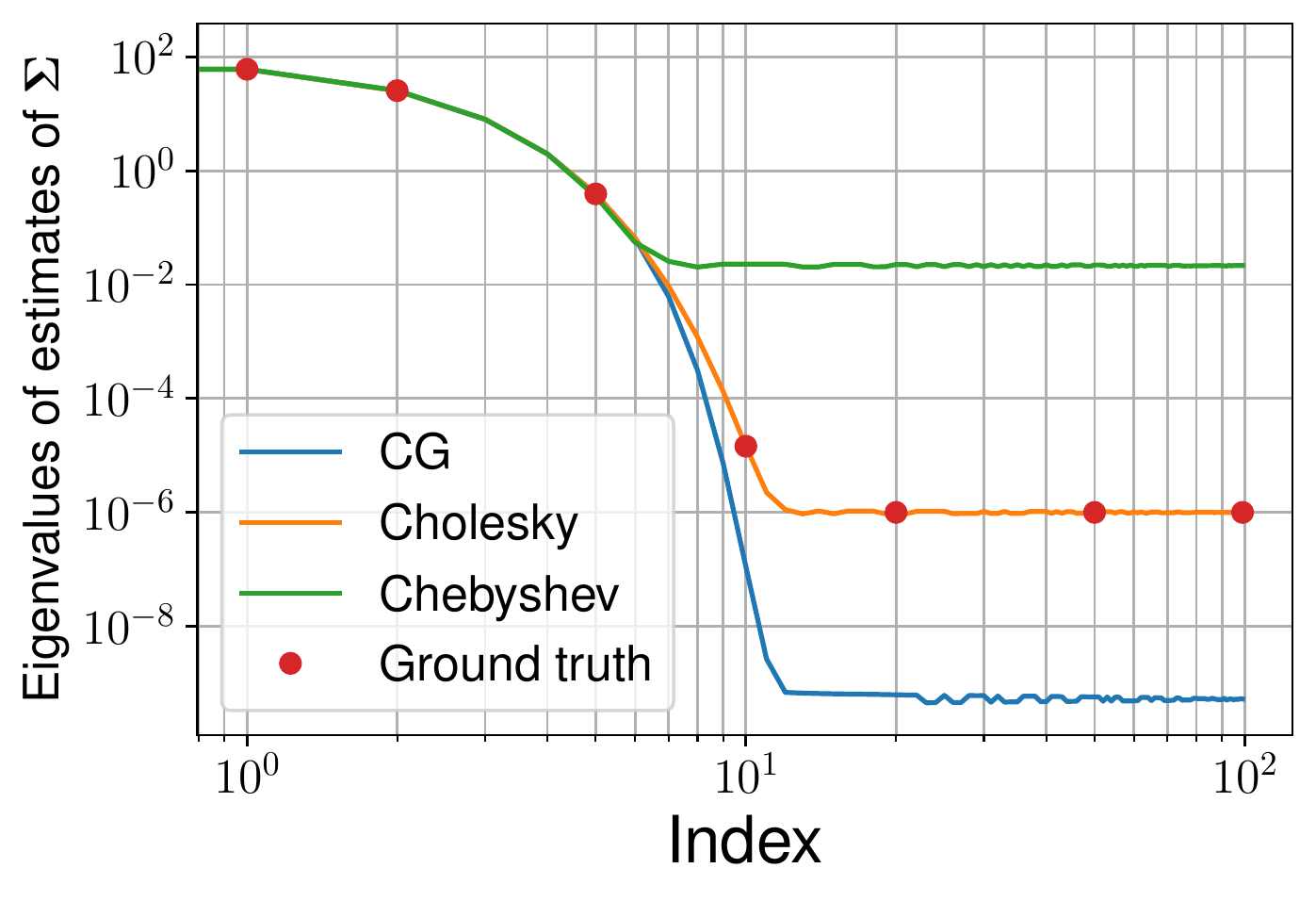}}}
  \mbox{{\includegraphics[scale=0.4]{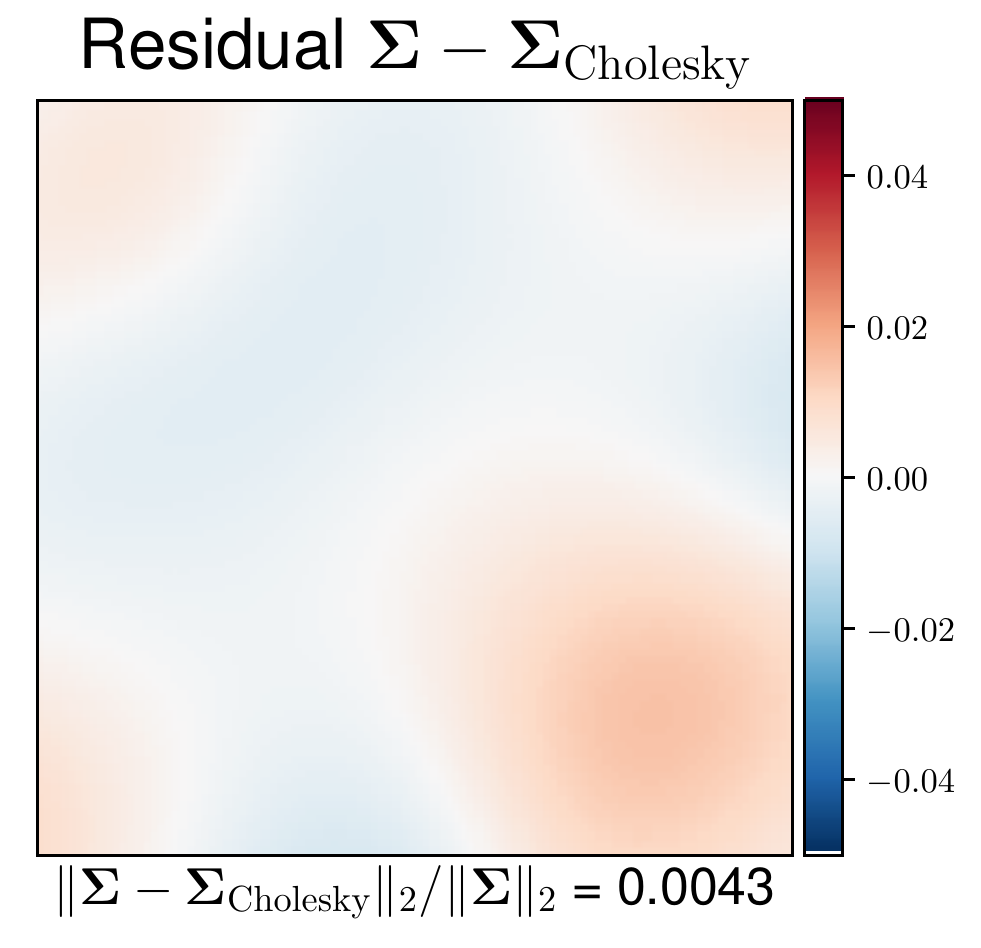}}}
  \mbox{{\includegraphics[scale=0.4]{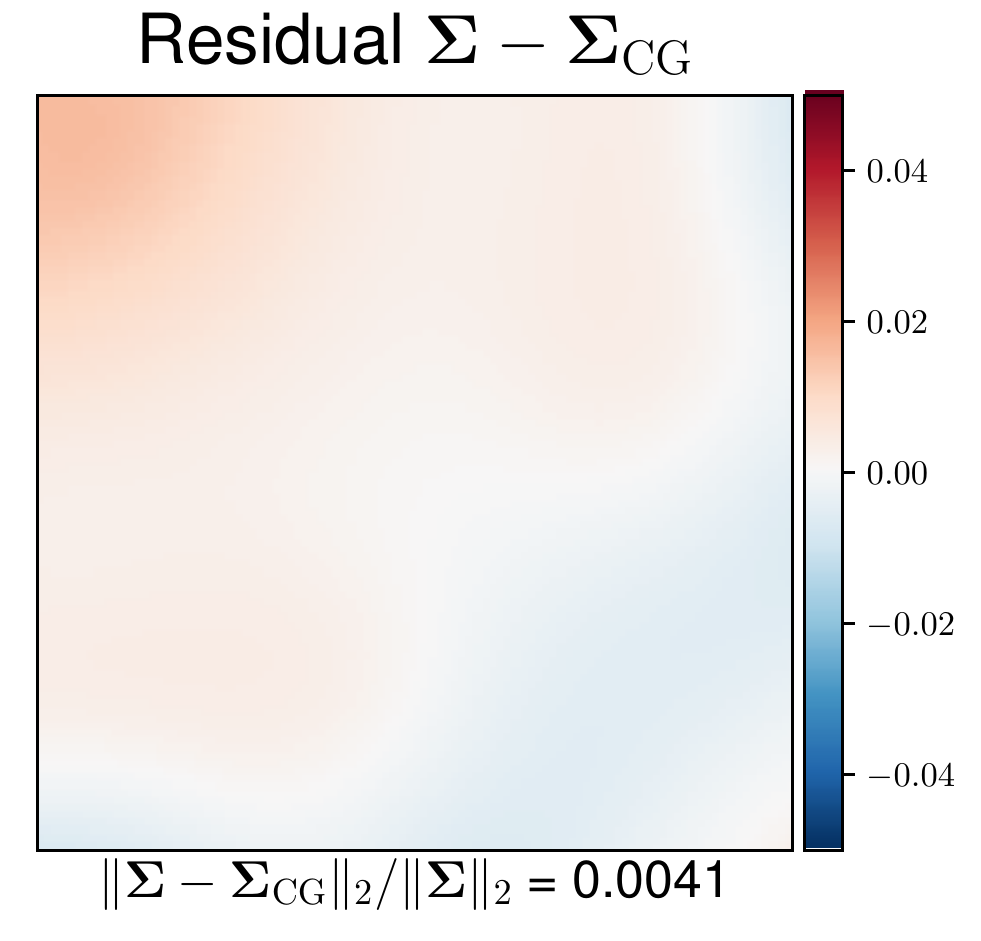}}}
  \mbox{{\includegraphics[scale=0.4]{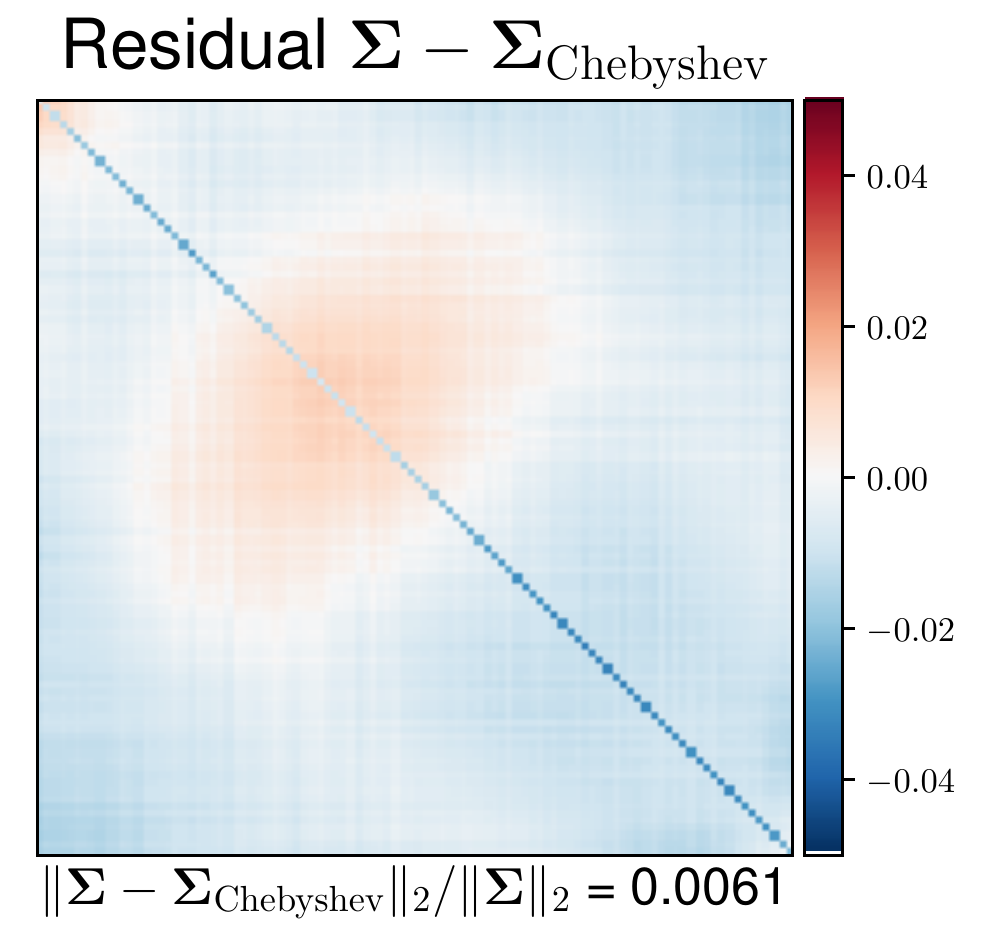}}}
\caption{Scenario 1. Results of the three considered samplers for the sampling from $\mathcal{N}(\B{0}_d,\B{\Sigma})$ in dimension $d = 100$ with $\B{\Sigma}$ detailed in \cref{eq:simu1_ex1}.}
  \label{fig:simu1_distrib_eigenvalues}
\end{figure}

\noindent\textbf{Comparison between factorization and iterative approaches.}
In this first sub-experiment, we compare so-called factorization approaches with iterative ones on two Gaussian sampling problems. We consider first the problem of sampling from $\mathcal{N}(\B{0}_d,\B{Q}^{-1})$ where the covariance matrix $\B{\Sigma} = \B{Q}^{-1}$ is chosen as a squared exponential matrix that is commonly used in applications involving Gaussian processes \cite{Higdon2007,MacKay2003,Rasmussen2003,Williams2006,Shi2000,Vasco1999}. Its coefficients are defined by
\begin{equation}
  \Sigma_{ij} = 2 \exp\pr{-\dfrac{(s_i-s_j)^2}{2a^2}} + \epsilon\delta_{ij}\eqsp, \quad \forall i,j \in [d]\eqsp, \label{eq:simu1_ex1}
\end{equation}
where $\{s_{i}\}_{i\in[d]}$ are evenly spaced scalars on $[-3,3]$, $\epsilon > 0$ and the Kronecker symbol $\delta_{ij} = 1$ if $i=j$ and zero otherwise.
In \cref{eq:simu1_ex1}, the parameters $a$ and $\epsilon$ have been set to $a = 1.5$ and $\epsilon = 10^{-6}$ which yields a distribution of the eigenvalues of $\B{\Sigma}$ such that the large ones are well separated while the small ones are clustered together near $10^{-6}$, see \cref{fig:simu1_distrib_eigenvalues} ($1$st row).
We compare the Cholesky sampler (\cref{algo:factorization_sampler}), the approximate inverse square root sampler using Chebyshev polynomials (\cref{algo:Chebychev}) and the conjugate gradient (CG) based sampler (\cref{algo:CG}). 
The sampler using Chebyshev polynomials needs $K_{\mathrm{cheby}}=23$ iterations on average while the CG iterations have been stopped once loss of conjugacy occurrs, following the guidelines of \cite{Parker2012}, that is at $K_{\mathrm{kryl}}=8$.
In all experiments, $T=10^5$ samples have been simulated in dimensions ranging from 1 to several thousands.

\cref{fig:simu1_distrib_eigenvalues} shows the results associated to these three direct samplers in dimension $d=100$.
The generated samples indeed follow a target Gaussian distribution admitting a covariance matrix close to $\B{\Sigma}$. 
This is attested by the small residuals observed between the estimated covariance and the true ones. Based on this criterion, all approaches successfully sample from $\mathcal{N}(\B{0}_d,\B{\Sigma})$.
This is emphasized by the spectrum of the estimated covariance matrices $\B{\widehat{\Sigma}}_T$ which coincides with the spectrum of $\B{\Sigma}$ for large eigenvalues. This observation ensures that most of the covariance information has been captured. However, note that only the Cholesky method is able to recover accurately all the eigenvalues, including the smallest ones.

\begin{figure}
  \centering
  \mbox{{\includegraphics[scale=0.4]{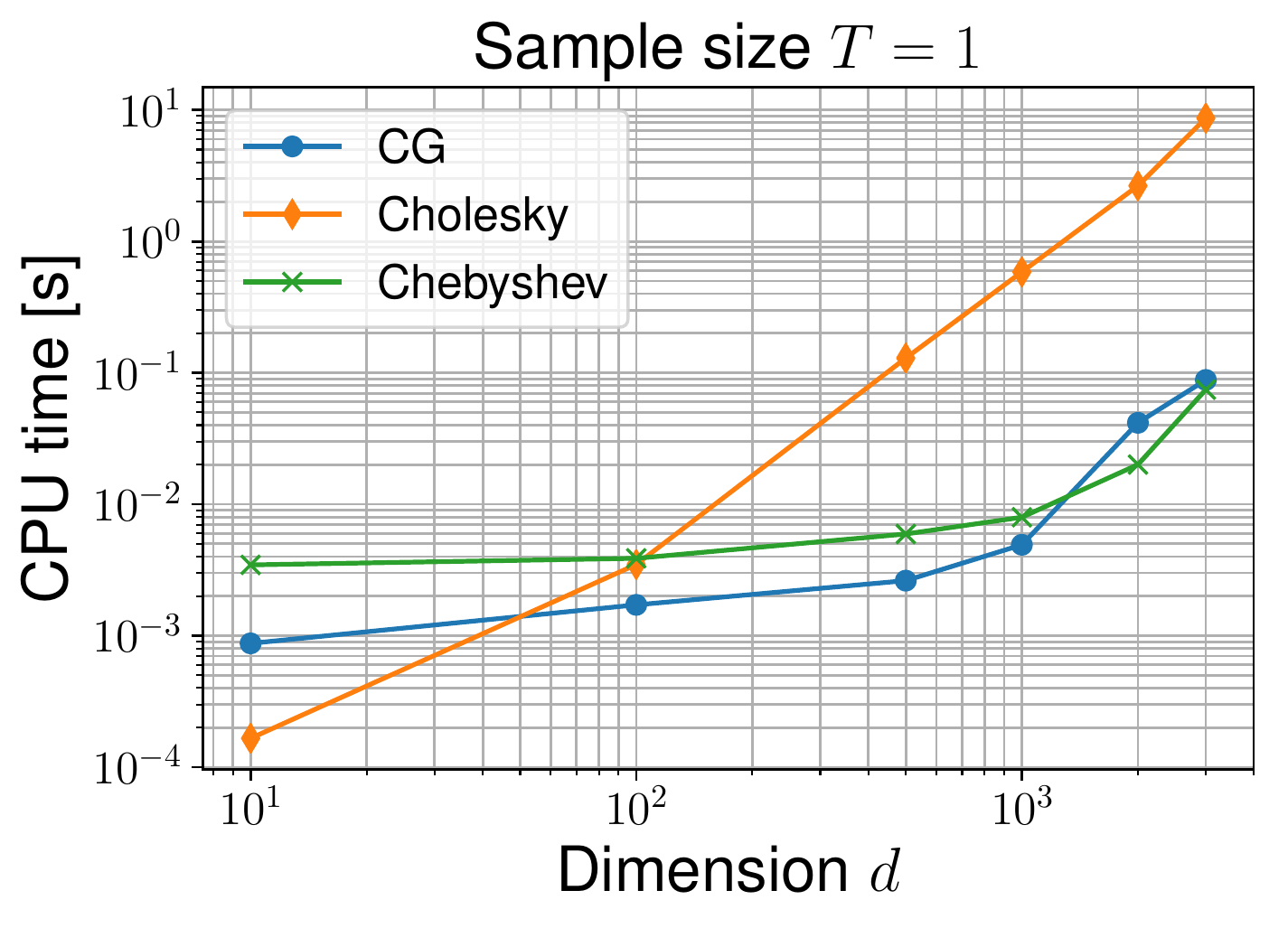}}}
  \mbox{{\includegraphics[scale=0.4]{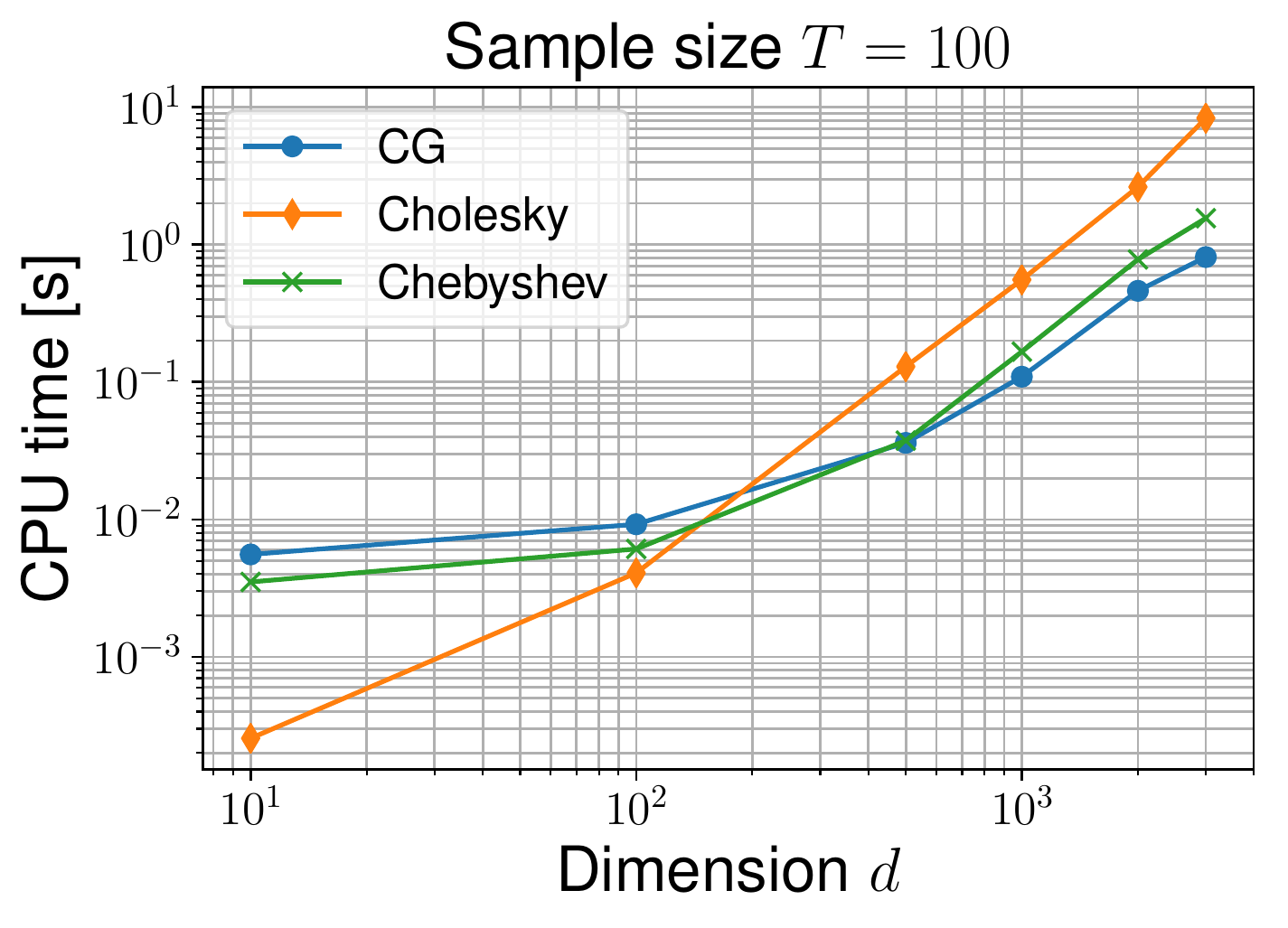}}}
  \mbox{{\includegraphics[scale=0.4]{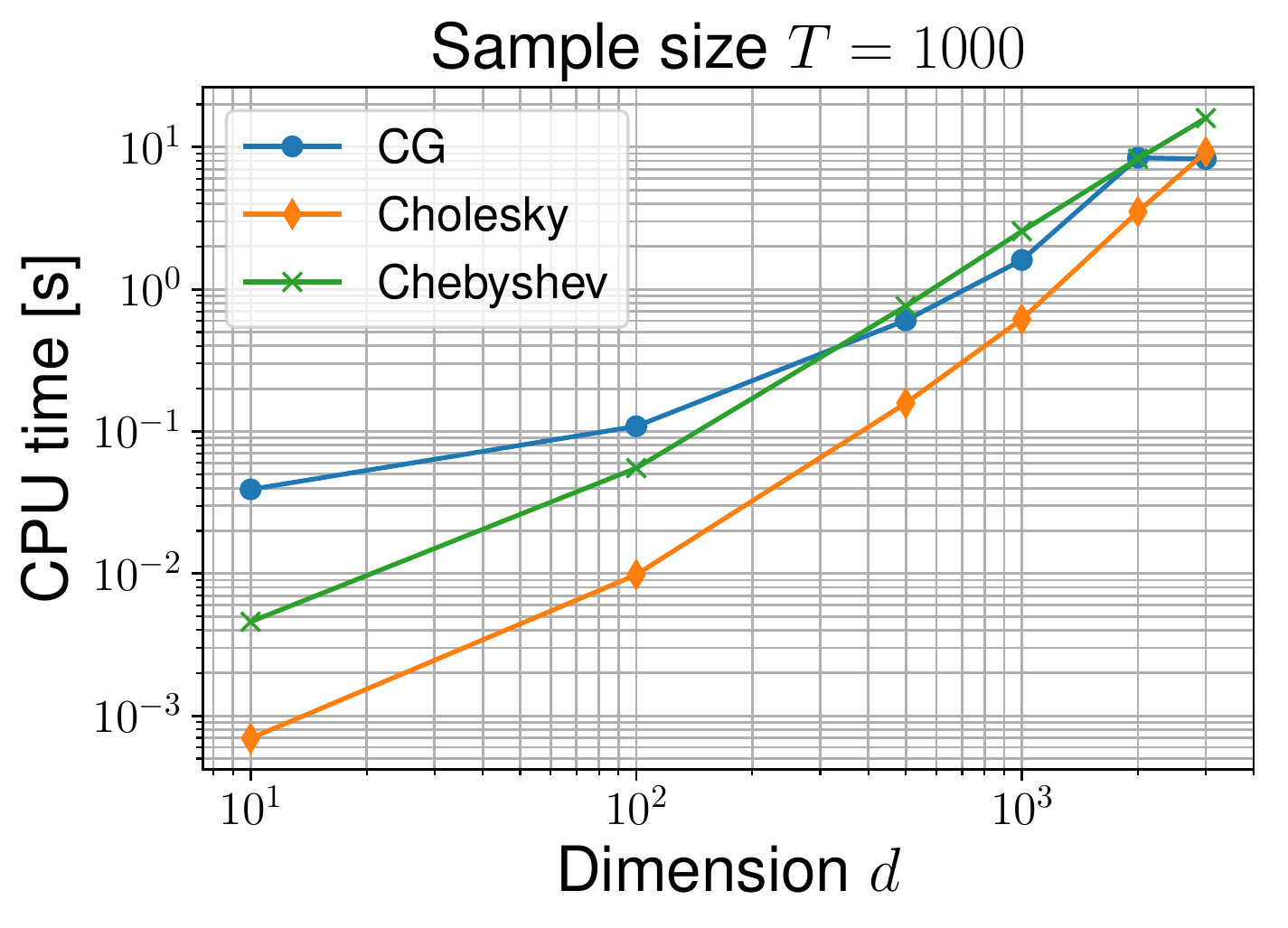}}}
  \mbox{{\includegraphics[scale=0.4]{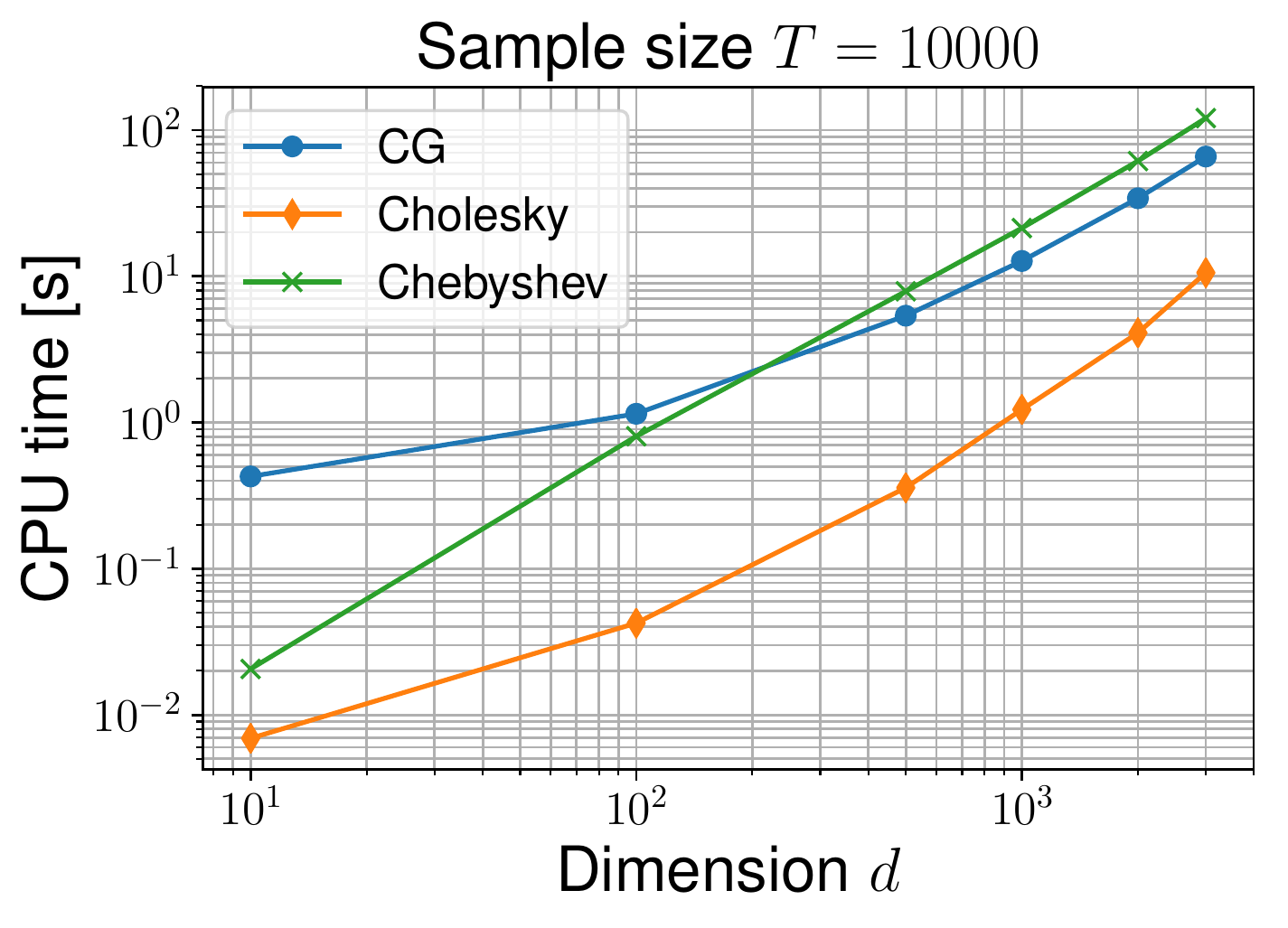}}}
\caption{Scenario 1. Comparison between the three considered direct samplers in terms of CPU time for the sampling from $\mathcal{N}(\B{0}_d,\B{\Sigma})$ with $\B{\Sigma}$ detailed in \cref{eq:simu1_ex1}.}
  \label{fig:simu1_time}
\end{figure}

\cref{fig:simu1_time} compares the previous direct samplers in terms of central processing unit (CPU) time. To generate only one sample ($T=1$), as expected, one can observe that the Cholesky sampler is faster than the two iterative samplers in small dimension $d$ and becomes computationally demanding as $d$ grows beyond a few hundreds.
Indeed, for small $d$ the cost implied by several iterations ($K_{\mathrm{cheby}}$ or $K_{\mathrm{kryl}}$) within each iterative sampler dominates the cost of the factorization in \cref{algo:factorization_sampler} while the contrary holds for large values of $d$.
Since the Cholesky factorization is performed only once, the Cholesky sampler becomes attractive over the two other approaches as the sample size $T$ increases.
However, as already pointed out in \cref{subsec:problem}, it is worth noting that pre-computing the Cholesky factor is no longer possible once the Gaussian sampling task involves a matrix $\B{\Sigma}$ which changes at each iteration of a Gibbs sampler, e.g., when considering a hierarchical Bayesian model with unknown hyperparameters (see Example \ref{ex:bayesian_ridge}).
We also point out that a comparison between direct samplers reviewed in \cref{sec:PIGauss_M_zero} and their related versions was conducted in \cite{Aune2013} in terms of CPU and GPU times.
In agreement with the findings reported here, this comparison essentially showed that the choice of the sampler in small dimension is not particularly important while iterative direct samplers become interesting in high-dimensional scenarios where Cholesky factorization becomes impossible.

\begin{figure}
  \centering
  \mbox{{\includegraphics[scale=0.45]{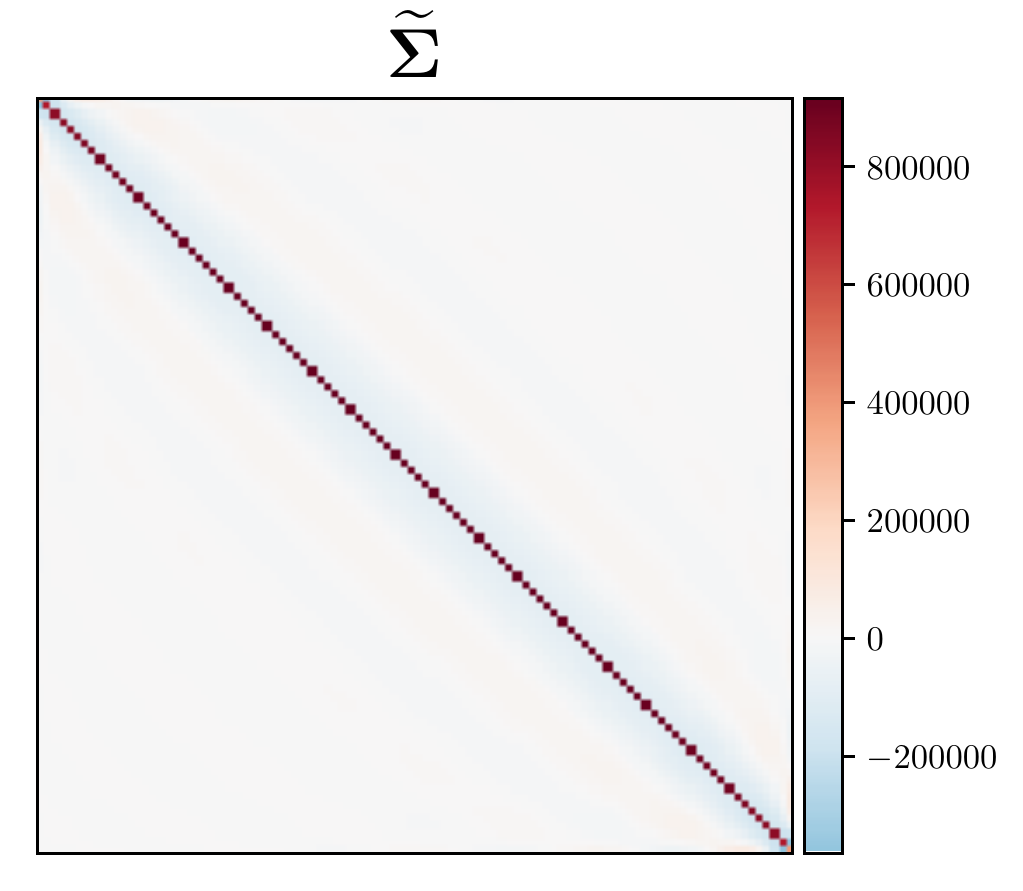}}}
  \mbox{{\includegraphics[scale=0.4]{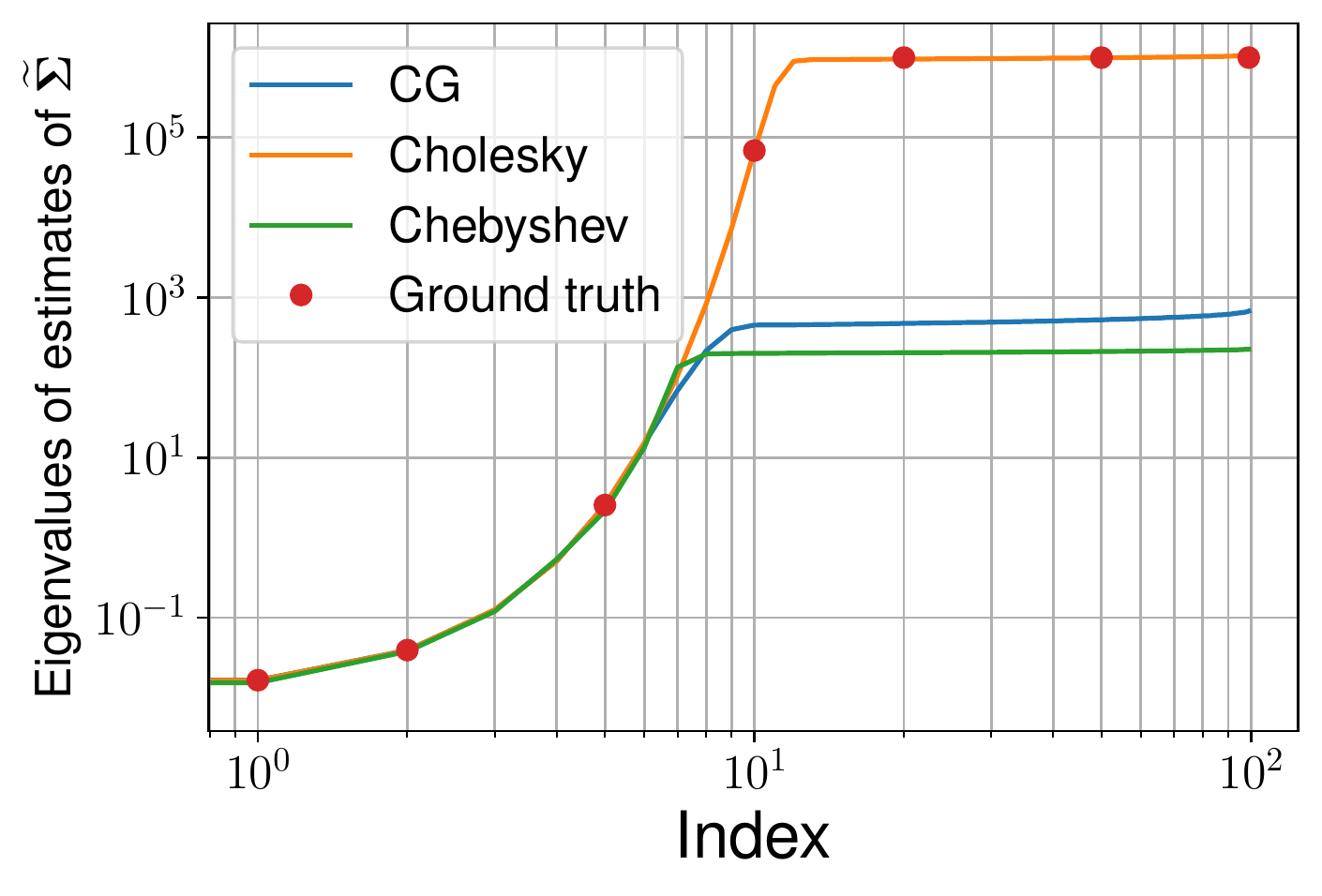}}}
  \mbox{{\includegraphics[scale=0.4]{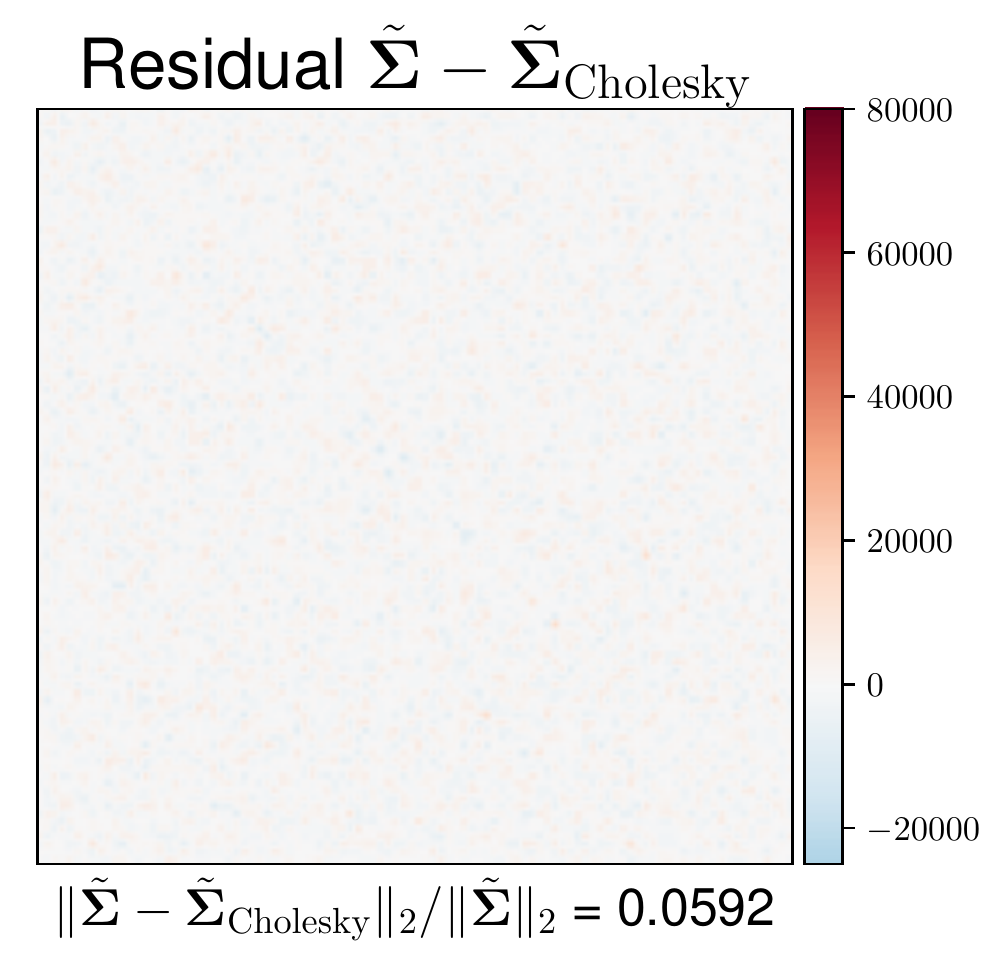}}}
  \mbox{{\includegraphics[scale=0.4]{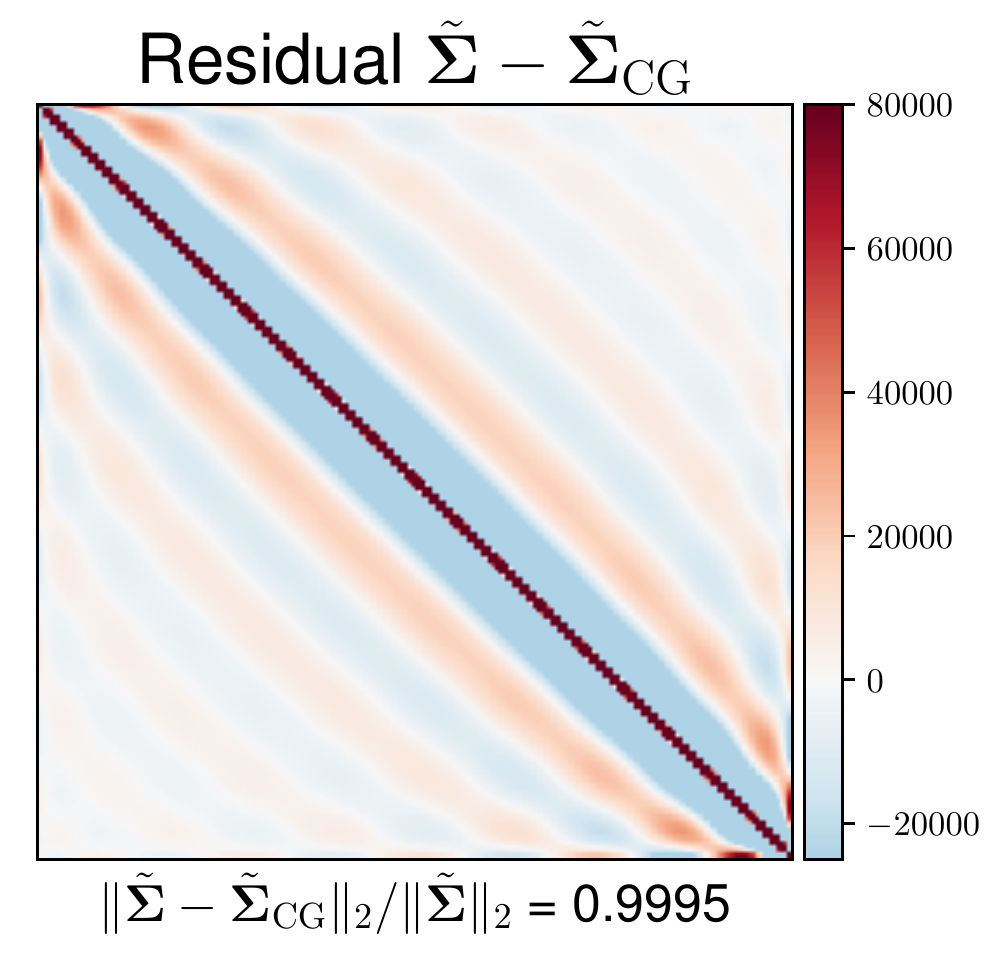}}}
  \mbox{{\includegraphics[scale=0.4]{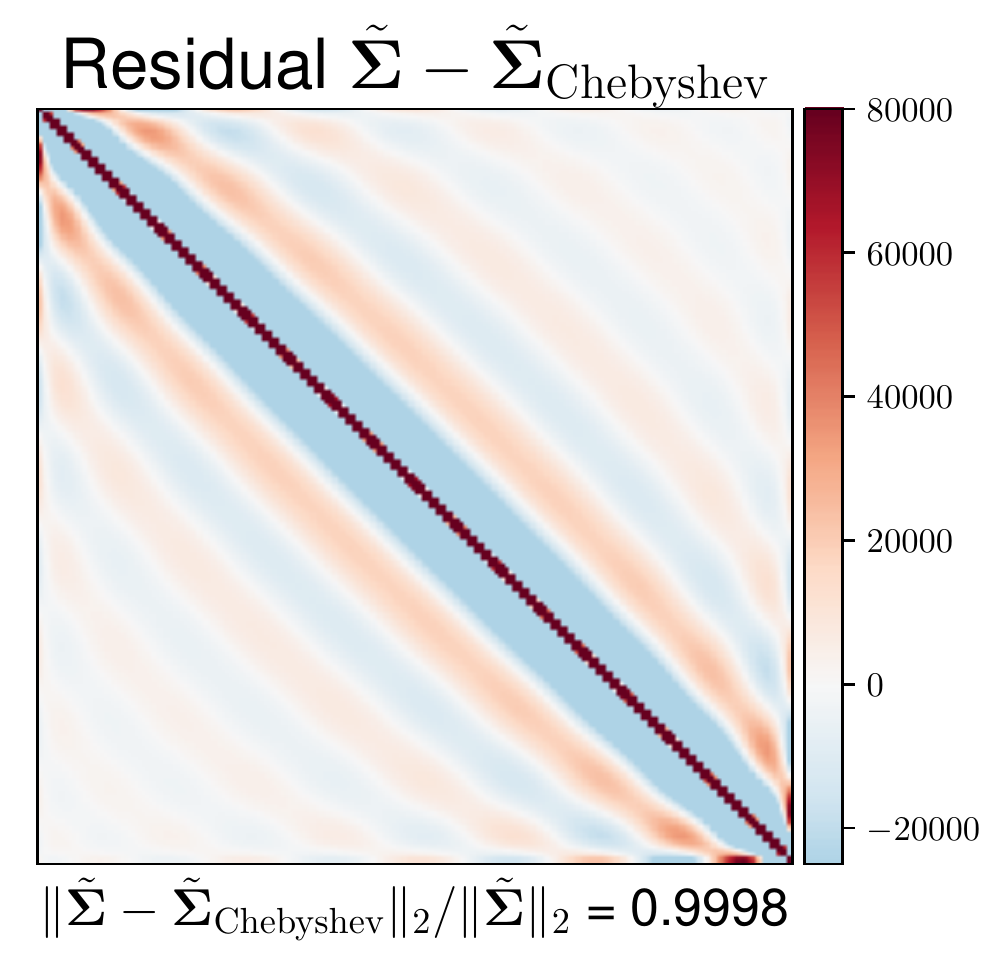}}}
\caption{Scenario 1. Results of the three considered samplers for the sampling from $\mathcal{N}(\B{0}_d,\widetilde{\B{\Sigma}})$ in dimension $d = 100$ .}
  \label{fig:simu1_distrib_eigenvalues_2}
\end{figure}

We complement our analysis by focusing on another sampling problem which now considers the matrix defined in \cref{eq:simu1_ex1} as a precision matrix instead of a covariance matrix: we now want to generate samples from  $\mathcal{N}(\B{0}_d,\widetilde{\B{\Sigma}})$ with $\widetilde{\B{\Sigma}} =  \B{\Sigma}^{-1}$. 
This sampling problem is expected to be more difficult since the largest eigenvalues of $\widetilde{\B{\Sigma}}$ are now clustered, see \Cref{fig:simu1_distrib_eigenvalues_2} ($1$st row).
\Cref{fig:simu1_distrib_eigenvalues_2} ($2$nd row) shows that the CG and Chebyshev samplers fail to capture covariance information as accurately as the Cholesky sampler.
The residuals between the estimated covariance and the true ones remain important on the diagonal: variances are inaccurately under-estimated. 
This observation is in line with \cite{Parker2012} who showed that the CG sampler is not suitable for the sampling from a Gaussian distribution whose covariance matrix has many clustered large eigenvalues, probably as a consequence of the bad conditioning of the matrix.
As far as the Chebyshev sampler is concerned, this failure was expected since the interval $[\lambda_l,\lambda_u]$ on which the function $x \mapsto x^{-1/2}$ has to be well-approximated becomes very large with an extent of about $10^6$. Of course the relative error between $\widetilde{\B{\Sigma}}$ and its estimate can be decreased by sufficiently increasing  the number of iterations $K_{\mathrm{cheby}}$ but this is possible only at a prohibitive computational cost.

\emph{On the choice of the metric to monitor convergence.} We saw on \cref{fig:simu1_distrib_eigenvalues} and \cref{fig:simu1_distrib_eigenvalues_2} that the covariance estimation error was small if the large values of the covariance matrix were captured and large in the opposite scenario.
Note that if the precision estimation error was chosen as a metric, we would have observed similar numerical results: if the largest eigenvalues of the precision matrix were not captured, the precision estimation error would have been large.
Regarding the CG sampler, Fox and Parker for instance in \cite{FoxParker14} highlighted this fact and illustrated it numerically (see equations (3.1) and (3.2) in that paper).

\begin{figure}
  \centering
  \mbox{{\includegraphics[scale=0.5]{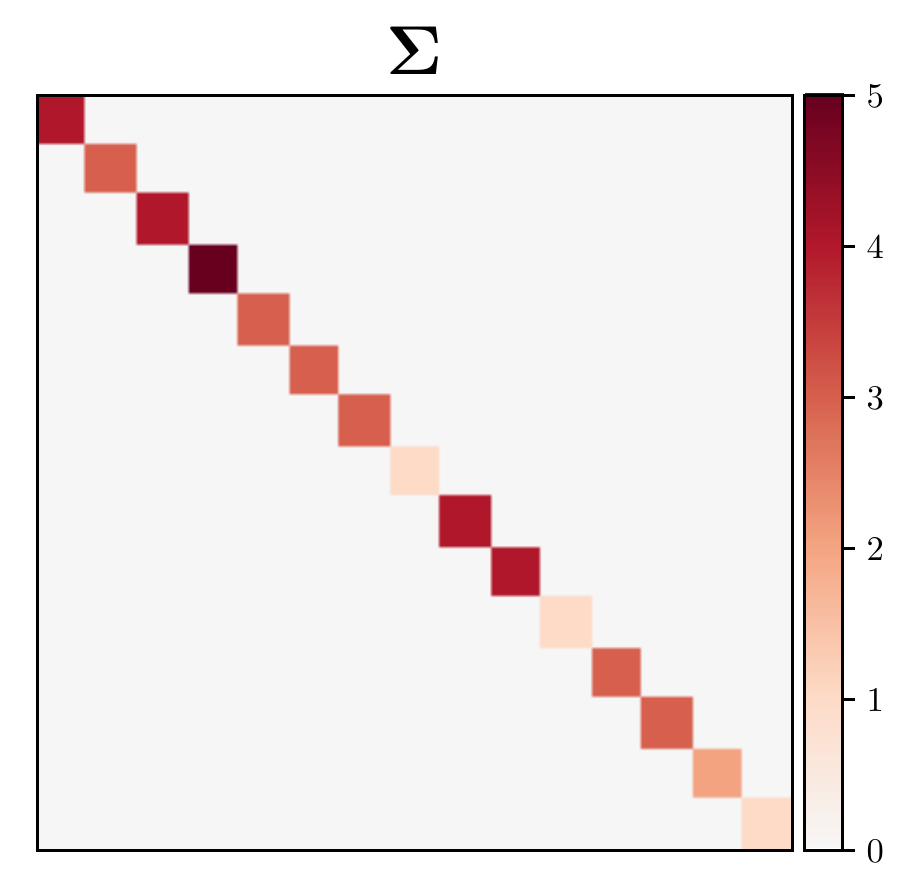}}}
  \mbox{{\includegraphics[scale=0.4]{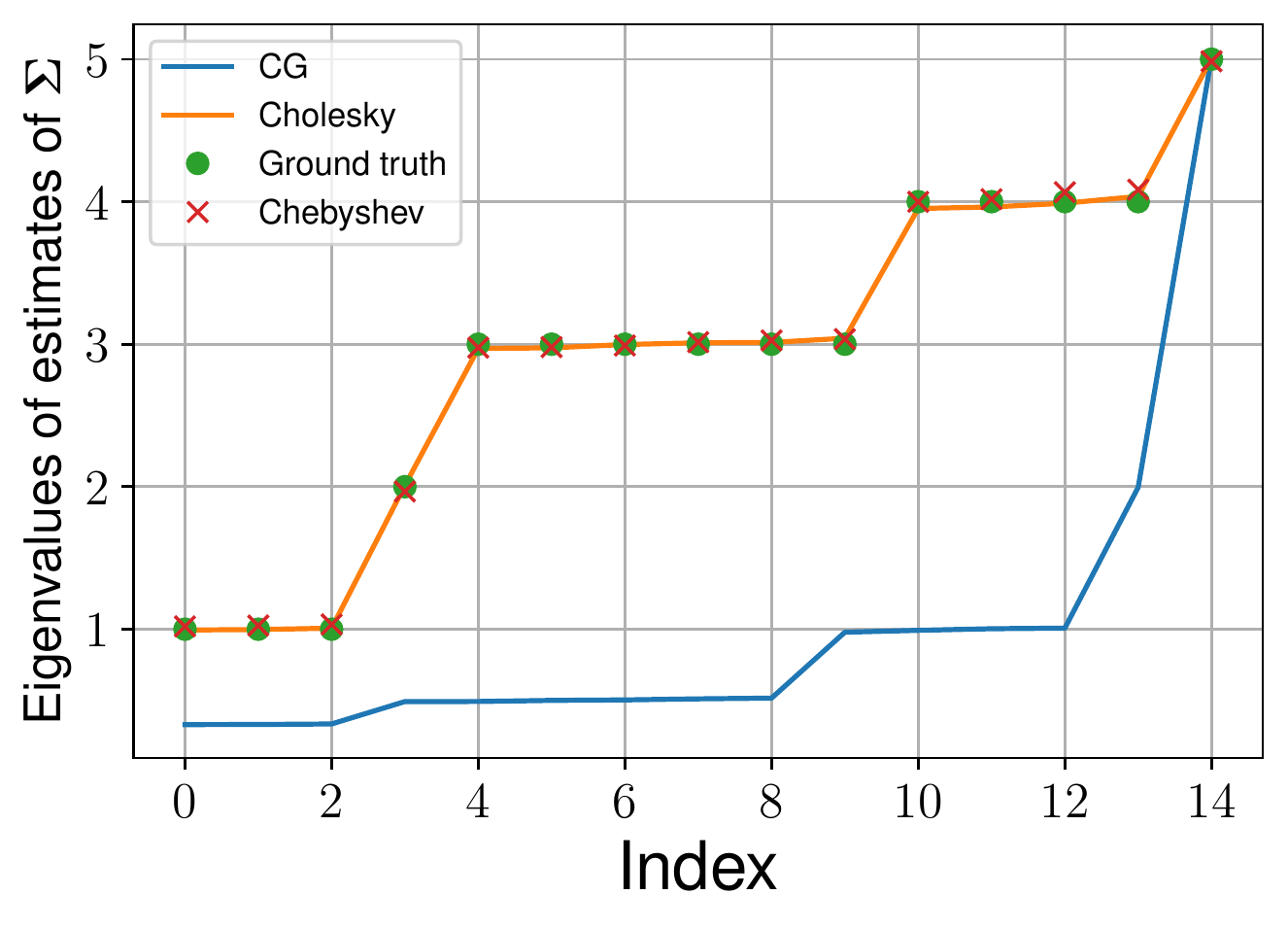}}}
  \mbox{{\includegraphics[scale=0.4]{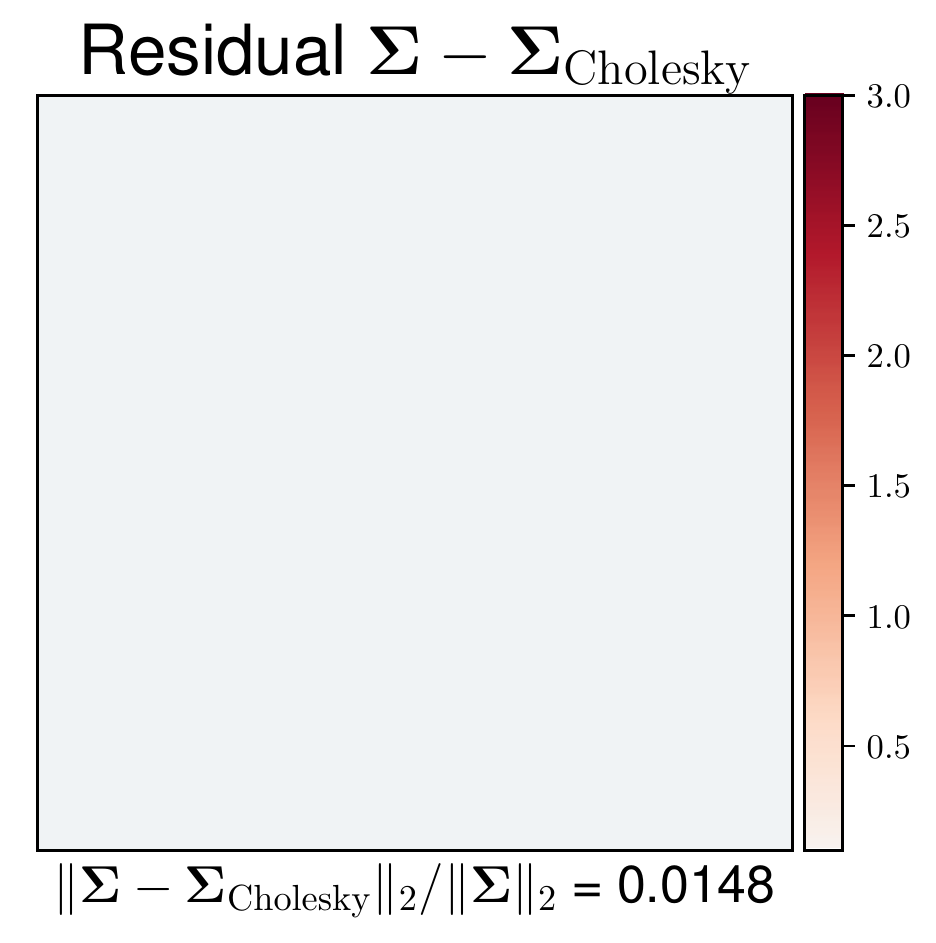}}}
  \mbox{{\includegraphics[scale=0.4]{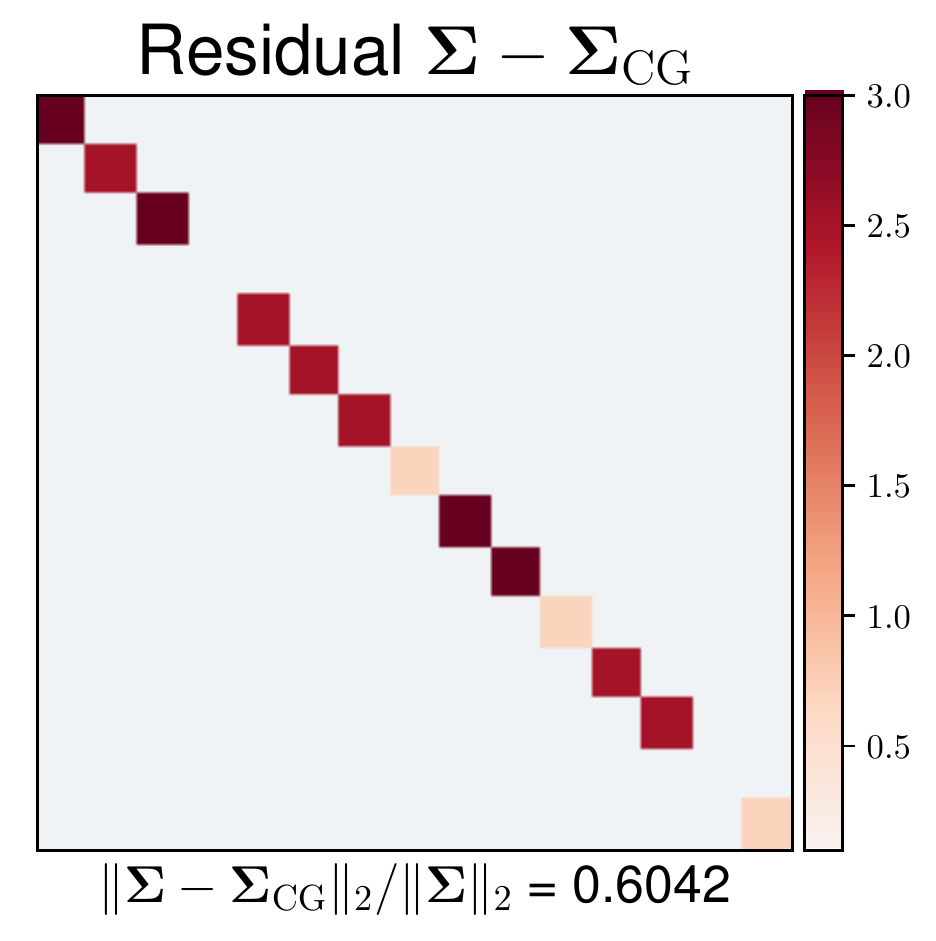}}}
  \mbox{{\includegraphics[scale=0.4]{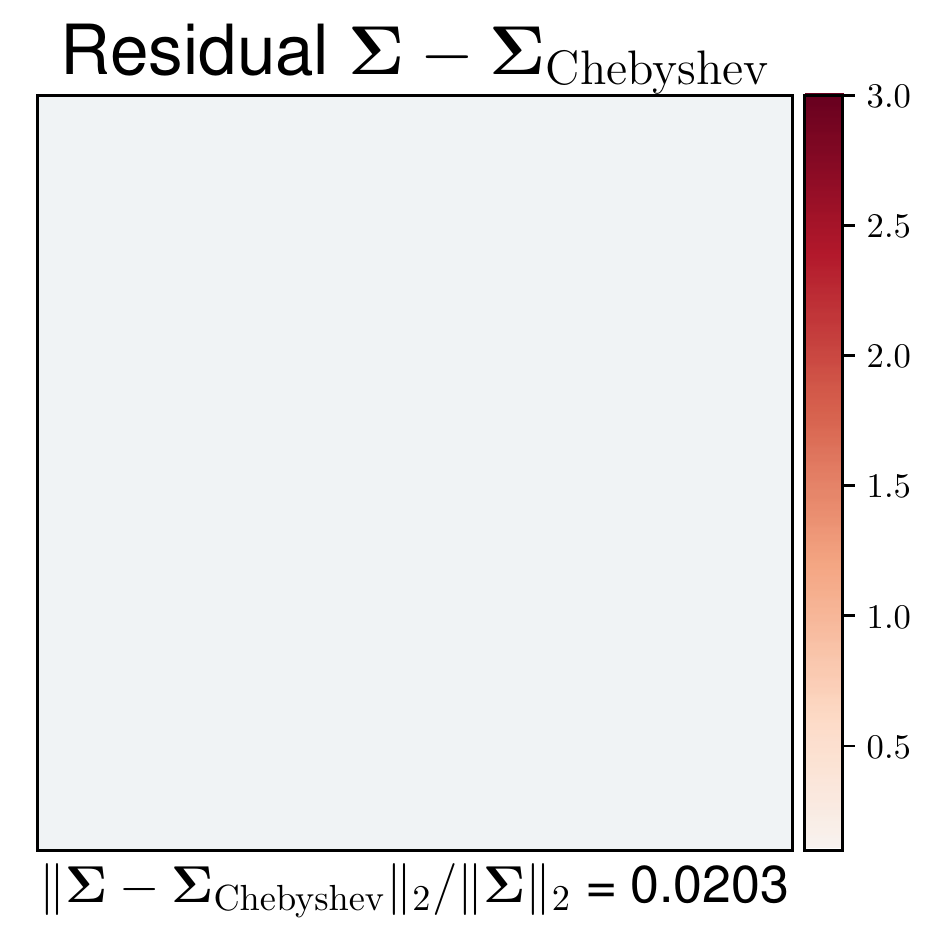}}}
\caption{Scenario 1. Results of the three considered direct samplers for the sampling from $\mathcal{N}(\B{0}_d,\B{\Sigma})$ with $\B{\Sigma}$ diagonal in dimension $d = 15$.}
  \label{fig:simu2}
\end{figure}

\noindent\textbf{Comparison between Chebyshev and CG-based samplers.} In order to discriminate the two iterative direct samplers \cref{algo:Chebychev} and \cref{algo:CG}, we consider a toy example in dimension $d=15$.
The covariance matrix $\B{\Sigma}$ is chosen as diagonal with diagonal elements drawn randomly from the discrete set $\llbracket1,5\rrbracket$.
As shown in \cref{fig:simu2}, $\B{\Sigma}$ has repeated and large eigenvalues.
Because of that, the CG sampler stopped at $K_{\mathrm{kryl}}=5$ (the number of distinct eigenvalues of $\B{\Sigma}$) and failed to sample accurately from $\mathcal{N}(\B{0}_d,\B{\Sigma})$ while the sampler based on Chebyshev polynomials yields samples of interest.
Hence, although the CG sampler is an attractive iterative option, its accuracy is known to be highly dependent on the distribution of the eigenvalues of $\B{\Sigma}$ which is in general unknown in high-dimensional settings.
Preconditioning techniques to spread out these eigenvalues can be used but might fail to reduce the error significantly as shown in \cite{Parker2012}. 

\subsubsection{Scenario 2}
\label{subsec:scenario2}
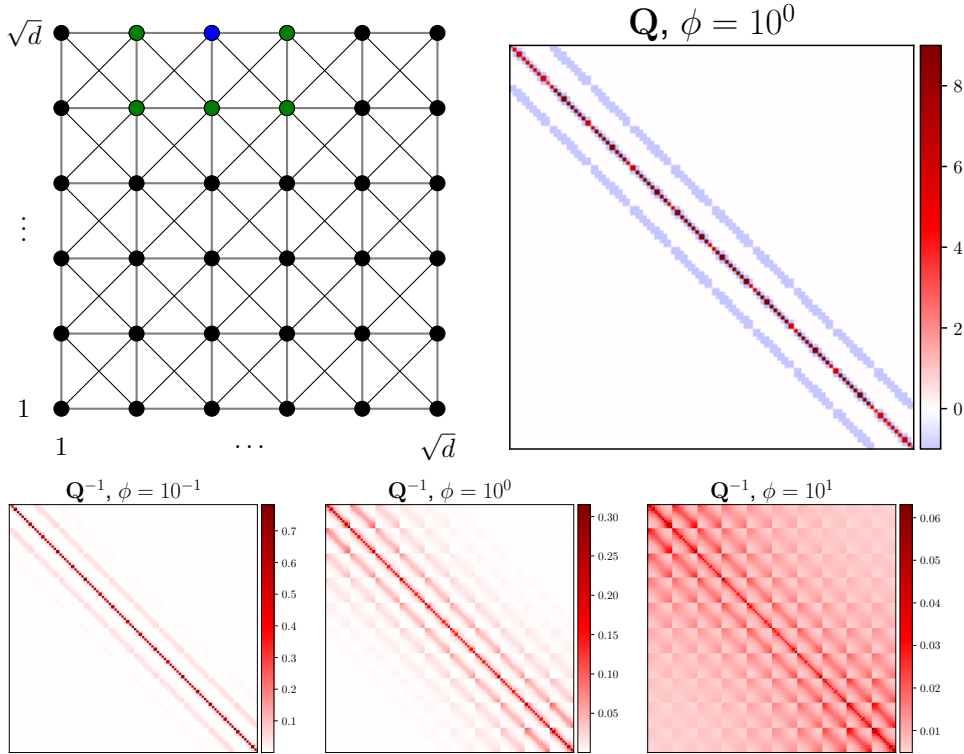
\begin{figure}
  \begin{tikzpicture}
    \draw[style=help lines,thick] (0,0) grid[step=1] (5,5);
    \foreach \x in {0,1,...,5}
    {
      \foreach \y in {0,1,...,5}
        {
          \node[draw,circle,inner sep=2pt,fill] at (1*\x,1*\y) {};
        }
    }
    \foreach \x in {0,1,...,4}
    {
      \foreach \y in {0,1,...,4}
      {
        \draw (\x,\y) -- (\x+1,\y+1);
      }
    }
        \foreach \x in {0,1,...,4}
    {
      \foreach \y in {1,2,...,5}
      {
        \draw (\x,\y) -- (\x+1,\y-1);
      }
    }
    \node[draw,circle,inner sep=2pt,fill=blue] at (2,5) {};
    \node[draw,circle,inner sep=2pt,fill=green] at (1,5) {};
    \node[draw,circle,inner sep=2pt,fill=green] at (3,5) {};
    \node[draw,circle,inner sep=2pt,fill=green] at (2,4) {};
    \node[draw,circle,inner sep=2pt,fill=green] at (1,4) {};
    \node[draw,circle,inner sep=2pt,fill=green] at (3,4) {};
    \node at (5,-0.5) {$\sqrt{d}$};
    \node at (2.5,-0.5) {$\hdots$};
    \node at (0,-0.5) {1};
    \node at (-0.5,0) {1};
    \node at (-0.5,2.5) {$\vdots$};
    \node at (-0.5,5) {$\sqrt{d}$};
  \end{tikzpicture}
  \mbox{{\includegraphics[scale=0.7]{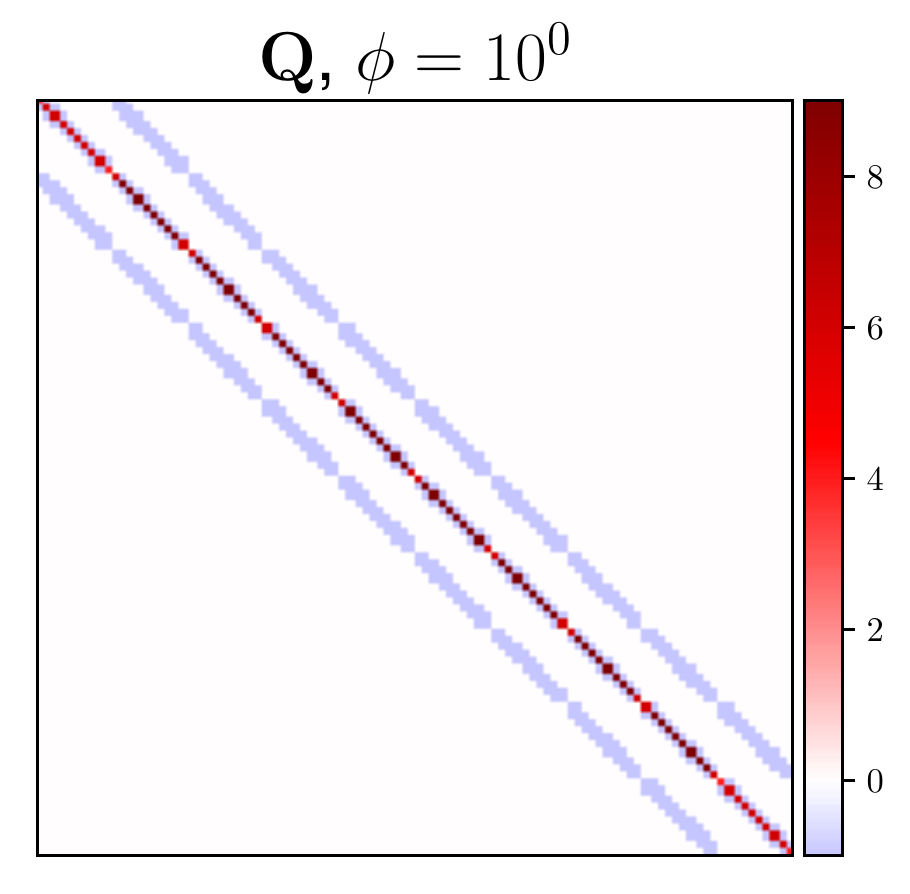}}}
  \mbox{{\includegraphics[scale=0.43]{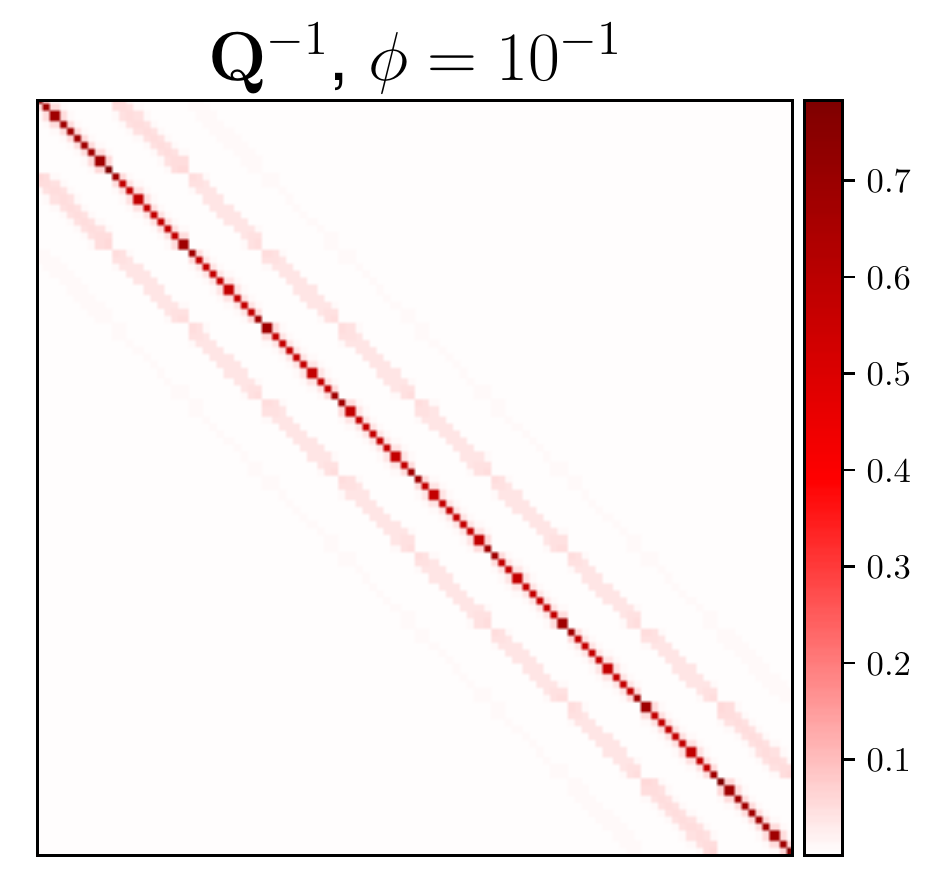}}}
  \mbox{{\includegraphics[scale=0.43]{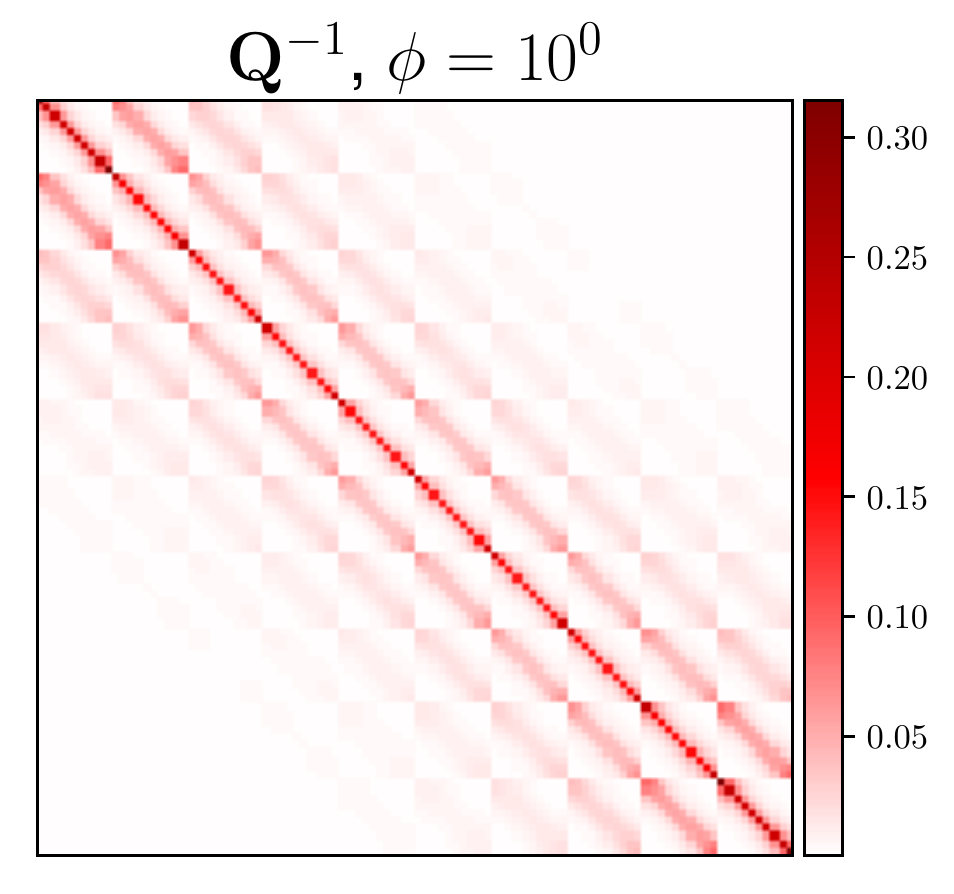}}}
  \mbox{{\includegraphics[scale=0.43]{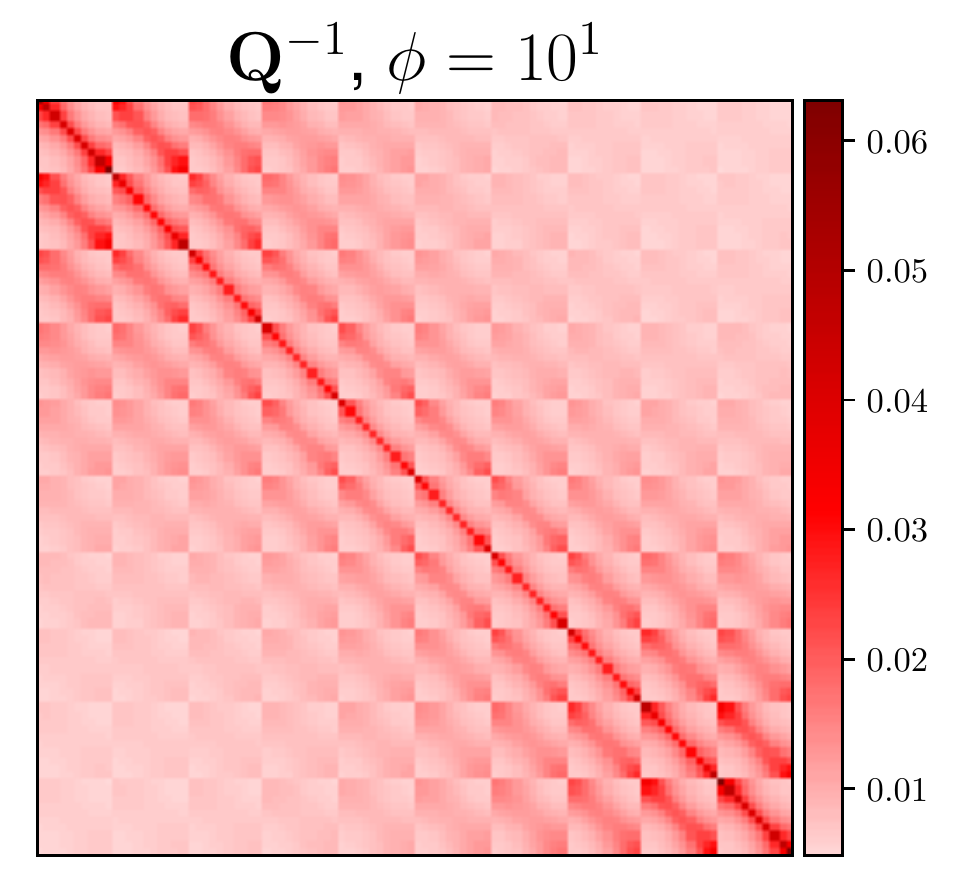}}}
\caption{Scenario 2. Illustrations of considered the Gaussian sampling problem: (top left) 2D-lattice ($d=36$) associated to the precision matrix $\B{Q}$ in \cref{eq:simu1_ex2}, (top right) $\B{Q}$ depicted for $\phi=1$ and (bottom) $\B{Q}^{-1}=\B{\Sigma}$ for $\phi \in \{0.1,1,10\}$. All the results are shown for $d=100$.
On the 2D-lattice, the green nodes stand for the neighbors of the blue node while the coordinates of the lattice correspond to the coordinates of $\Bs{\theta} \in \mathbb{R}^d$ with non-zero correleations, that is $1\leq i,j\leq \sqrt{d}$.}
\label{fig:scenario_2_illutration_grid}
\end{figure}

Now we turn to Question 2 and compare iterative to MCMC samplers. 
In this scenario, we consider a precision matrix $\B{Q}$ which is commonly used to build Gaussian Markov random fields (GMRFs) \cite{Rue2005}.
Before defining this matrix, we introduce some notations.
Let $\mathcal{G} = (\mathcal{V},\mathcal{E})$ an undirected 2-dimensional graph, see \cref{fig:scenario_2_illutration_grid}, where $\mathcal{V}$ stands for the set of $d$ nodes in the graph and $\mathcal{E}$ for the edges between the nodes.
We say that nodes $i$ and $j$ are neighbors and write $i \sim j$ if there is an edge connecting these two nodes.
The number of neighbors of node $i$ is denoted $n_i$; it is also called the {\em degree}.
Using these notations, we set $\B{Q}$ to be a second order locally linear precision matrix \cite{Higdon2007,Rue2005} associated to the two-dimensional lattice shown in \cref{fig:scenario_2_illutration_grid}, which writes
\begin{equation}
  Q_{ij} = \epsilon\delta_{ij} + \left\{
                \begin{array}{ll}
                  \phi n_i \ \text{if} \ i=j\\
                  -\phi \ \text{if} \ i \sim j\\
                  0 \ \text{otherwise}
                \end{array}
              \right., \forall i,j \in [d]\eqsp,\label{eq:simu1_ex2}
\end{equation}
where we set $\epsilon = 1$ (actually $\epsilon > 0$ suffices) and $\phi>0$ to ensure that $\B{Q}$ is a non-singular matrix yielding a non-intrinsic Gaussian density w.r.t. the $d$-dimensional Lebesgue measure, see \cref{subsec:special_instances}. 
Note that this precision matrix is band-limited with bandwidth of the order $\mathcal{O}(\sqrt{d})$ \cite{Rue2005} preluding the possible embedding of \cref{algo:multi_band} within the samplers considered in this scenario.
Related instances of this precision matrix have also been considered in \cite{Ilic2009,Parker2012,Fox2017} in order to show the benefits of both direct and MCMC samplers.
In the sequel, we consider the sampling from $\mathcal{N}(\B{0}_d,\B{Q}^{-1})$ for three different scalar parameters $\phi \in \{0.1,1,10\}$ leading to three covariance matrices $\B{Q}^{-1}$ with different correlation structures, see \cref{fig:scenario_2_illutration_grid}.
This will be of interest since it is known that the efficiency of Gibbs samplers is highly dependent on the correlation between the components of the Gaussian vector $\Bs{\theta} \in \mathbb{R}^d$ \cite{Robert2004,Rue2005}.

For this scenario, we set $d = 100$ in order to provide complete diagnostics for evaluating the accuracy of the samples generated by each algorithm. 
We implemented the four MCMC samplers based on exact matrix splitting (see \cref{table:matrix_splitting}) without considering a burn-in period (i.e., $T_{\mathrm{bi}} = 0$). These MCMC algorithms are compared with the direct samplers based on Cholesky factorization and Chebyshev polynomials, see \cref{sec:PIGauss_M_zero}.
Since the matrix $\B{Q}$ is strictly diagonally dominant, that is $|Q_{ii}| > \sum_{j\neq i}|Q_{ij}|$ for all $i \in [d]$, the convergence of the MCMC sampler based on Jacobi splitting is ensured \cite{Golub1989,Fox2017}.
Based on this convergence property, we can use an optimal value for the parameter $\omega$ appearing in the MCMC sampler based on successive over-relaxation (SOR) splitting, see \cref{appendix:C}. 

\cref{fig:scenario_2_results} shows the relative error between the estimated covariance matrix and the true one w.r.t. the number of samples generated with i.i.d. samplers from \cref{sec:PIGauss_M_zero} and MCMC samplers from \cref{sec:PIGauss_M_nonzero}.
Regarding MCMC samplers, no burn-in has been considered here to emphasize that these algorithms do not yield i.i.d. samples from the first iteration compared to samplers reviewed in \cref{sec:PIGauss_M_zero}.
This behavior is particularly noticeable when $\phi = 10$ where one can observe that both Gauss-Seidel, Jacobi and Richardson samplers need far more samples than Chebyshev and Cholesky samplers to reach the same precision in terms of covariance estimation.
Interestingly, this claim does not hold for ``accelerated'' Gibbs samplers such as the SOR (accelerated version of the Gauss-Seidel sampler) and the Chebyshev accelerated SSOR samplers.
Indeed, for $\phi=1$, one can note that the latter sampler is as efficient as i.i.d. samplers.
On the other hand, when $\phi=10$, these two accelerated Gibbs samplers manage to achieve lower relative covariance estimation error than the Chebyshev sampler when the number of iterations increases.
This behavior is due to the fact that the Chebyshev sampler involves a truncation procedure and as such provides approximate samples from $\mathcal{N}(\Bs{\mu},\B{Q}^{-1})$, compared to exact MCMC schemes which produce asymptotically exact samples.
These numerical findings match observations made by \cite{Fox2017} who experimentally showed that accelerated Gibbs approaches can be considered as serious contenders for the fastest samplers in high-dimensional settings compared to i.i.d. methods.

\cref{table:scenario_2_table} complements these numerical findings by reporting the spectral radius of the iteration operator $\B{M}^{-1}\B{N}$ associated to each MCMC sampler.
This radius is particularly important since it is directly related to the convergence factor of the corresponding MCMC sampler \cite{Fox2017}. 
In order to provide quantitative insights about the relative performance of each sampler, \cref{table:scenario_2_table} also shows the number of samples $T$ and corresponding CPU time such that the relative error between the covariance matrix $\B{\Sigma}$ and its estimate $\B{\widehat{\Sigma}}_T$ computed from $T$ samples generated by each algorithm is lower than $5 \times 10^{-2}$, i.e., a relative error of 5$\%$.
Thanks to the small bandwdith ($b=11$) of $\B{Q}$, the covariance matrix $\B{M}^{\top} + \B{N}$ of the vector $\tilde{\B{z}}$ appearing in step 3 of \cref{algo:matrix_splitting} is also band-limited with $b=11$ for both Jacobi and Richardson splitting.
Hence \cref{algo:multi_band} specifically dedicated to band matrix can be used within \cref{algo:matrix_splitting} for these two splitting strategies.
Although the convergence of these samplers is slower than that of the Gauss-Seidel sampler, their CPU times become roughly equivalent.
Note that this computational gain is problem dependent and cannot be ensured in general. Cholesky factorisation appears to be much faster in all cases when the same constant covariance is used for many samples. Next scenario will precisely consider high dimensional scenarios where Cholesky factorization is not possible anymore.
\begin{figure}
  \mbox{{\includegraphics[scale=0.19]{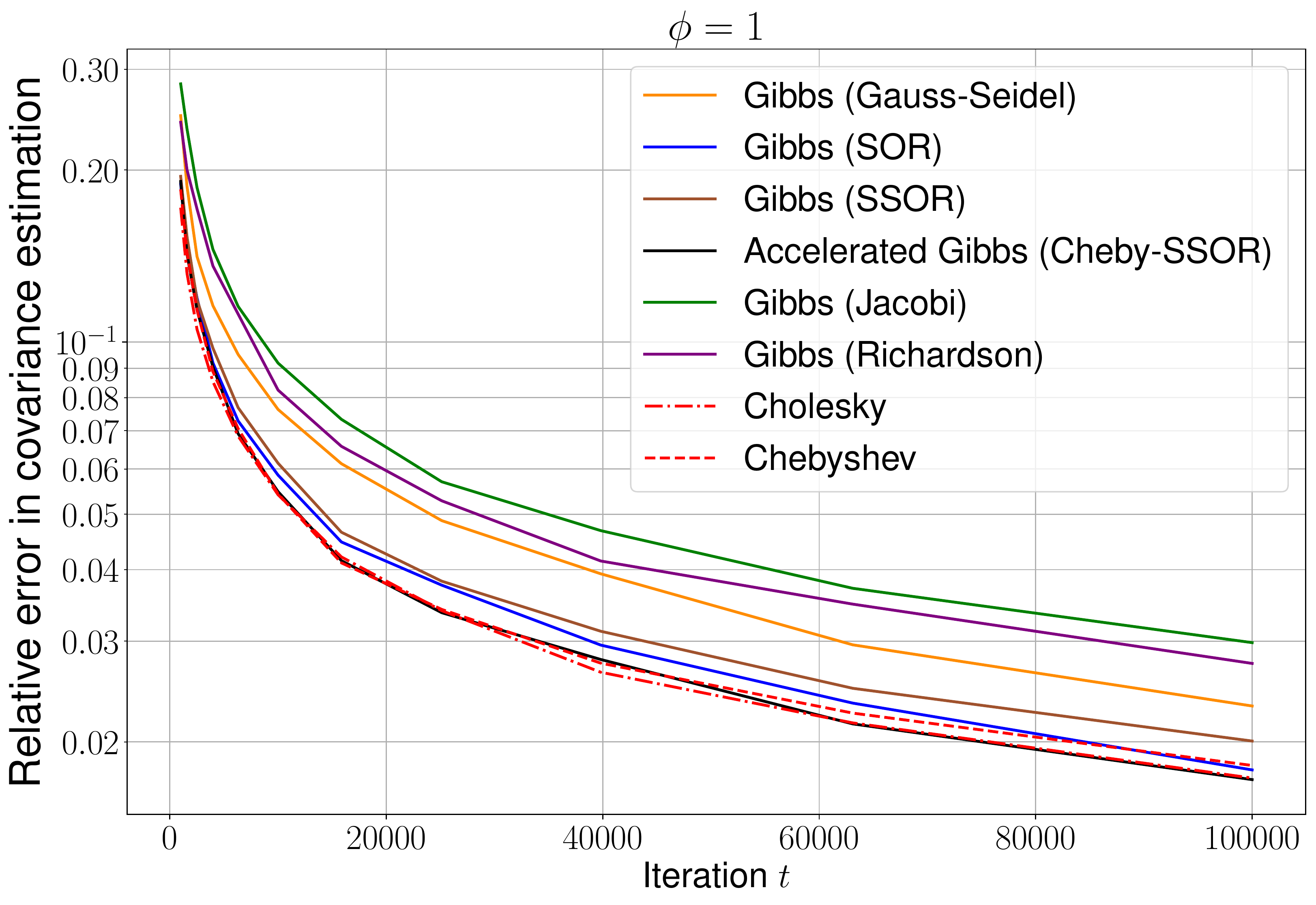}}}
  \mbox{{\includegraphics[scale=0.19]{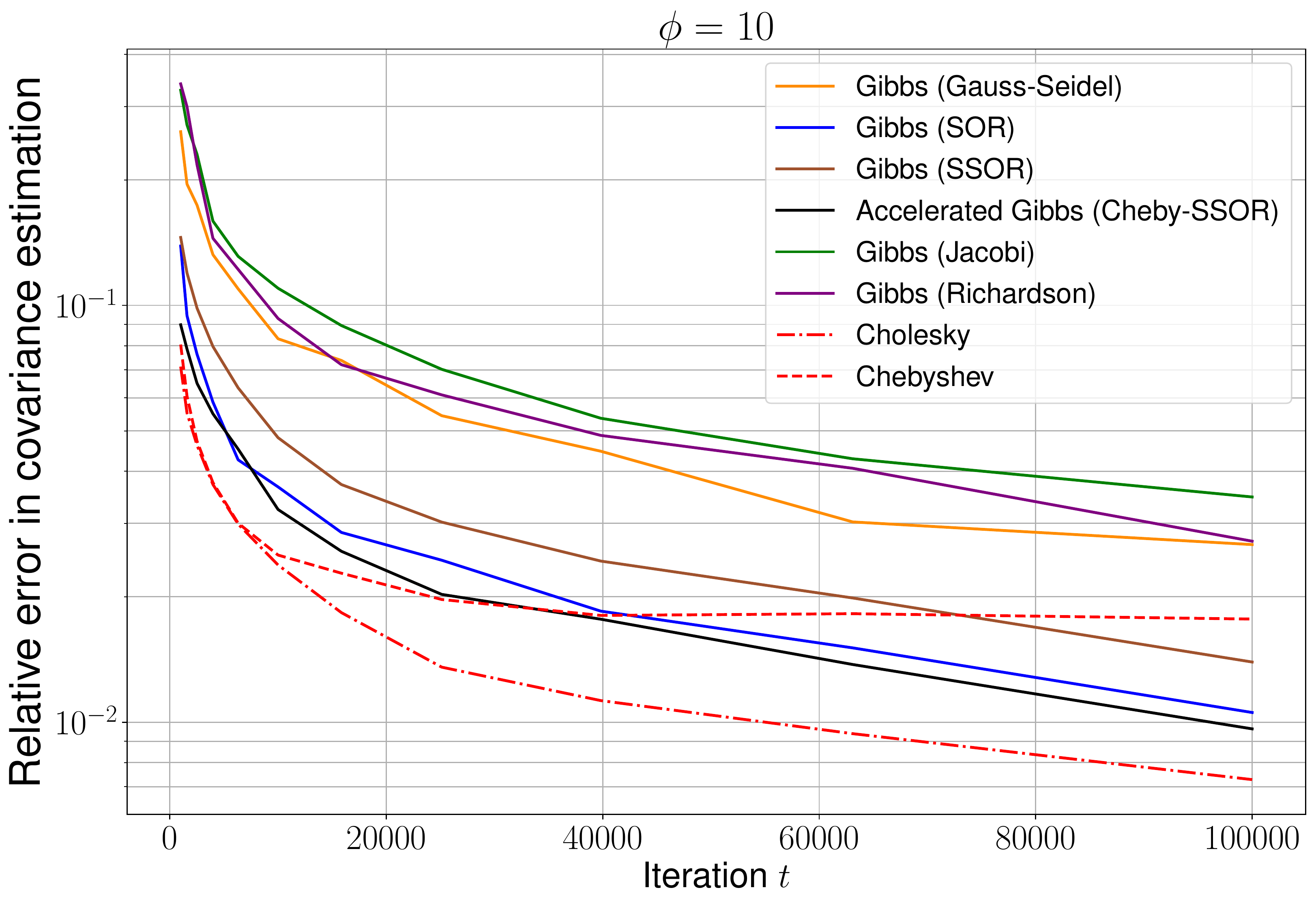}}}
  \caption{Scenario 2. Relative error associated to the estimation of the covariance matrix $\B{Q}^{-1}$ defined by $\|\B{Q}^{-1} - \mathrm{var}(\boldsymbol{\theta}^{(1:t)})\|_2/\|\B{Q}^{-1}\|_2$  w.r.t. the number of iterations $t$ in \cref{algo:matrix_splitting}, with $d=100$ (left: $\phi=1$, right: $\phi=10$). 
  We also highlighted the relative error obtained with an increasing number of samples generated independently from Cholesky and Chebyshev samplers.
  The results have been averaged over 30 independent runs.
  The standard deviations are not shown for readability reasons.}
\label{fig:scenario_2_results}
\end{figure}
\begin{table}
  \caption{Scenario 2. Comparison between Cholesky, Chebyshev and MS-based Gibbs samplers for $d=100$. 
  The samplers have been run until the relative error between the covariance matrix $\B{Q}^{-1}$ and its estimate is lower than $5 \times 10^{-2}$.
  For Richardson, SOR, SSOR and Cheby-SSOR samplers, the tuning parameter $\omega$ is the optimal one, see \cref{appendix:C}.
  The results have been averaged over 30 independent runs.}
  \label{table:scenario_2_table}
  \begin{center}
    \begin{tabular}{|cl|c|c|c|c|c|} 
      \hline
       \multicolumn{2}{|c|}{\textbf{Sampler}} & $\phi$ & $\omega$ & $\rho(\B{M}^{-1}\B{N})$ & $T$ & \textbf{CPU time [s]}\\
      \hline 
      \multicolumn{2}{|c|}{\multirow{3}{*}{Cholesky}} & 0.1 & -  & - & $6.3 \times 10^4$ & 0.29\\
      && 1 & -  & - & $1.3 \times 10^4$ & 0.06\\
      && 10 & -  & - & $2.9 \times 10^3$ & 0.01\\
      \hline
      \multicolumn{2}{|c|}{\multirow{3}{*}{Chebyshev ($K=21$)}} & 0.1 & -  & - & $6.4 \times 10^4$ & 2.24\\
      && 1 & -  & - & $1.3 \times 10^4$ & 0.44\\
      && 10 & -  & - & $2.5 \times 10^3$ & 0.19\\
      \hline      
      \multirow{18}{*}{\rotatebox{90}{MCMC MS-based samplers}} & \multicolumn{1}{|c|}{} & 0.1 & 0.6328  & 0.3672 & $6.7 \times 10^4$ & 5.44\\
      &\multicolumn{1}{|c|}{Richardson}& 1 & 0.1470  & 0.8530 & $3.8 \times 10^4$ & 3.03\\
      &\multicolumn{1}{|c|}{}& 10 & 0.0169 & 0.9831 & $4 \times 10^4$ & 3.31\\
      \cline{2-7}
      &\multicolumn{1}{|c|}{} & 0.1 & -  & 0.4235 & $6.8 \times 10^4$ & 5.72\\
      &\multicolumn{1}{|c|}{Jacobi}& 1 & - & 0.8749 & $3.9 \times 10^4$ & 3.24\\
      &\multicolumn{1}{|c|}{}& 10 & - & 0.9856 & $4.6 \times 10^4$ & 3.69\\
      \cline{2-7}
      &\multicolumn{1}{|c|}{} & 0.1 & -  & 0.1998 & $6.5 \times 10^4$ & 8.48\\
      &\multicolumn{1}{|c|}{Gauss-Seidel}& 1 & -  & 0.7677 & $2.5 \times 10^4$ & 3.34\\
      &\multicolumn{1}{|c|}{}& 10 & -  & 0.9715 & $2.5 \times 10^4$ & 3.32\\
      \cline{2-7}
      &\multicolumn{1}{|c|}{} & 0.1 & 1.0494  & 0.1189 & $6.4 \times 10^4$ & 8.40\\
      &\multicolumn{1}{|c|}{SOR}& 1 & 1.3474  & 0.4726 & $1.6 \times 10^4$ & 1.31\\
      &\multicolumn{1}{|c|}{}& 10 & 1.7110  & 0.7852 & $5.4 \times 10^3$ & 0.71\\
      \cline{2-7}
      &\multicolumn{1}{|c|}{} & 0.1 & 0.9644  & 0.0936 & $6.4 \times 10^4$ & 19.65\\
      &\multicolumn{1}{|c|}{SSOR}& 1 & 1.3331  & 0.4503 & $1.6 \times 10^4$ & 4.91\\
      &\multicolumn{1}{|c|}{}& 10 & 1.7101  & 0.9013 & $9.3 \times 10^3$ & 2.86\\
      \cline{2-7}
      &\multicolumn{1}{|c|}{} & 0.1 & 0.9644  & 0.0246 & $6.3 \times 10^4$ & 9.17\\
      &\multicolumn{1}{|c|}{Cheby-SSOR}& 1 & 1.3331  & 0.1485 & $1.3 \times 10^4$ & 1.89\\
      &\multicolumn{1}{|c|}{}& 10 & 1.7101  & 0.5213 & $4.5 \times 10^3$ & 0.65\\
      \hline
    \end{tabular}    
  \end{center}
\end{table}

\subsubsection{Scenario 3}\label{subsubsec:scenario3}
Finally, we deal with Question 3 above to assess the benefits of samplers which take advantage of the decomposition structure $\B{Q} = \B{Q}_1 + \B{Q}_2$ of the precision matrix.
As motivated in \cref{subsec:summary}, we will focus here on exact data augmentation approaches detailed in \cref{subsec:data_aug} and compare the latter to iterative samplers which produce uncorrelated samples, such as those reviewed in \cref{sec:PIGauss_M_zero}.

To this purpose, we consider Gaussian sampling problems in high dimensions $d \in [10^4,10^6]$ for which Cholesky factorization is both computationally and memory prohibitive when a standard computer is used.
This sampling problem commonly appears in image processing \cite{Giovannelli2015,Orieux2010,Marnissi2018,Vono2019} and arises from the linear inverse problem, usually called {\em deconvolution} or {\em deblurring} in image processing:
\begin{equation}
  \B{y} = \B{S}\Bs{\theta} + \Bs{\varepsilon}\eqsp,
\end{equation}
where $\B{y} \in \mathbb{R}^d$ refers to a blurred and noisy observation, $\Bs{\theta} \in \mathbb{R}^d$ is the unknown original image rearranged in lexicographic order, $\Bs{\varepsilon} \sim \mathcal{N}(\B{0}_d,\B{\Gamma})$ with $\B{\Gamma} = \mathrm{diag}(\gamma_1,\hdots,\gamma_d)$ stands for a synthetic structured noise such that $\gamma_i \sim 0.7\delta_{\kappa_1} + 0.3\delta_{\kappa_2}$, for all $i \in [d]$.
In the sequel, we set $\kappa_1 = 13$ and $\kappa_2 = 40$.
The matrix $\B{S} \in \mathbb{R}^{d \times d}$ stands for a circulant convolution matrix associated to the space-invariant box blurring kernel $\frac{1}{9}\B{1}_{3 \times 3}$ where $\B{1}_{p \times p}$ stands for the $p \times p$-matrix filled with ones.
We adopt a smoothing conjugate prior distribution on $\Bs{\theta}$ \cite{Molina1989,Molina2006,Likas2004}, already introduced in \cref{subsec:special_instances} and \cref{fig:circulant}, which writes $\mathcal{N}(\B{0}_d, (\frac{\xi_0}{d}\B{1}_{d \times d} +  \xi_1\B{\Delta}^{\top}\B{\Delta})^{-1})$ where $\B{\Delta}$ is the discrete two-dimensional Laplacian operator; $\xi_0 = 1$ ensures that this prior is non-intrinsic while $\xi_1 = 1$ controls the smoothing.
Bayes' rule then yields the Gaussian posterior distribution
\begin{equation}
  \Bs{\theta} \mid \B{y} \sim \mathcal{N}\pr{\Bs{\mu},\B{Q}^{-1}} \label{eq:scenario3_target}
\end{equation}
where
\begin{align}
  \B{Q} &= \B{S}^{\top}\B{\Delta}^{-1}\B{S} + \dfrac{\xi_0}{d}\B{1}_{d \times d} + \xi_1\B{\Delta}^{\top}\B{\Delta} \label{eq:scenario3_Q}\\
  \Bs{\mu} &= \B{Q}^{-1}\B{S}^{\top}\B{\Delta}^{-1}\B{y}\eqsp.
\end{align}
Sampling from \cref{eq:scenario3_target} is challenging since the size of the precision matrix forbids its computation. Moreover, the presence of the matrix $\B{\Gamma}$ rules out the diagonalization of $\B{Q}$ in the Fourier basis and therefore the direct use of \cref{algo:circulant}.
In addition, resorting to  MCMC samplers based on exact matrix splitting to sample from \cref{eq:scenario3_target} raises several difficulties.
First, both Richardson and Jacobi-based samplers involve a sampling step with an unstructured covariance matrix, see \cref{table:matrix_splitting}.
This step can be performed with one of the direct samplers reviewed in \cref{sec:PIGauss_M_zero} but this implies an additional computational cost.
On the other hand, although Gauss-Seidel and SOR-based MCMC samplers involve a simple sampling step, they require to have access to the lower triangular part of \cref{eq:scenario3_Q}.
In this high-dimensional scenario, the precision matrix cannot be easily computed on a standard desktop computer and this lower triangular part must be found with surrogate approaches.
One possibility consists in computing each non-zero coefficient of this triangular matrix following the matrix-vector products $\B{e}_i^{\top}\B{Q}\B{e}_j$ for all $i,j \in [d]$ such that $j \leq i$ where we recall that $\B{e}_i$ is the $i$-th canonical vector of $\mathbb{R}^d$.
These quantities can be pre-computed when $\B{Q}$ remains constant along the $T$ iterations but, again, becomes computationally prohibitive when $\B{Q}$ depends on unknown hyperparameters to be estimated within a Gibbs sampler. 

Nonetheless, since the precision matrix \cref{eq:scenario3_Q} can be decomposed as $\B{Q} = \B{Q}_1 + \B{Q}_2$ with $\B{Q}_1 = \B{S}^{\top}\B{\Gamma}^{-1}\B{S}$ and $\B{Q}_2 = \frac{\xi_0}{d}\B{1}_{d \times d} + \xi_1\B{\Delta}^{\top}\B{\Delta}$, we can apply \cref{algo:exact_DA} to sample efficiently from \cref{eq:scenario3_target}.
This algorithm is particularly interesting in this example since the three sampling steps involve two diagonal and one circulant precision matrices, respectively.
For the two first ones, one can use \cref{algo:multi_diag} while \cref{algo:circulant} can be resorted to sample from the last one.

In the sequel, we compare \cref{algo:exact_DA} with the CG direct sampler defined by \cref{algo:CG}.
Since we consider high-dimensional scenarios, the covariance estimate in \cref{eq:cov_estimate} cannot be used to assess the convergence of these samplers.
Instead, we compare the respective efficiency of the considered samplers by computing the effective sample size ratio per second (ESSR).
For an MCMC sampler, the ESSR gives an estimate of the equivalent number of i.i.d. samples that can be drawn in one second, see \cite{Kass1998,Liu2001}.
It is defined as
\begin{equation}
  \mathrm{ESSR}(\vartheta) = \dfrac{1}{T_1}\dfrac{\mathrm{ESS}(\vartheta)}{T} =  \dfrac{1}{T_1\pr{1 + 2 \displaystyle\sum_{t=1}^\infty \rho_t(\vartheta)}} \label{eq:ESS}
\end{equation}
where $T_1$ is the CPU time in seconds required to draw one sample and $\rho_t(\vartheta)$ is the lag-$t$ autocorrelation of a scalar parameter $\vartheta$.
A variant of the ESSR has for instance been used in \cite{Gilavert2015} in order to measure the efficiency of an MCMC variant of the PO algorithm (\cref{algo:PO}).
For a direct sampler providing i.i.d. draws, the ESSR \cref{eq:ESS} simplifies to $1/T_1$ and represents the number of samples obtained in one second. 
In both cases, the larger the ESSR, the more computationally efficient is the sampler.

\begin{figure}
  \mbox{{\includegraphics[scale=0.45]{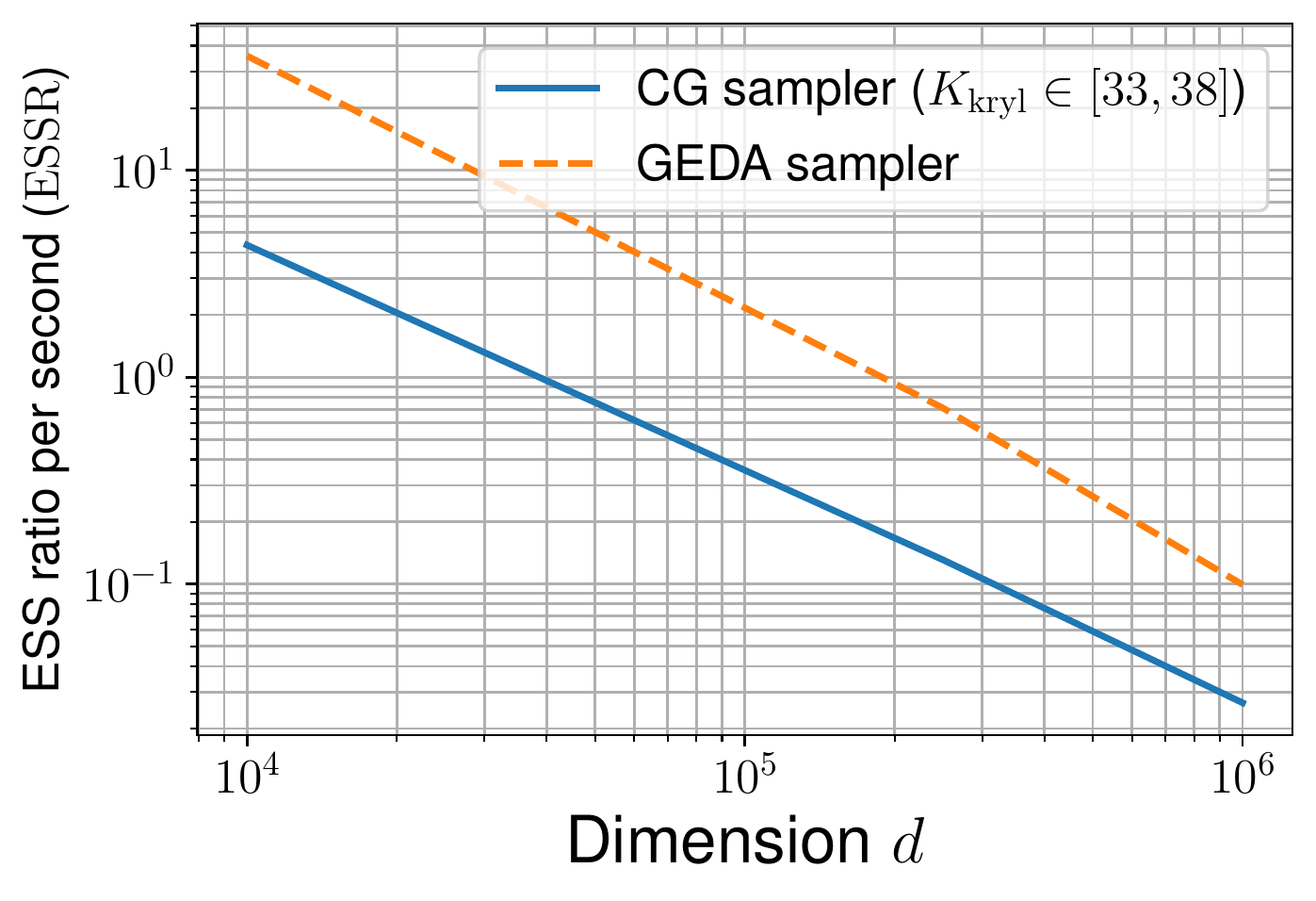}}}
  \mbox{{\includegraphics[scale=0.45]{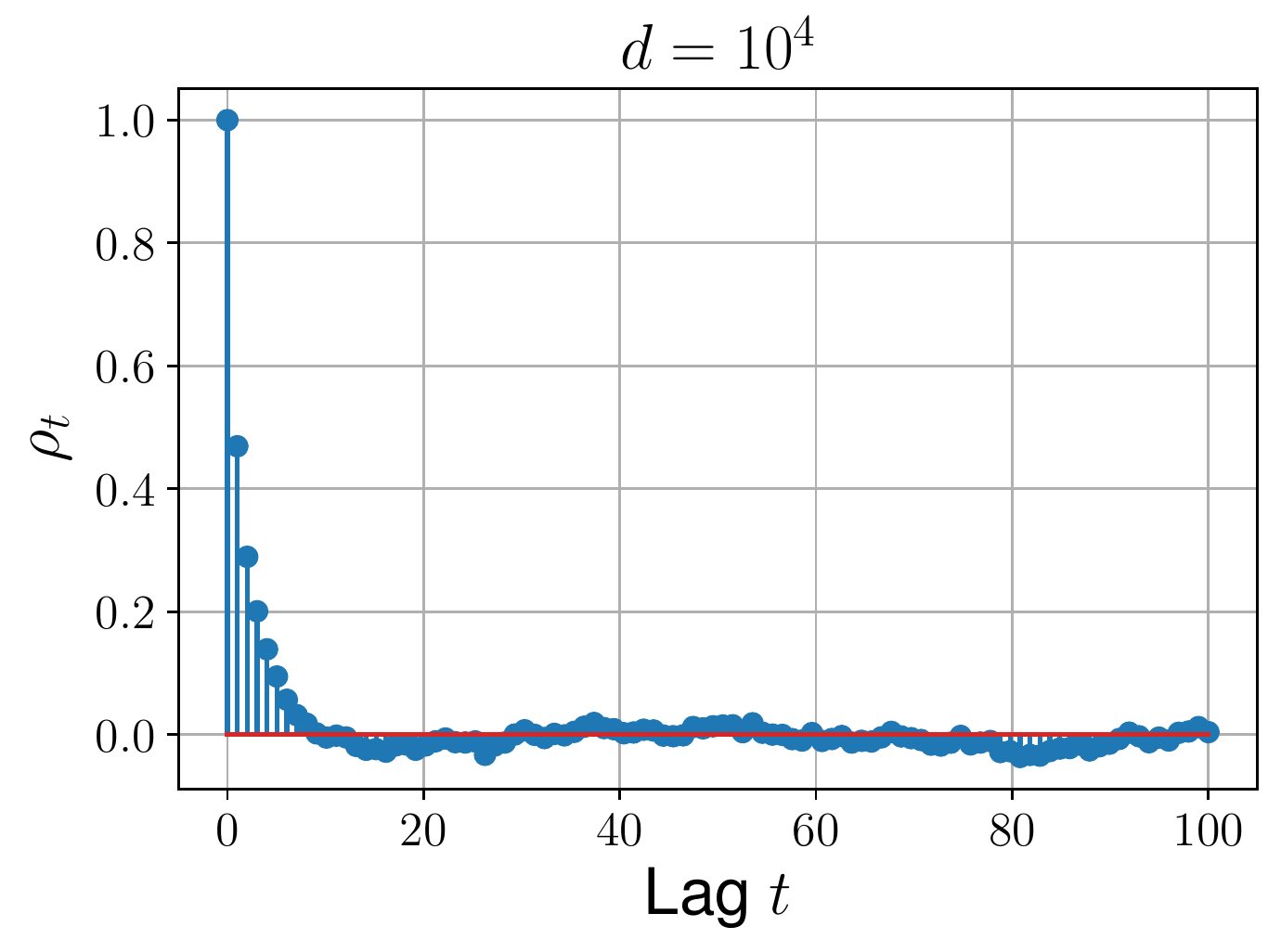}}}
  \caption{Scenario 3. (left) ESS ratio per second (ESSR); (right) autocorrelation function $\rho_t$ shown for $d=10^4$.
  For both figures, we used the slowest component of $\Bs{\theta}$ as the scalar summary $\vartheta$.}
\label{fig:scenario_3_results}
\end{figure}

\cref{fig:scenario_3_results} shows the ESSR associated to the two considered algorithms for $d \in [10^4,10^6]$.
The latter has been computed by choosing the ``slowest'' component of $\Bs{\theta}$ as the scalar summary $\vartheta$, that is the one with the largest variance.
As in the statistical software \textsc{Stan} \cite{Stan2017}, we truncated the infinite sum in \cref{eq:ESS} at the first negative $\rho_t$.
One can note that for the various high-dimensional problems considered here, the GEDA sampler exhibits good mixing properties which, combined with its low computational cost per iteration: it yields a larger ESSR than the direct CG sampler.
Hence, in this specific case, building on both the decomposition $\B{Q} = \B{Q}_1 + \B{Q}_2$ of the precision matrix and an efficient MCMC sampler is highly beneficial compared to directly using $\B{Q}$ in \cref{algo:CG}.
Obviously, this gain in computational efficiency w.r.t. direct samplers is not guaranteed in general since GEDA relies on an appropriate decomposition $\B{Q} = \B{Q}_1 + \B{Q}_2$.

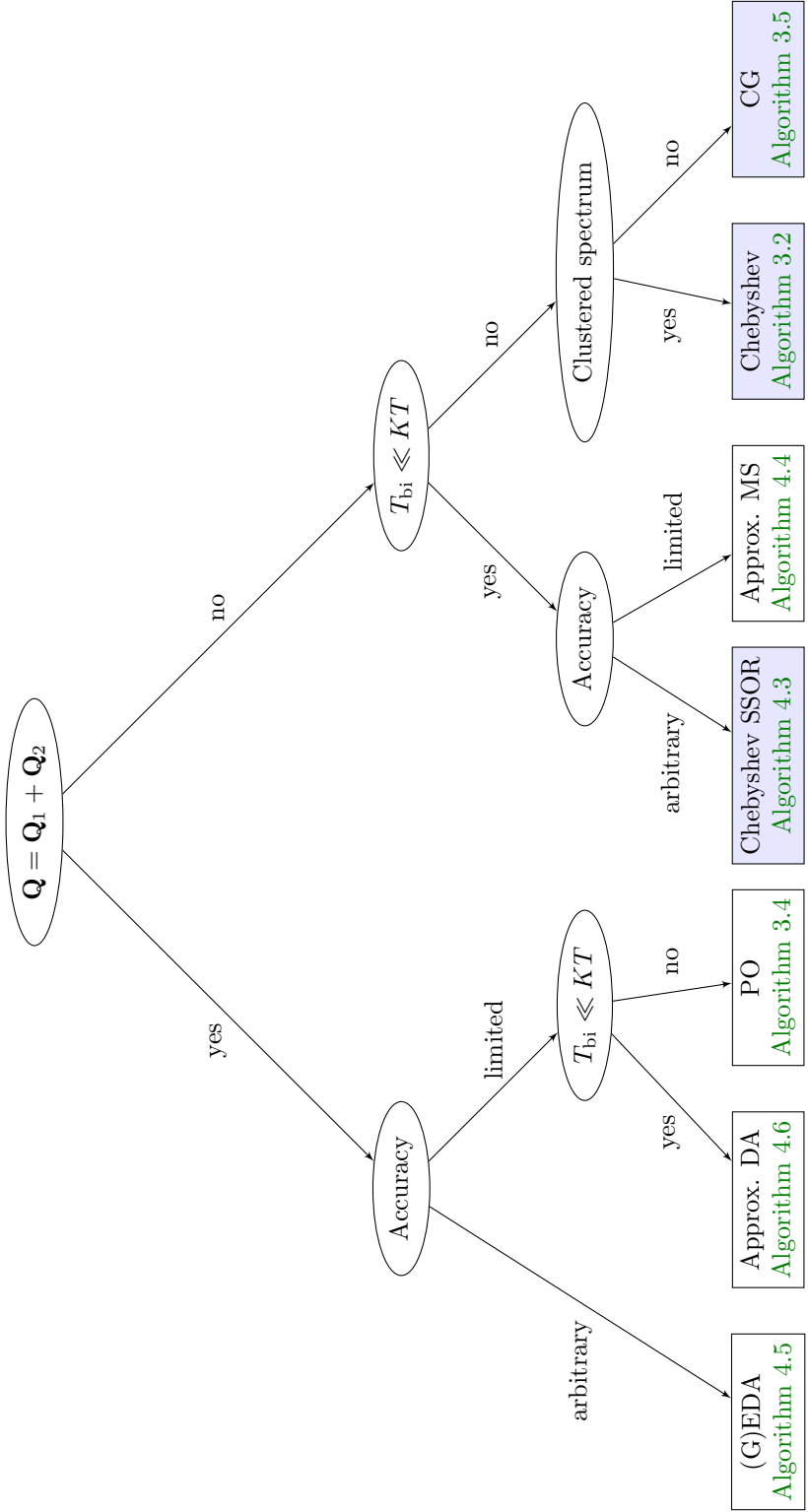
\begin{sidewaysfigure}
\tikzstyle{line} = [draw, -latex']
\begin{tikzpicture}[node distance=1.5cm,
                  align=center]

  \node (node0) [ellipse,draw=black] {$\B{Q} = \B{Q}_1 + \B{Q}_2$};

  \node (node00) [ellipse,draw=black,below left of=node0, node distance=7cm] {Accuracy};
  \path [line] (node0) -- node [midway,align=center,left=0.3em] {yes} (node00);
  
  \node (node01) [ellipse,draw=black,below right of=node0, node distance=7cm] {$T_{\mathrm{bi}} \ll K T$};
  \path [line] (node0) -- node [midway,align=center,right=0.3em] {no} (node01);
  
  \node (node010) [ellipse,draw=black,below left of=node01, node distance=3.5cm] {Accuracy};
  \path [line] (node01) -- node [midway,align=center,left=0.3em] {yes} (node010);
  
  \node (node011) [ellipse,draw=black,below right of=node01, node distance=3.5cm] {Clustered spectrum};
  \path [line] (node01) -- node [midway,align=center,right=0.3em] {no} (node011);
  
  \node (node0111) [rectangle,fill=blue!10,draw=black,below right of=node011, node distance=3.5cm] {CG \\ \cref{algo:CG}};
  \path [line] (node011) -- node [midway,align=center,right=0.3em] {no} (node0111);
  
  \node (node0110) [rectangle,fill=blue!10,draw=black,left of=node0111, node distance=3cm] {Chebyshev \\ \cref{algo:Chebychev}};
  \path [line] (node011) -- node [midway,align=center,left=0.3em] {yes} (node0110);
  
  \node (node0101) [rectangle,draw=black,left of=node0110, node distance=3cm] {Approx. MS \\ \cref{algo:approx_matrix_splitting}};
  \path [line] (node010) -- node [midway,align=center,right=0.3em] {limited} (node0101);
  
  \node (node0100) [rectangle,fill=blue!10,draw=black,left of=node0101, node distance=3cm] {Chebyshev SSOR \\ \cref{algo:acceleratedSSOR}};
  \path [line] (node010) -- node [midway,align=center,left=0.3em] {arbitrary} (node0100);
  
  \node (node001) [ellipse,draw=black,below right of=node00, node distance=3.5cm] {$T_{\mathrm{bi}} \ll K T$};
  \path [line] (node00) -- node [midway,align=center,right=0.3em] {limited} (node001);
  
  \node (node0011) [rectangle,draw=black,left of=node0100, node distance=3cm] {PO \\
  \cref{algo:PO}};
  \path [line] (node001) -- node [midway,align=center,right=0.3em] {no} (node0011);
  
  \node (node0010) [rectangle,draw=black,left of=node0011, node distance=3cm] {Approx. DA \\
  \cref{algo:approx_DA}};
  \path [line] (node001) -- node [midway,align=center,left=0.3em] {yes} (node0010);
  
  \node (node000) [rectangle,draw=black,left of=node0010, node distance=3cm] {(G)EDA \\ \cref{algo:exact_DA}};
  \path [line] (node00) -- node [midway,align=center,left=0.3em] {arbitrary} (node000);
  
  \end{tikzpicture}
\caption{General guidelines to choose the most appropriate sampling approach based on those reviewed in this paper. 
When computation of order $\Theta(d^3)$ and storage of $\Theta(d^2)$ elements are not prohibitive, the sampler of choice is obviously \cref{algo:factorization_sampler}.
The accuracy of a sampler is referred to \emph{arbitrary} when it is expected to obtain samples with arbitrary accuracy more efficiently than vanilla Cholesky sampling; the accuracy is said to be \emph{limited} when this is not the case.
The integer parameters $T_{\mathrm{bi}}$, $T$ and $K$ respectively refer to the burn-in period of MCMC approaches, the number of desired samples from $\mathcal{N}(\Bs{\mu},\B{Q}^{-1})$ and the truncation parameter associated to samplers of \cref{sec:PIGauss_M_zero}.
The nodes in blue color highlight that non-asymptotic convergence bounds were available when this review was written.
}
\label{fig:guidelines}
\end{sidewaysfigure}

\subsection{Guidelines to choose the appropriate Gaussian sampling approach}
\label{subsec:discussion}

In this section, we provide the reader with some insights about how to choose the most appropriate sampler for a given Gaussian simulation task when vanilla Cholesky sampling cannot be envisioned.
These guidelines are formulated as simple binary questions of a decision tree, see \cref{fig:guidelines}, to determine which class of samplers is of potential interest. 
We choose to start\footnote{Note that alternative decision trees could be built by considering other issues as the primary decision level.} from the existence of a natural decomposition $\B{Q} = \B{Q}_1 + \B{Q}_2$ since some existing approaches are specifically dedicated to this scenario.
Then, we discriminate existing sampling approaches based on several criteria which have been discussed throughout this review such as the prescribed accuracy, the number of desired samples or the eigenvalue spectrum of the precision matrix $\B{Q}$.
Regarding sampling accuracy, we highlight sampling approaches that are expected or not to yield samples with arbitrary accuracy more efficiently than vanilla Cholesky sampling, see \cref{table:complexity}.
MCMC approaches introduced in \cref{sec:PIGauss_M_nonzero} and iterative i.i.d. samplers of \cref{sec:PIGauss_M_zero} are distinguished depending on the number of desired samples $T$ and their relative efficiency measured via the burn-in period $T_{\mathrm{bi}}$ for MCMC samplers and via a truncation parameter $K \in \mathbb{N}^*$ for i.i.d. samplers.
The last guidelines follow from remarks highlighted in \cref{sec:PIGauss_M_zero} and \cref{sec:PIGauss_M_nonzero} and from the numerical results in \cref{subsec:results}.

As already mentioned in \cref{subsec:problem}, we emphasize that this review only aims at referring to the main approaches dedicated to high-dimensional Gaussian sampling which arises in many different contexts.
Therefore, it remains difficult to enunciate precise rules for each context. Thus, the guidelines in \cref{fig:guidelines} correspond to general principles to guide the  practitioner towards an appropriate class of sampling approaches which is reviewed and complemented by additional references provided in this paper. 

\section{Conclusion}
\label{sec:conclusion}

Given the ubiquity of the Gaussian distribution and the huge number of related contributions, this paper aimed at proposing an up-to-date review of the main approaches dedicated to high-dimensional Gaussian sampling in a single venue.
To this purpose, we first presented the Gaussian sampling problem at stake as well as its specific and already-reviewed instances. Then we pointed out the main difficulties when the associated covariance matrix is not structured and the dimension of the problem increases.  
We reviewed two main classes of approaches from the literature, namely approaches derived from numerical linear algebra and those based on MCMC sampling.
In order to help practitioners in choosing the most appropriate algorithm for a given sampling task, we compared the reviewed methods by highlighting and illustrating their respective pros and cons. Eventually, we provided general insights about how to select one of the most appropriate samplers by proposing a decision tree, see \cref{fig:guidelines}.
On top of that, we also unified most of the reviewed MCMC approaches under a common umbrella by building upon a stochastic counterpart of the celebrated proximal point algorithm that is well known in optimization.
This permitted to shed a new light on existing sampling approaches and draw further links between them.
To promote reproducibility, this article is completed by a companion package written in \textsc{Python}\xspace named $\mathsf{PyGauss}$\footnote{\url{http://github.com/mvono/PyGauss}}; it implements all the reviewed approaches.

\section*{Acknowledgments}
The authors would like to thank Dr. Jérôme Idier (LS2N, France) and Prof. Jean-Christophe Pesquet (CentraleSup\'elec, France) for relevant feedbacks on a earlier version of this paper. They are also grateful to the Editor and two anonymous reviewers whose comments helped to significantly improve the quality of the paper.

\section*{Funding}
Part of this work has been supported by the ANR-3IA Artificial and Natural Intelligence Toulouse Institute (ANITI) under grant agreement ANITI ANR-19-PI3A-0004. P. Chainais is supported by his 3IA Chair Sherlock funded by the ANR, under grant agreement ANR-20-CHIA-0031-01, Centrale Lille, the I-Site Lille-Nord-Europe and the region Hauts-de-France.

\newpage
\appendix

\section{Guide to notations}
\label{appendix_notations}
The following table lists and defines all the notations used in this paper.

  \begin{tabular}{ll}

    \multirow{2}{*}{$\mathcal{N}(\cdot\mid\Bs{\mu},\B{\Sigma})$} & Multivariate Gaussian probability distribution \\
    & with mean $\Bs{\mu}$ and covariance matrix $\B{\Sigma}$. \\[0.5em]

    $\B{M}^{\top}$ & Transpose of matrix $\B{M}$. \\[0.5em]

    $f(d) = \mathcal{O}(d)$ & Order of the function $f$ when $d \rightarrow \infty$ up to constant factors. \\[0.5em]
    
    \multirow{2}{*}{$f(d) = \Theta(d)$} & There exists $C_1,C_2 \in \mathbb{R}$ such that $C_1 d \le f(d) \le C_2 d$ \\ & when $d \rightarrow \infty$. \\[0.5em]

    $\mathrm{det}(\B{M})$ & Determinant of the matrix $\B{M}$. \\[0.5em]

    $\triangleq$ & By definition. \\[0.5em]

    $\B{0}_d$ & Null vector on $\mathbb{R}^d$. \\[0.5em]

    $\B{0}_{d \times d}$ & Null matrix on $\mathbb{R}^{d \times d}$. \\[0.5em]

    $\B{I}_d$ & Identity matrix on $\mathbb{R}^{d \times d}$. \\[0.5em]

    $\nr{\cdot}$ & The $L^2$ norm. \\[0.5em]

    \multirow{2}{*}{$\mathrm{diag}(\B{v})$} & The $d \times d$ diagonal matrix \\
    & with diagonal elements $\B{v} = (v_1,\hdots,v_d)^{\top}$. \\[0.5em]
    
    $\B{Q} = \B{M} - \B{N}$ & Matrix splitting decomposition of the precision matrix $\B{Q}$. \\[0.5em]

  \end{tabular}

\section{Details and proofs associated to \cref{subsec:PIGauss}}
\label{appendix_sec_2_subsec_PPA}
First, we briefly recall some useful definitions associated to monotone operators.
For more information about the theory of monotone operators in Hilbert spaces, we refer the interested reader to the book \cite{Bauschke2017}.\\[0.5em]

\noindent\textbf{General definitions.}

\begin{definition}[Operator]
  Let the notation $2^{\mathbb{R}^d}$ stands for the family of all subsets of $\mathbb{R}^d$.
  An operator or multi-valued function $\mathsf{K}: \mathbb{R}^d \rightarrow 2^{\mathbb{R}^d}$ maps every point in $\mathbb{R}^d$ to a subset of $\mathbb{R}^d$.
\end{definition}

\begin{definition}[Graph]
  Let $\mathsf{K}: \mathbb{R}^d \rightarrow 2^{\mathbb{R}^d}$.
  The \textit{graph} of $\mathsf{K}$ is defined by
  \begin{equation}
    \mathrm{gra}(\mathsf{K}) = \{(\Bs{\theta},\B{u}) \in \mathbb{R}^d \times \mathbb{R}^d \mid \B{u} \in \mathsf{K}(\Bs{\theta})\}.
  \end{equation}
\end{definition}

\begin{definition}[Monotone operator]
  Let $\mathsf{K}: \mathbb{R}^d \rightarrow 2^{\mathbb{R}^d}$.
  $\mathsf{K}$ is said to be monotone if
  \begin{equation}
    \forall (\Bs{\theta},\B{u}) \in \mathrm{gra}(\mathsf{K}) \text{ and } \forall (\B{y},\B{p}) \in \mathrm{gra}(\mathsf{K}), \langle \Bs{\theta}-\B{y} , \B{u}-\B{p}\rangle \geq 0.
  \end{equation}
\end{definition}

\begin{definition}[Maximal monotone operator]
  Let $\mathsf{K}: \mathbb{R}^d \rightarrow 2^{\mathbb{R}^d}$ be monotone. 
  Then $\mathsf{K}$ is maximal monotone if there exists no monotone operator $\mathsf{P}$ : $\mathbb{R}^d \rightarrow 2^{\mathbb{R}^d}$ such that $\mathrm{gra}(\mathsf{P})$ properly contains $\mathrm{gra}(\mathsf{K})$, i.e., for every $(\Bs{\theta},\B{u}) \in \mathbb{R}^d \times \mathbb{R}^d$,
  \begin{equation}
    (\Bs{\theta},\B{u}) \in \mathrm{gra}(\mathsf{K}) \Leftrightarrow \forall (\B{y},\B{p}) \in \mathrm{gra}(\mathsf{K}), \langle \Bs{\theta}-\B{y} , \B{u}-\B{p}\rangle \geq 0.
  \end{equation}
\end{definition}

\begin{definition}[Nonexpansiveness]
  Let $\mathsf{K}: \mathbb{R}^d \rightarrow 2^{\mathbb{R}^d}$.
  Then $\mathsf{K}$ is nonexpansive if it is Lipschitz continuous with constant 1, i.e, for every $(\Bs{\theta},\B{y}) \in \mathbb{R}^d \times \mathbb{R}^d$,
  \begin{equation}
    \nr{\mathsf{K}(\B{y})-\mathsf{K}(\Bs{\theta})} \leq \nr{\B{y}-\Bs{\theta}},
  \end{equation}
  where $\nr{\cdot}$ is the standard Euclidean norm.
\end{definition}

\begin{definition}[Domain]
  Let $\mathsf{K}: \mathbb{R}^d \rightarrow 2^{\mathbb{R}^d}$.
  The \textit{domain} of $\mathsf{K}$ is defined by
  \begin{equation}
    \mathrm{dom}(\mathsf{K}) = \{\Bs{\theta}\in \mathbb{R}^d \mid \mathsf{K}(\Bs{\theta})\neq \varnothing\}.
  \end{equation}
\end{definition}

\text{}\\[0.5em]\noindent\textbf{The PPA.}
For $\lambda > 0$, let define the Moreau-Yosida resolvent operator associated to $\mathsf{K}$ as the operator $\mathsf{L}$ defined by
\begin{equation}
  \mathsf{L} = (\mathrm{Id} + \lambda \mathsf{K})^{-1},
\end{equation}
where $\mathrm{Id}$ is the identity operator.
The monotonicity of $\mathsf{K}$ implies that $\mathsf{L}$ is nonexpansive and its maximal monotonicity yields $\mathrm{dom}(\mathsf{L}) = \mathbb{R}^d$ \cite{Minty1962}, where the notation ``$\mathrm{dom}$'' stands for the domain of the operator $\mathsf{L}$.
Therefore, solving the problem \cref{eq:zeros_operators} is equivalent to solve the fixed point problem for all $\Bs{\theta} \in \mathbb{R}^d$,
\begin{equation}
  \Bs{\theta} = \mathsf{L}(\Bs{\theta}).
\end{equation}
This result suggests that finding the zeros of $\mathsf{K}$ can be achieved by building a sequence of iterates $\{\Bs{\theta}^{(t)}\}_{t\in\mathbb{N}}$ such that for $t \in \mathbb{N}$,
\begin{equation}
  \Bs{\theta}^{(t+1)} = (\mathrm{Id} + \lambda \mathsf{K})^{-1}(\Bs{\theta}^{(t)}).
\end{equation}
This iteration corresponds to the PPA with an arbitrary monotone operator $\mathsf{K}$.\\[0.5em]

\noindent\textbf{Proof of \cref{eq:ADMM_x_PPA_2}.}
Applying the PPA with $\B{R} = \B{W} - \rho^{-1}\B{A}^{\top}\B{A}$ to \cref{eq:ADMM_x} leads to 
{\small
\begin{align*}
  \Bs{\theta}^{(t+1)} &= \underset{\Bs{\theta}\in\mathbb{R}^d}{\arg \min}\ g_2(\Bs{\theta}) + \dfrac{1}{2\rho}\nr{\B{A}\Bs{\theta} - \B{z}^{(t+1)} + \B{u}^{(t)}}^2 + \dfrac{1}{2}\nr{\Bs{\theta}-\Bs{\theta}^{(t)}}^2_{\B{R}} \\
  &= \underset{\Bs{\theta}\in\mathbb{R}^d}{\arg \min}\ g_2(\Bs{\theta}) + \dfrac{1}{2\rho}\nr{\B{A}\Bs{\theta} - \B{z}^{(t+1)} + \B{u}^{(t)}}^2 + \dfrac{1}{2}\left\langle \B{R} \pr{\Bs{\theta}-\Bs{\theta}^{(t)}}, \Bs{\theta}-\Bs{\theta}^{(t)}\right\rangle \\
  &= \underset{\Bs{\theta}\in\mathbb{R}^d}{\arg \min}\ g_2(\Bs{\theta}) + \dfrac{1}{2}\pr{\Bs{\theta}^{\top}\br{\dfrac{1}{\rho}\B{A}^{\top}\B{A} + \B{R}}\Bs{\theta} - 2\Bs{\theta}^{\top}\br{\dfrac{1}{\rho}\B{A}^{\top}\bbr{\B{z}^{(t+1)}-\B{u}^{(t)}} + \B{R}^{\top}\Bs{\theta}^{(t)}}} \\
  &= \underset{\Bs{\theta}\in\mathbb{R}^d}{\arg \min}\ g_2(\Bs{\theta}) + \dfrac{1}{2}\pr{\Bs{\theta}^{\top}\B{W}\Bs{\theta} - 2\Bs{\theta}^{\top}\br{\B{W}\Bs{\theta}^{(t)} + \dfrac{1}{\rho}\B{A}^{\top}\bbr{\B{z}^{(t+1)}-\B{u}^{(t)} - \B{A}\Bs{\theta}^{(t)}}}} \\
  &= \underset{\Bs{\theta}\in\mathbb{R}^d}{\arg \min}\ g_2(\Bs{\theta}) + \dfrac{1}{2}\nr{\Bs{\theta} - \pr{\Bs{\theta}^{(t)} + \dfrac{1}{\rho}\B{W}^{-1}\B{A}^{\top}\br{\B{z}^{(t+1)}-\B{u}^{(t)} - \B{A}\Bs{\theta}^{(t)}}}}^2_{\B{W}}.
\end{align*}
}

\section{Details associated to \cref{subsec:scenario2}}
\label{appendix:C}

The optimal value of the tuning parameter $\omega$ for the two matrix splitting schemes SOR and Richardson are given by 
\begin{equation}
  \omega_{\mathrm{SOR}}^* = \dfrac{2}{1 + \sqrt{1 - \rho\pr{\B{I}_d - \B{D}^{-1}\B{Q}}^2}}\eqsp,
\end{equation}
where $\B{D}$ stands for the diagonal part of $\B{Q}$.
Regarding the MCMC sampler based on Richardson splitting, we used the optimal value
\begin{equation}
  \omega_{\mathrm{Richardson}}^* = \dfrac{2}{\lambda_{\mathrm{min}}(\B{Q}) + \lambda_{\mathrm{max}}(\B{Q})}\eqsp,
\end{equation}
where $\lambda_{\mathrm{min}}(\B{Q})$ and $\lambda_{\mathrm{max}}(\B{Q})$ are the minimum and maximum eigenvalues of $\B{Q}$, respectively.
Finally, for the samplers based on SSOR splitting including \cref{algo:acceleratedSSOR}, we used the optimal relaxation parameter
\begin{equation}
  \omega_{\mathrm{SSOR}}^* = \dfrac{2}{1 + \sqrt{2(1 - \rho\pr{\B{D}^{-1}(\B{L}+\B{L}^{\top})})}}\eqsp,
\end{equation}
where $\B{L}$ is the strictly lower triangular part of $\B{Q}$.

\bibliographystyle{siamplain}
\bibliography{biblio}
\end{document}